\documentclass[a4paper,fleqn,usenatbib]{mnras}

\usepackage{newtxtext,newtxmath}

\usepackage[T1]{fontenc}
\usepackage{ae,aecompl}

\usepackage{graphicx}	

\usepackage{amsmath}	
\usepackage{amsfonts}
\usepackage{amsbsy}
\usepackage{amssymb}
\usepackage{makecell}
\usepackage{gensymb}
\usepackage{float}
\usepackage{enumerate}
\usepackage{lastpage}
\usepackage{natbib}
\usepackage{threeparttable}
\usepackage{array}
\usepackage{booktabs}
\usepackage{multirow}
\usepackage{longtable}
\usepackage{lscape}
\usepackage{tablefootnote}
\usepackage{url}

\usepackage{subfigure}

\title[NH$_2$D  distribution in MSF regions]{Spatial distribution of NH$_2$D  in massive star-forming regions}

\author[Yuqiang Li et al.]{Yuqiang Li$^{1,2}$,\thanks{E-mail: liyuqiang@shao.ac.cn} Junzhi Wang$^{3}$\thanks{E-mail: junzhiwang@gxu.edu.cn}, Juan Li$^{1}$\thanks{E-mail: lijuan@shao.ac.cn}, Shu Liu$^{4}$, Kai Yang$^{5}$, Siqi Zheng$^{1,2}$, Zhe Lu$^{6}$\\
$^{1}$Shanghai Astronomical Observatory, Chinese Academy of Sciences,80 Nandan Road, Shanghai 200030, China\\
$^{2}$University of Chinese Academy of Sciences, 19A Yuquanlu, Beijing 100049, China \\
$^{3}$Guangxi Key Laboratory for Relativistic Astrophysics, Department of Physics, Guangxi University, Nanning 530004, PR China\\
$^{4}$CAS Key Laboratory of FAST, National Astronomical Observatories, Chinese Academy of Sciences, Beijing 100012, China\\
$^{5}$School of Astronomy and Space Science, Nanjing University, 163 Xianlin Avenue, Nanjing 210023, China\\
$^{6}$Department of Electrical and Electronic Engineering, Guilin University of Technology at Nanning, 530001, China.\\
}

\date{Accepted XXX. Received YYY; in original form ZZZ}

\pubyear{2022}

\begin{document}
\label{firstpage}
\pagerange{\pageref{firstpage}--\pageref{lastpage}}
\maketitle

\begin{abstract}
To understand the relation between NH$_2$D and its physical environment, we mapped ortho-NH$_2$D $1_{11}^s-1_{01}^a$ at 85.9 GHz toward 24 Galactic late-stage massive star-forming regions with Institut de Radioastronomie Millim${\rm \acute{e}}$trique (IRAM) 30-m telescope. Ortho-NH$_2$D $1_{11}^s-1_{01}^a$  was detected in 18 of 24 sources.  Comparing with the distribution of H$^{13}$CN 1-0 as a dense gas tracer and radio recombination line H42$\alpha$, ortho-NH$_2$D $1_{11}^s-1_{01}^a$ present complex and diverse spatial distribution in these targets. 11 of the 18 targets,  present a different distribution between ortho-NH$_2$D  $1_{11}^s-1_{01}^a$ and H$^{13}$CN 1-0, while no significant difference between these two lines can be found in other 7 sources, mainly due to limited spatial resolution and sensitivity.  Moreover, with H42$\alpha$ tracing massive young stellar objects, ortho-NH$_2$D $1_{11}^s-1_{01}^a$ seems to show a relatively weak emission near the massive young stellar objects. 

\end{abstract}


\section{Introduction}
Deuterium chemistry is an active process in molecular clouds \citep{2008A&A...492..703C}. The deuterated molecules, such as DCN, DCO$^+$, N$_2$D$^+$ and NH$_2$D, can be formed at different evolution stages of star-forming \citep{2008A&A...492..703C}. Due to the relatively low zero-point energy, these deuterated molecules are generated more easily in molecular clouds \citep{1982A&A...111...76H}. Therefore, the deuterium fractionation ($D_{\rm frac}$), as the ratio between deuterated and normal molecule, is often higher than the D/H ratio of $\approx 1.5 \times 10^{-5}$ \citep{1995ApJ...451..335L,2003ApJ...587..235O}.

Deuterated ammonia (NH$_2$D, NHD$_2$ and ND$_3$) can form in different ways: 1) The deuterated ions (D) take the place of hydrogen (H) in the ammonia in the gas phase \citep{2001ApJ...553..613R,2008A&A...492..703C,2012A&ARv..20...56C,2014prpl.conf..859C}. 2) The reaction between deuterated ions and NH$_3$. For cold regions (T<30K), H$_2$D$^+$ react with NH$_3$, while in warm regions (T=30-80 K), the  CH$_2$D$^+$ and C$_2$HD$^+$ react with NH$_3$ to generate deuterated ammonia \citep{2000A&A...361..388R,2000A&A...364..780R,2012A&ARv..20...56C,2014prpl.conf..859C}. 3) The reactions with H, D and N atoms are also an effective approach \citep{1989MNRAS.240P..25B,2015MNRAS.446..449F}. 4) The chemical reaction of deuteration could happen on the surface of dust grains if the reactions of the gas phase are unavailable \citep{1990ApJ...362L..29T,2003ApJ...593L..51C,2008A&A...492..703C}. 5) In addition, shocks may be associated with the enhancement of deuterated species \citep{2002ApJ...569..322L}.

As a single deuterated form molecule, NH$_2$D was detected extensively in the molecular cloud. $D_{\rm frac}$(NH$_3$) was observed to be $\sim 1-0.1$ in the cold environment  \citep[e.g.][]{2003A&A...403L..25H,2007A&A...470..221C,2017A&A...600A..61H} and  $\sim 0.1-0.001$ in the warm environment  \citep[e.g.][]{2010A&A...517L...6B,2011A&A...530A.118P,2015A&A...581A..48G}. Besides, with the model calculations, deuterium fractionation is sensitive to its environment changes, such as temperature and density \citep{2000A&A...361..388R,2000A&A...364..780R}. 

According to the high-resolution radio observation in late-stage massive star-forming regions, a series of the researches have reported the relation between NH$_2$D and its physical environment. With the IRAM Plateau de Bure Interferometer (PdBI), \cite{2010A&A...517L...6B}  observed the ortho-NH$_2$D $1_{11}^s-1_{01}^a$ at 85.9263  GHz in the ultracompact H\uppercase\expandafter{\romannumeral2} region (UCH\uppercase\expandafter{\romannumeral2} region) IRAS 20293+3952,  through which they reported that the $D_{\rm frac}$(NH$_3$) in young stellar objects (YSOs) was lower than in pre-protostellar cores, and the $D_{\rm frac}$(NH$_3$) could be related to the evolutionary stage in the pre-protostellar phase. With  AtacamaLarge Millimeter/Submillimeter Array (ALMA) observation of Orion KL Hot Core and Compact Ridge for  NH$_2$D at 239.848 GHz, \cite{2013ApJ...777...85N}  found that the deuterium fractionation in Orion KL Hot Core was lower than that in Compact Ridge, suggesting that the hot core originated in a slightly
warmer environment result in low deuterium fractionation. \cite{2018ApJ...869..121M} reported that the $D_{\rm frac}$(NH$_3$) of Sgr B2(N2) is $\approx 10$ times higher than that of Sgr B2(N1) with VLA observations of the NH$_2$D $4_{14}^s-4_{04}^a$ transitions at 25.0 GHz and ATCA observations of the NH$_2$D $2_{12}^a-2_{02}^s$ transitions at 49.9 GHz and the NH$_2$D $3_{13}^s-3_{03}^a$ transitions at 43.0 GHz, so that they attributed the abundance variations to the shock and fast warmup in Sgr B2(N2). However, the researches mentioned above are only based on one or two sources \citep{2010A&A...517L...6B,2013ApJ...777...85N,2018ApJ...869..121M}.  Consequently, a relatively large sample of massive star-forming regions is needed to understand the relation between NH$_2$D and its physical environment.

NH$_2$D $1_{11}^a-1_{01}^s$ at 110.1535 GHz with single point observations using Institut de Radioastronomie Millim${\rm \acute{e}}$trique (IRAM) 30-m telescope toward 50 late-stage massive star-forming regions was reported with 72\% detection rate \citep{2022MNRAS.512.4934L}. With stronger emission \citep[e.g.][]{2001ApJ...554..933S,2021A&A...649A..21W}, ortho-NH$_2$D $1_{11}^s-1_{01}^a$ at 85.9263 GHz is more suitable than para-NH$_2$D $1_{11}^a-1_{01}^s$ at 110.1535 GHz to derive  NH$_2$D spatial distribution. Therefore, we mapped the ortho-NH$_2$D $1_{11}^s-1_{01}^a$ at 85.9263 GHz in late-stage high mass star-forming regions with IRAM 30-m to obtain spatial distribution information of NH$_2$D.

With IRAM 30-m telescope observation, this work presents the results of ortho-NH$_2$D $1_{11}^s-1_{01}^a$ at 85.9263 GHz mapping toward a relatively large sample of 18 Galactic late-stage massive star-forming regions with known 6.7 GHz CH$_3$OH masers. The observations are described in Section \ref{observations}, the main results are reported in Section \ref{results}, a discussion is presented in Section \ref{discussion}, and a brief summary is given in Section \ref{summary}.

\section{Observations}
\label{observations}
The 24 targets, selected from \cite{2014ApJ...783..130R}, are late-stage massive star-forming regions with 6.7 GHz CH$_3$OH masers with HC$_3$N 12-11 higher than 1 K ($T_{\rm mb}$)  which is based on IRAM 30-m observations. The observations were performed with the Institut de Radioastronomie Millim${\rm \acute{e}}$trique (IRAM) 30-m telescope at Pico Veleta, Spain during 2019 July, 2019 October, 2019 November, 2020 December and 2021 January. The on-the-fly (OTF) mode was used in our observations with the frequency coverage about 8 GHz bandwidth. The system temperatures were 80-150 K during our observation. We used the E090 receiver to cover 84 GHz to 92 GHz and the Fourier Transform Spectrometers (FTS) backend with a 195 kHz channel spacing, which corresponds to 0.68 km s$^{-1}$ at 86 GHz. The map centers and map area are listed in table \ref{source}.

The antenna temperature ($T_{\rm A}^{\ast}$) is converted to the main beam brightness temperature ($T_{\rm mb}$), using $T_{\rm mb}$ = $T_{\rm A}^{\ast}\cdot$ $F_{\rm eff}/$$B_{\rm eff}$, the forward efficiency $F_{\rm eff}$ is 0.95 and the beam efficiency $B_{\rm eff}$ is 0.81. The used lines consist of ortho-NH$_2$D $1_{11}^s-1_{01}^a$ at 85.9263 GHz, H$^{13}$CN 1-0 at 86.3387 GHz and H42$\alpha$ at 85.6884 GHz. Data reduction was conducted with GILDAS software\footnote{\url{http://www.iram.fr/IRAMFR/GILDAS}}. The angular resolution of the IRAM 30-m telescope is about 29.3 arcsec at 84 GHz and about 26.7 arcsec at 92 GHz, while all sources were re-gridded to steps of 9 arcsecs, about 1/3 beam size. First-order baseline was used for all spectra. The typical rms are around 50 mK at 0.7 km s$^{-1}$ velocity resolution. In the following analysis, we assume that the beam-filling factor is unity.

\section{Results}
\label{results}

The ortho-NH$_2$D $1_{11}^s-1_{01}^a$ at 85.9263 GHz was detected among 18 targets of our sample, while it was not detected in the rest six targets: G009.62+00.19, G011.49-014.8, G012.80-00.20, G043.16+00.01, G133.94+01.06 and G192.60-00.04. For ortho-NH$_2$D detected targets, the H$^{13}$CN 1-0 was also detected, while H42$\alpha$  was detected only in nine targets.

11 of 18 sources, present a different distribution between ortho-NH$_2$D $1_{11}^s-1_{01}^a$ and H$^{13}$CN 1-0 with asymmetrically and resolvable distributed spatial structure. Due to limited sensitivity, the structure of ortho-NH$_2$D $1_{11}^s-1_{01}^a$ at 85.9263 GHz was unresolved in G005.88-00.39 and G188.94+00.88 (see Figure \ref{app1} and Figure \ref{app18}). For 4 sources, G011.91-00.61, G023.44-00.18, G034.39+00.22 and G049.48-00.38 (see Figure \ref{app2}, Figure \ref{app4}, Figure \ref{app6} and Figure \ref{app11}), no significant difference between these two lines can be found. It is mainly due to the limited spatial resolution. Only one target, G081.75+00.59 (see Figure \ref{app13}), presents a relatively similar distribution between ortho-NH$_2$D $1_{11}^s-1_{01}^a$ at 85.9263 GHz and H$^{13}$CN 1-0  and resolvable distributed spatial structure.

The detailed results of the three lines are given in Section \ref{NH2D distribution}-\ref{H42alpha maps result}, the ortho-NH$_2$D $1_{11}^s-1_{01}^a$, H$^{13}$CN 1-0 and H42$\alpha$ distribution details of each source are described in Section \ref{Individual target detail}.

\subsection{NH$_2$D spatial distribution}
\label{NH2D distribution}
\begin{figure}
\centering
\subfigure[]{\includegraphics[width=0.5\textwidth]{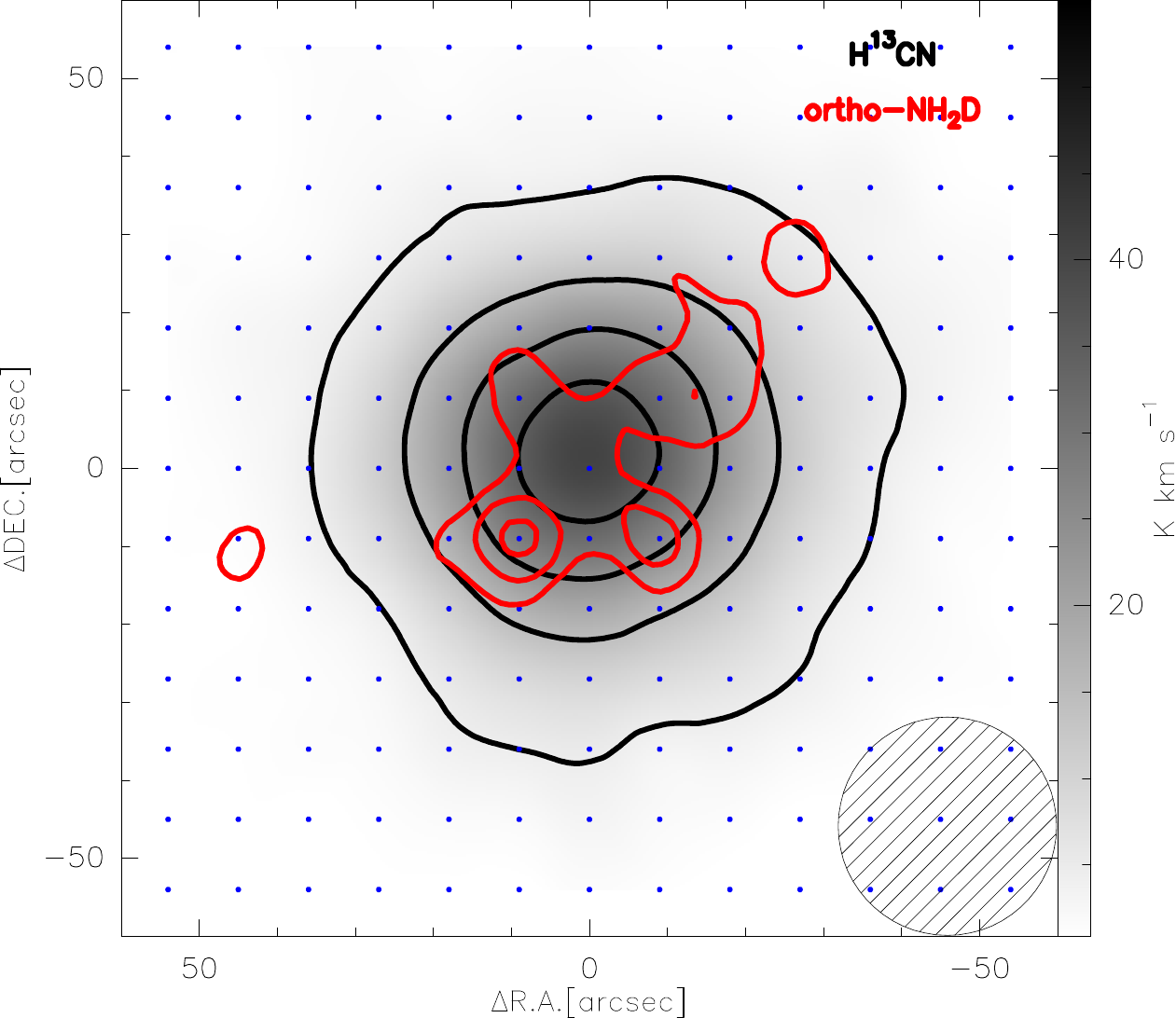}}\subfigure[]{\includegraphics[width=0.5\textwidth]{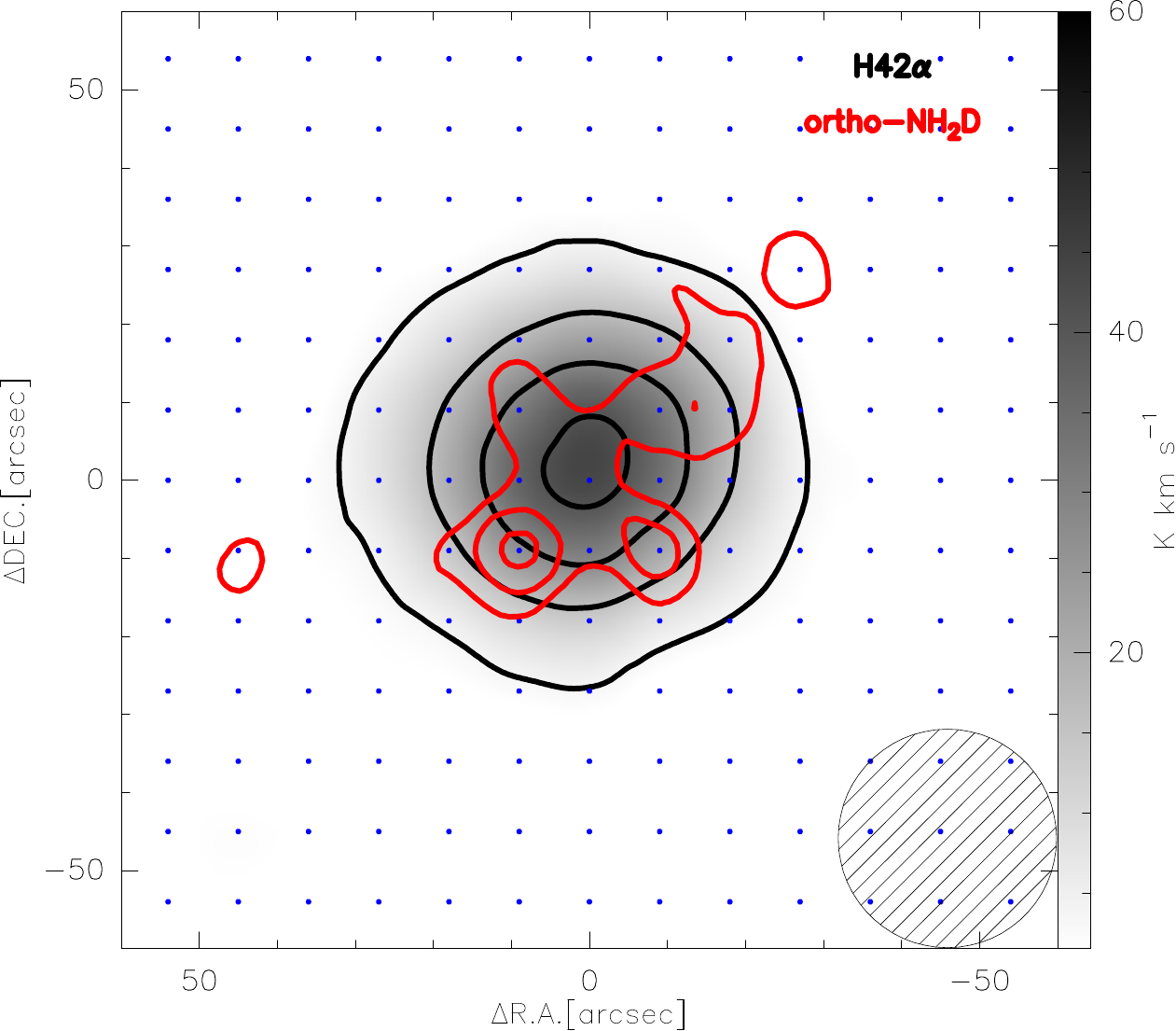}}\\
\subfigure[]{\includegraphics[width=0.5\textwidth]{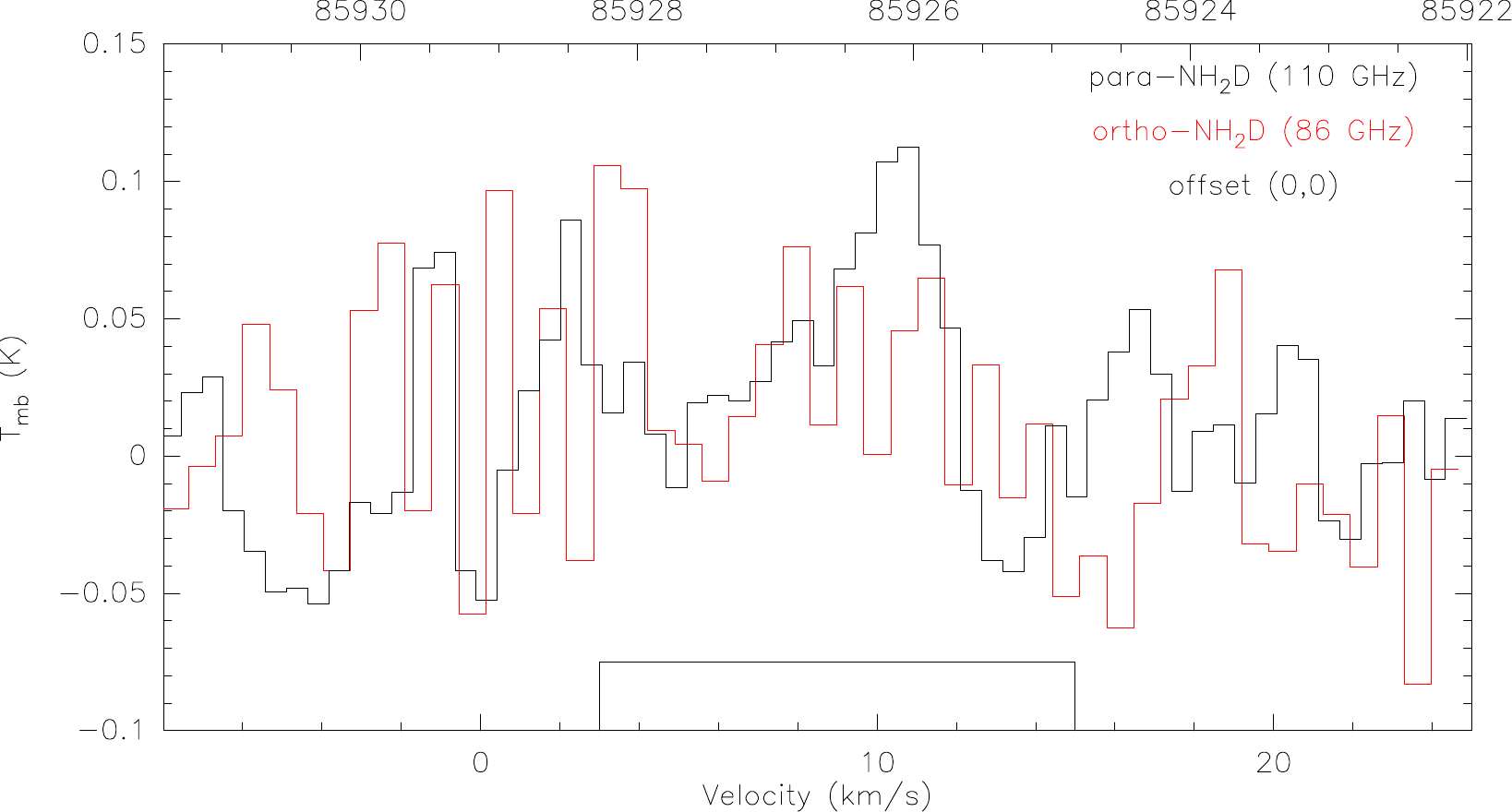}}\subfigure[]{\includegraphics[width=0.5\textwidth]{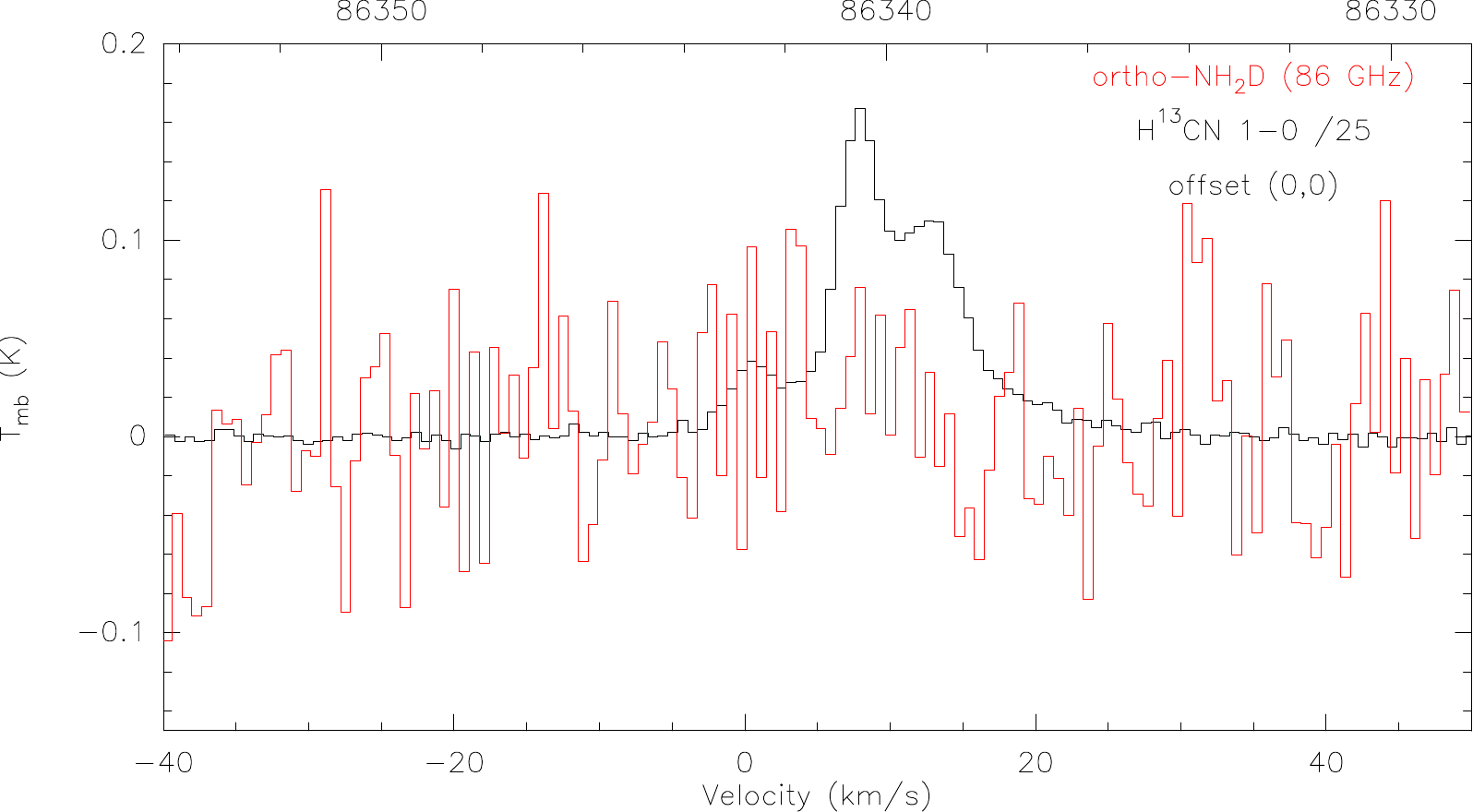}}
\caption{(a): NH$_2$D at 85.9263 GHz velocity integrated intensity contour (red contour) overlaid on H$^{13}$CN 1-0 velocity integrated intensity image (gray scale and black contour) in G005.88-00.39. The contour levels start at 2$\sigma$ in steps of 1$\sigma$ for NH$_2$D, while the contour levels start at 18$\sigma$ in steps of 36$\sigma$ for H$^{13}$CN 1-0. The gray scale starts at 3$\sigma$. (b): NH$_2$D at 85.9263 GHz velocity integrated intensity contour (red contour) overlaid on H42$\alpha$ velocity integrated intensity image (Gray scale and black contour) in G005.88-00.39. The contour levels start at 2$\sigma$ in steps of 1$\sigma$ for NH$_2$D, while the contour levels start at 7$\sigma$ in steps of 24$\sigma$ for H42$\alpha$. The gray scale starts at 3$\sigma$. (c): Spectra of ortho-NH$_2$D at 85.9263 GHz and para-NH$_2$D at 110.1535 GHz in G005.88-00.39. The para-NH$_2$D was observed by IRAM-30m with position-switching mode. The red spectra is ortho-NH$_2$D and the black is para-NH$_2$D. The offset for two spectra is (0$^{\prime \prime}$,0$^{\prime \prime}$). (d): Spectra of ortho-NH$_2$D at 85.9263 GHz and H$^{13}$CN 1-0 in G005.88-00.39. The red spectra is ortho-NH$_2$D and the black is H$^{13}$CN 1-0. The offset for two spectra is (0$^{\prime \prime}$,0$^{\prime \prime}$).}
\label{app1}
\end{figure}

\begin{figure}
\centering
\subfigure[]{\includegraphics[width=0.5\textwidth]{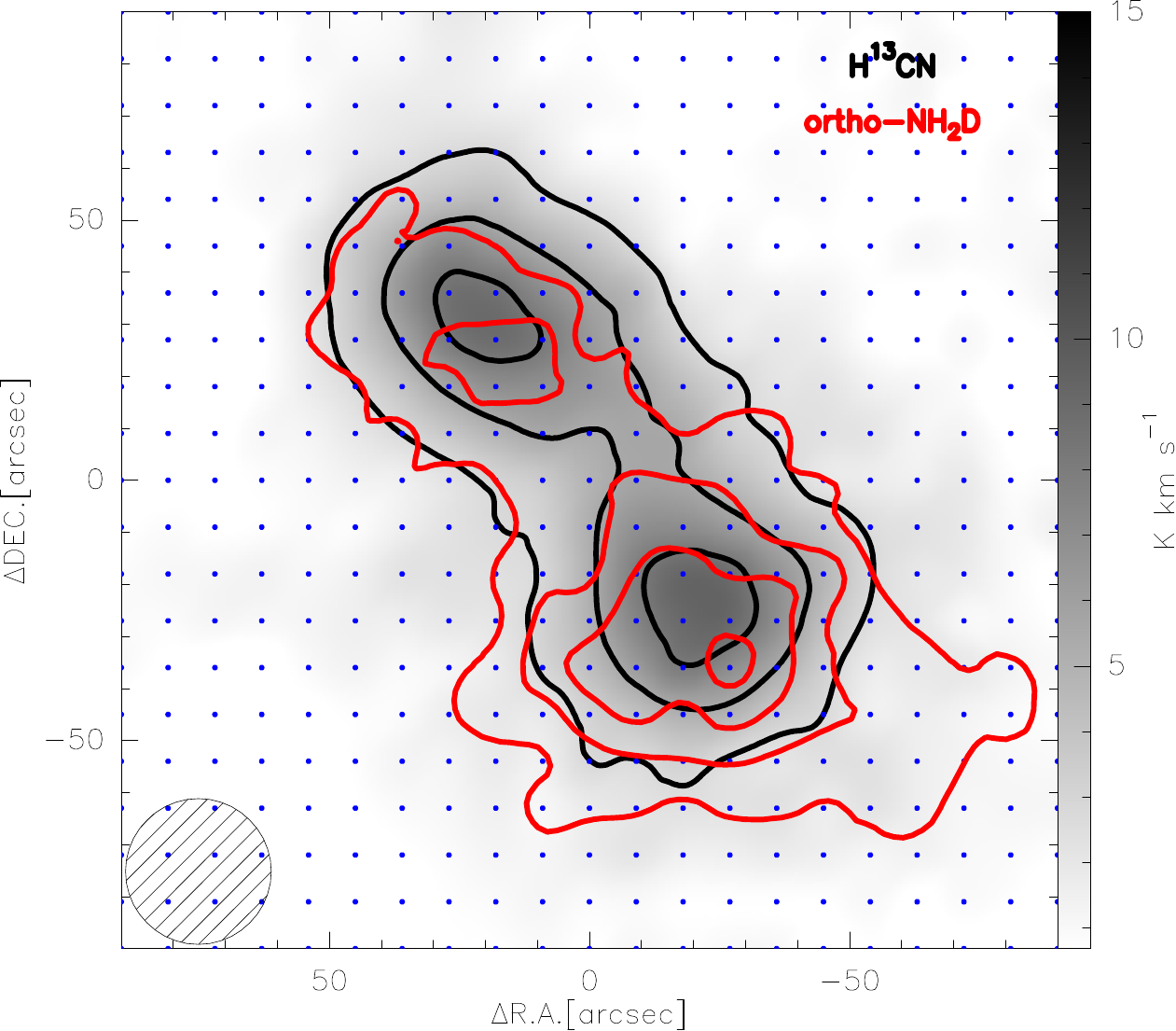}}\subfigure[]{\includegraphics[width=0.5\textwidth]{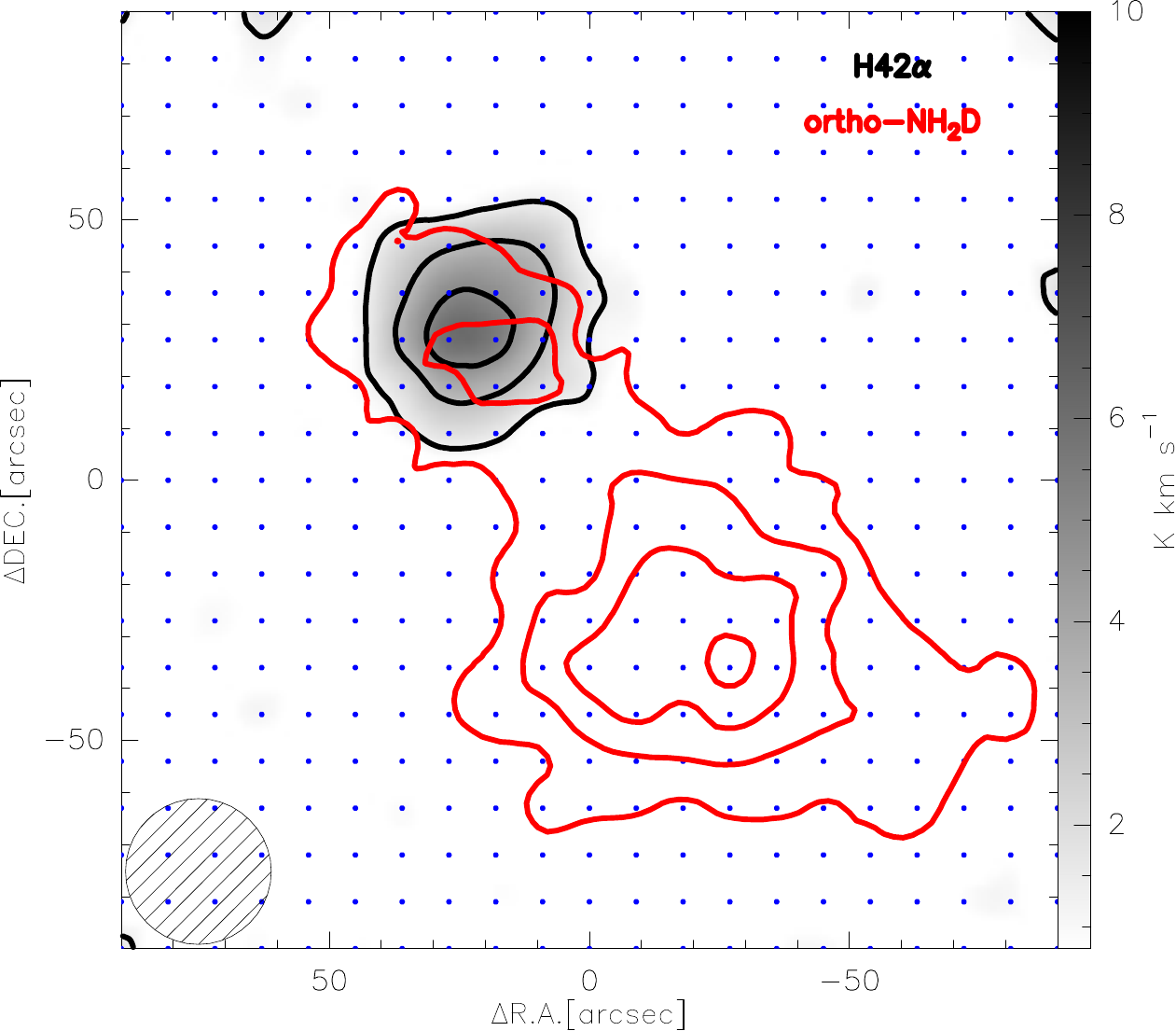}}
\subfigure[]{\includegraphics[width=0.5\textwidth]{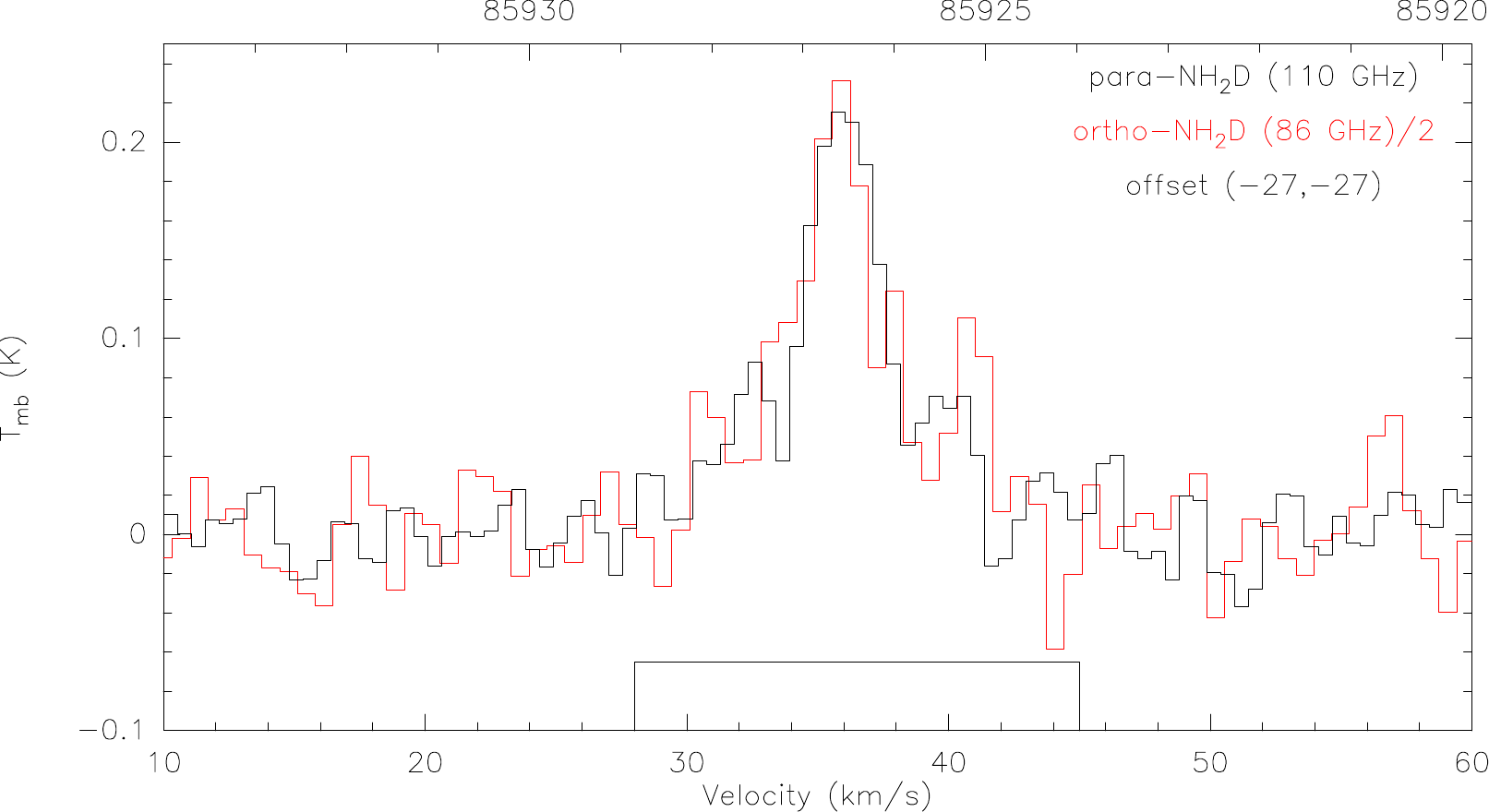}}\subfigure[]{\includegraphics[width=0.5\textwidth]{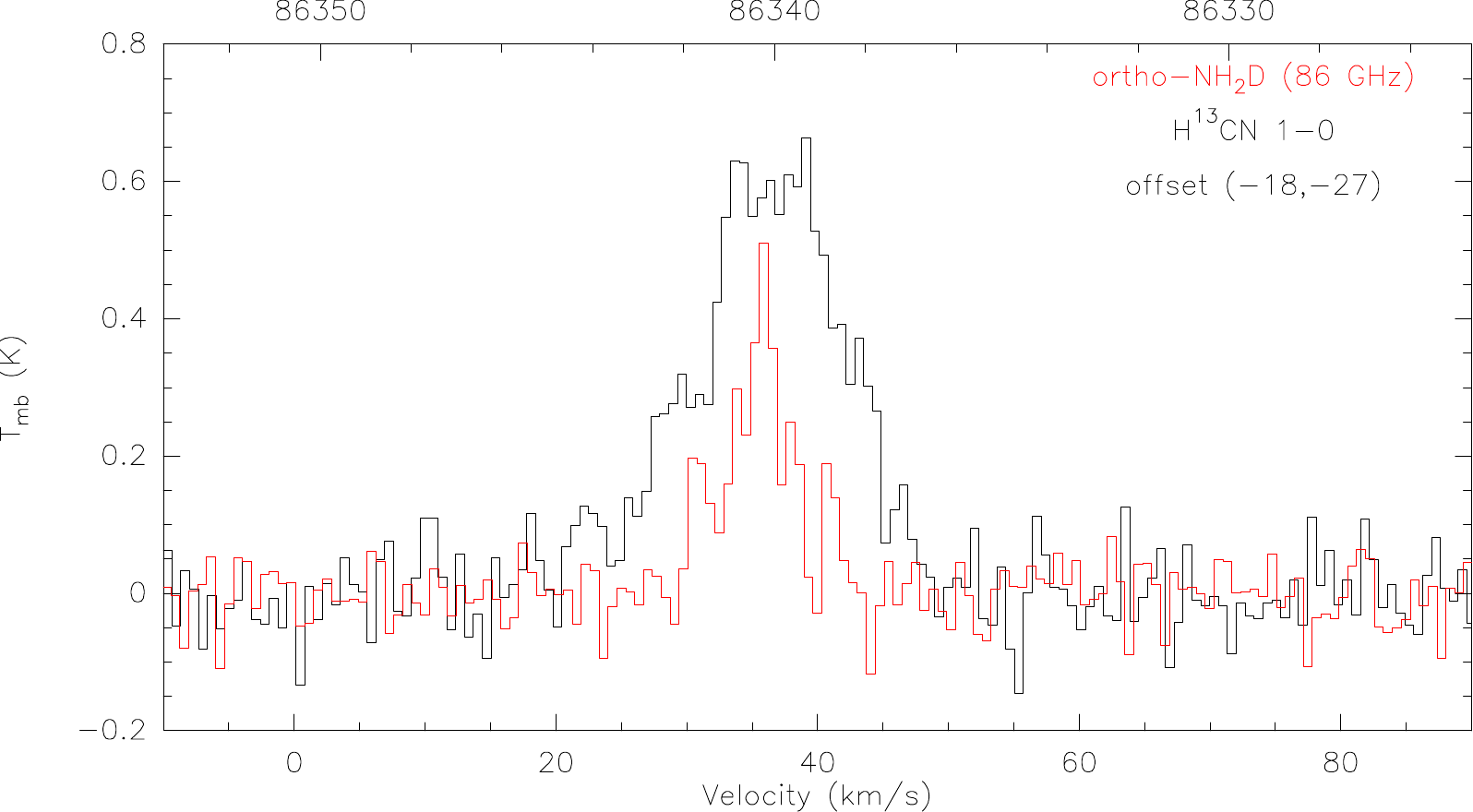}}
\subfigure[]{\includegraphics[width=0.5\textwidth]{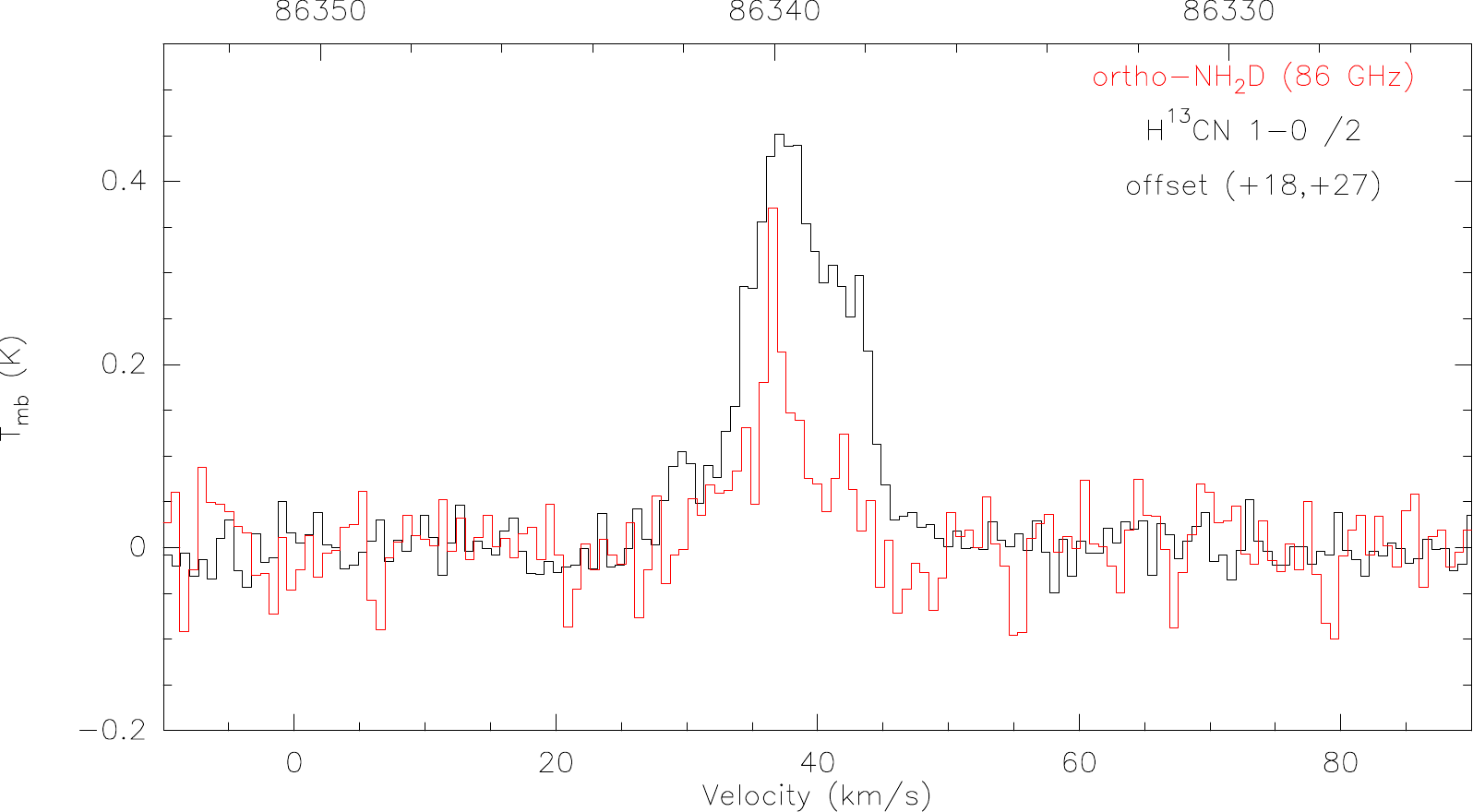}}
\caption{(a): NH$_2$D at 85.9263 GHz velocity integrated intensity contour (red contour) overlaid on H$^{13}$CN 1-0 velocity integrated intensity image (Gray scale and black contour) in G011.91-00.61. The contour levels start at 5$\sigma$ in steps of 4$\sigma$ for NH$_2$D, while the contour levels start at 12$\sigma$ in steps of 12$\sigma$ for H$^{13}$CN 1-0. The gray scale starts at 3$\sigma$. (b): NH$_2$D at 85.9263 GHz velocity integrated intensity contour (red contour) overlaid on H42$\alpha$ velocity integrated intensity image (Gray scale and black contour) in G011.91-00.61. The contour levels start at 5$\sigma$ in steps of 4$\sigma$ for NH$_2$D, while the contour levels start at 5$\sigma$ in steps of 7$\sigma$ for H42$\alpha$. The gray scale starts at 3$\sigma$. (c): Spectra of ortho-NH$_2$D at 85.9263 GHz and para-NH$_2$D at 110.1535 GHz in G011.91-00.61. The para-NH$_2$D was observed by IRAM-30m with position-switching mode. The red spectra is ortho-NH$_2$D and the black is para-NH$_2$D. The offset for two spectra is (-27$^{\prime \prime}$,-27$^{\prime \prime}$). The black box in the picture indicates the range of flux integration of ortho-NH$_2$D at 85.9263 GHz. (d): Spectra of ortho-NH$_2$D at 85.9263 GHz and H$^{13}$CN 1-0 in G011.91-00.61. The red spectra is ortho-NH$_2$D and the black is H$^{13}$CN 1-0. The offset for two spectra is (-18$^{\prime \prime}$,-27$^{\prime \prime}$). (e): Spectra of ortho-NH$_2$D at 85.9263 GHz and H$^{13}$CN 1-0 in G011.91-00.61.  The red spectra is ortho-NH$_2$D and the black is H$^{13}$CN 1-0. The offset for two spectra is (+18$^{\prime \prime}$,+27$^{\prime \prime}$).}
\label{app2}
\end{figure}

\begin{figure}
\centering
\subfigure[]{\includegraphics[width=0.5\textwidth]{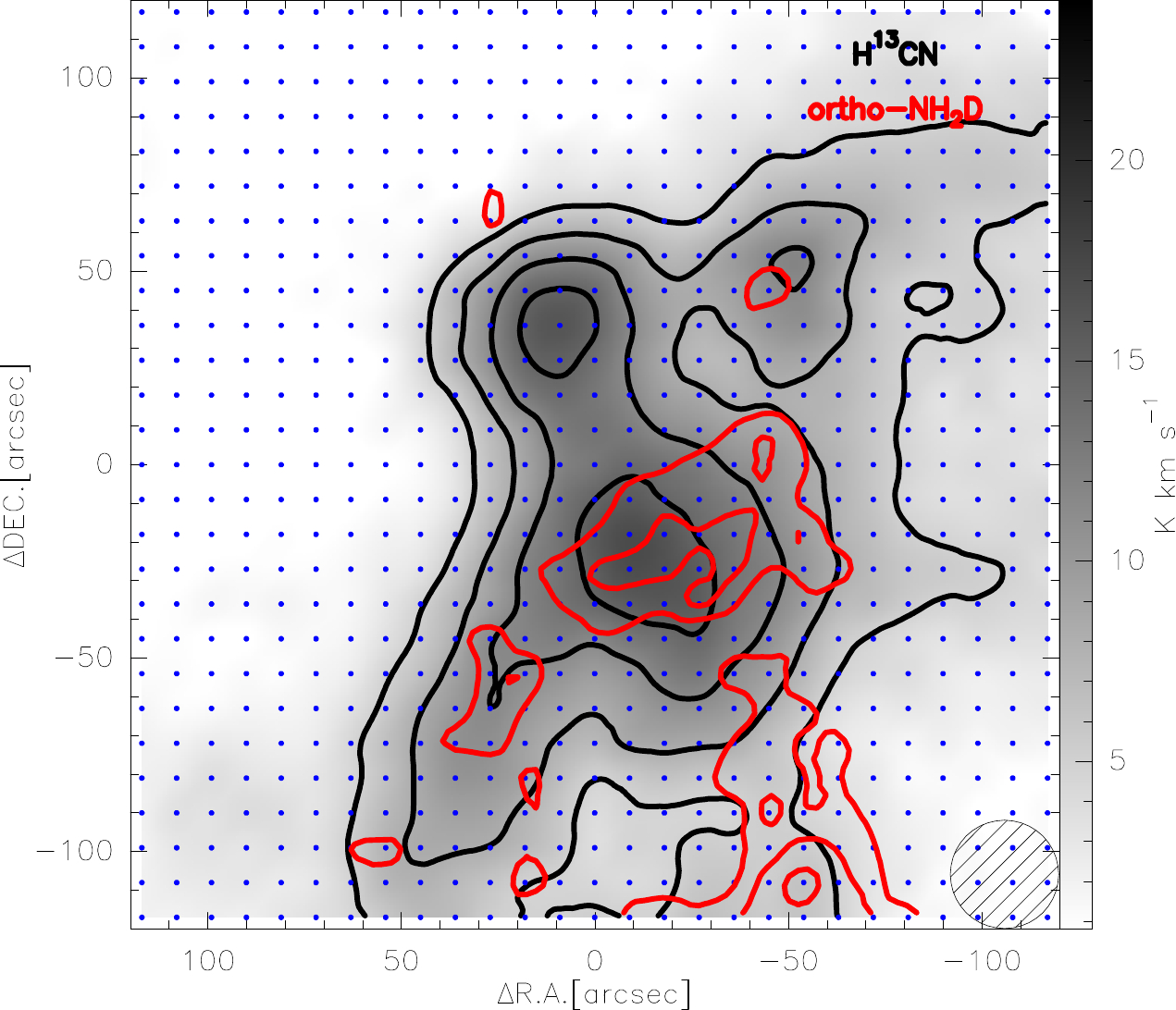}}\subfigure[]{\includegraphics[width=0.5\textwidth]{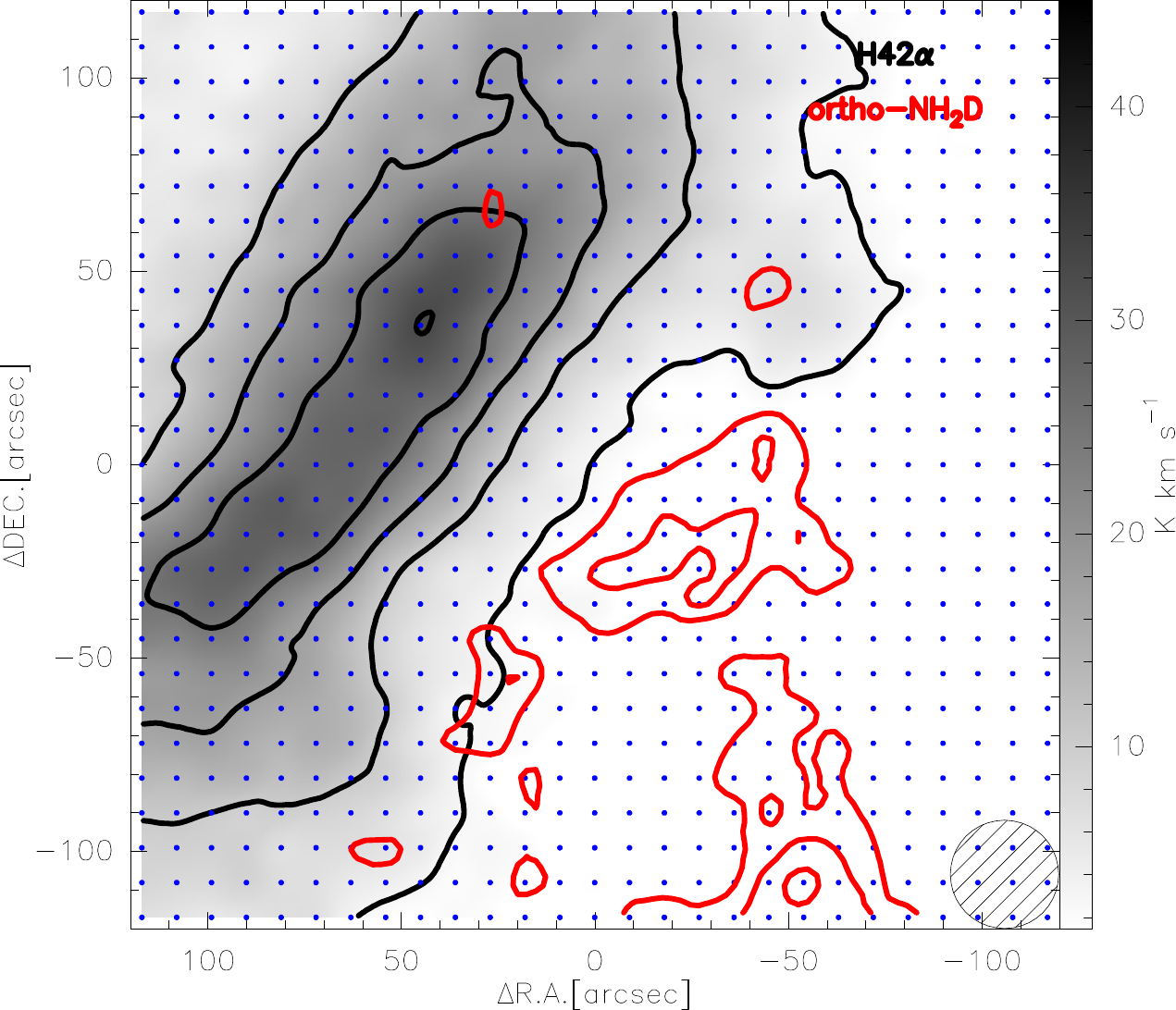}}
\subfigure[]{\includegraphics[width=0.5\textwidth]{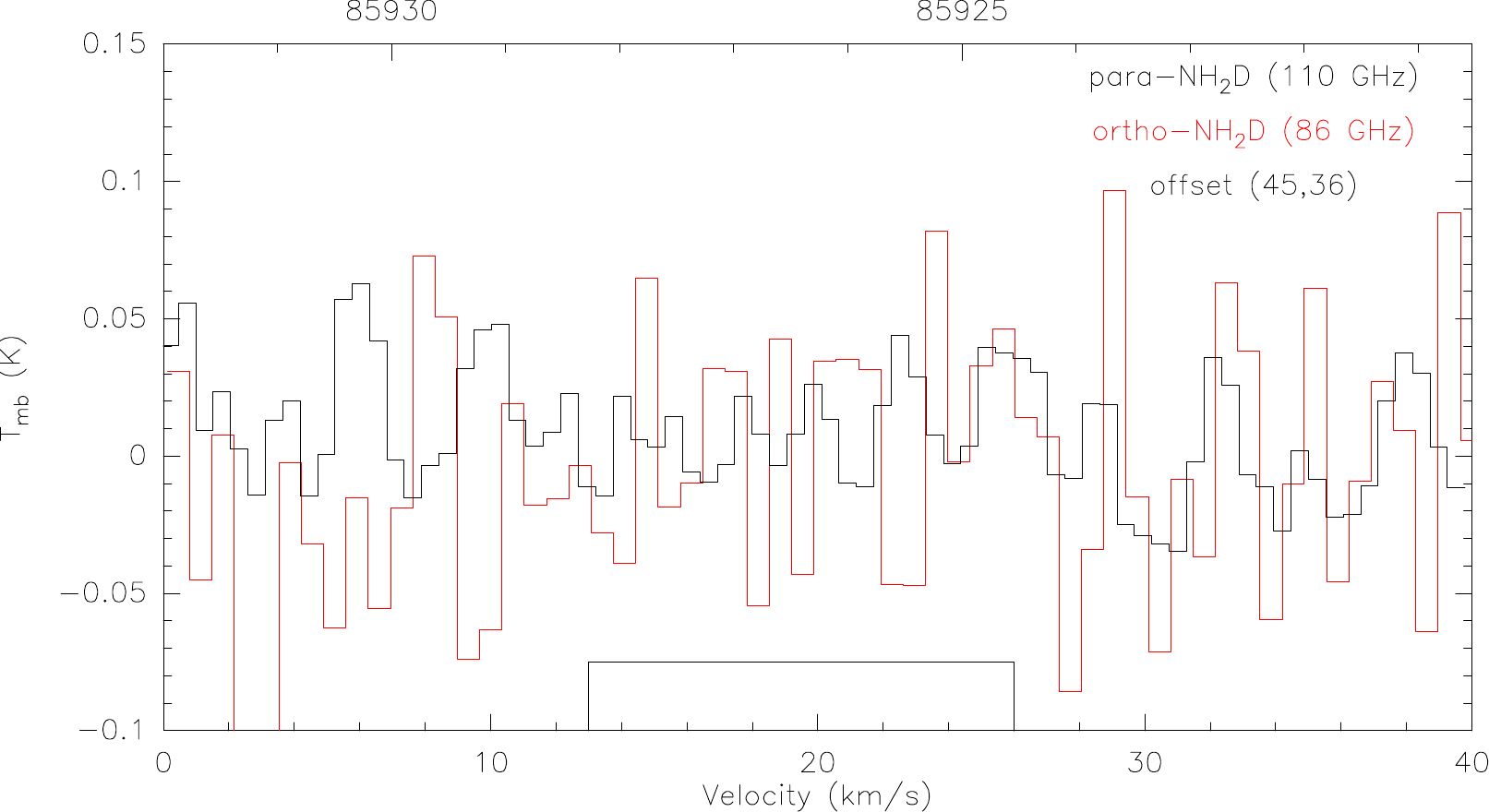}}\subfigure[]{\includegraphics[width=0.5\textwidth]{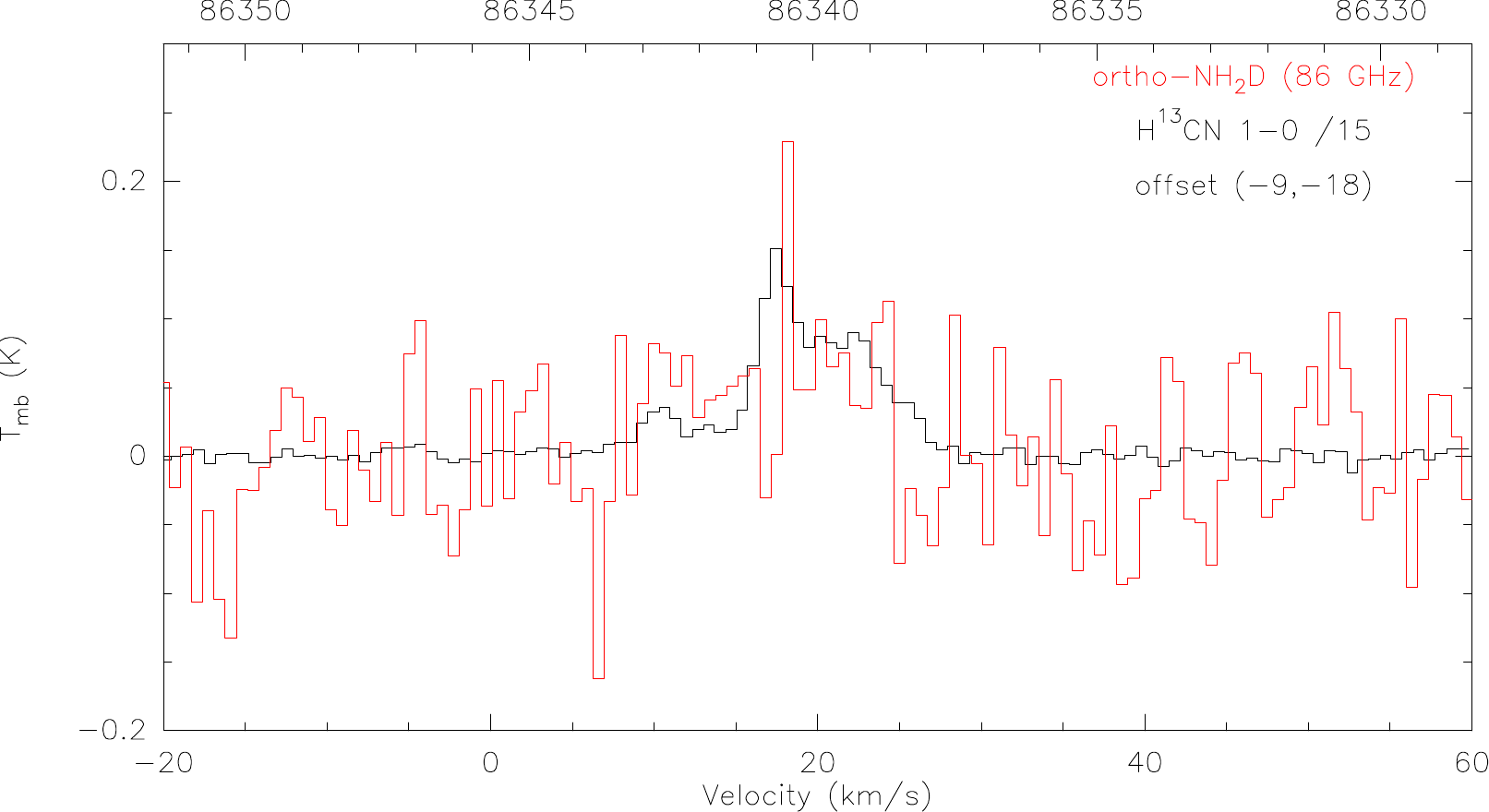}}
\subfigure[]{\includegraphics[width=0.5\textwidth]{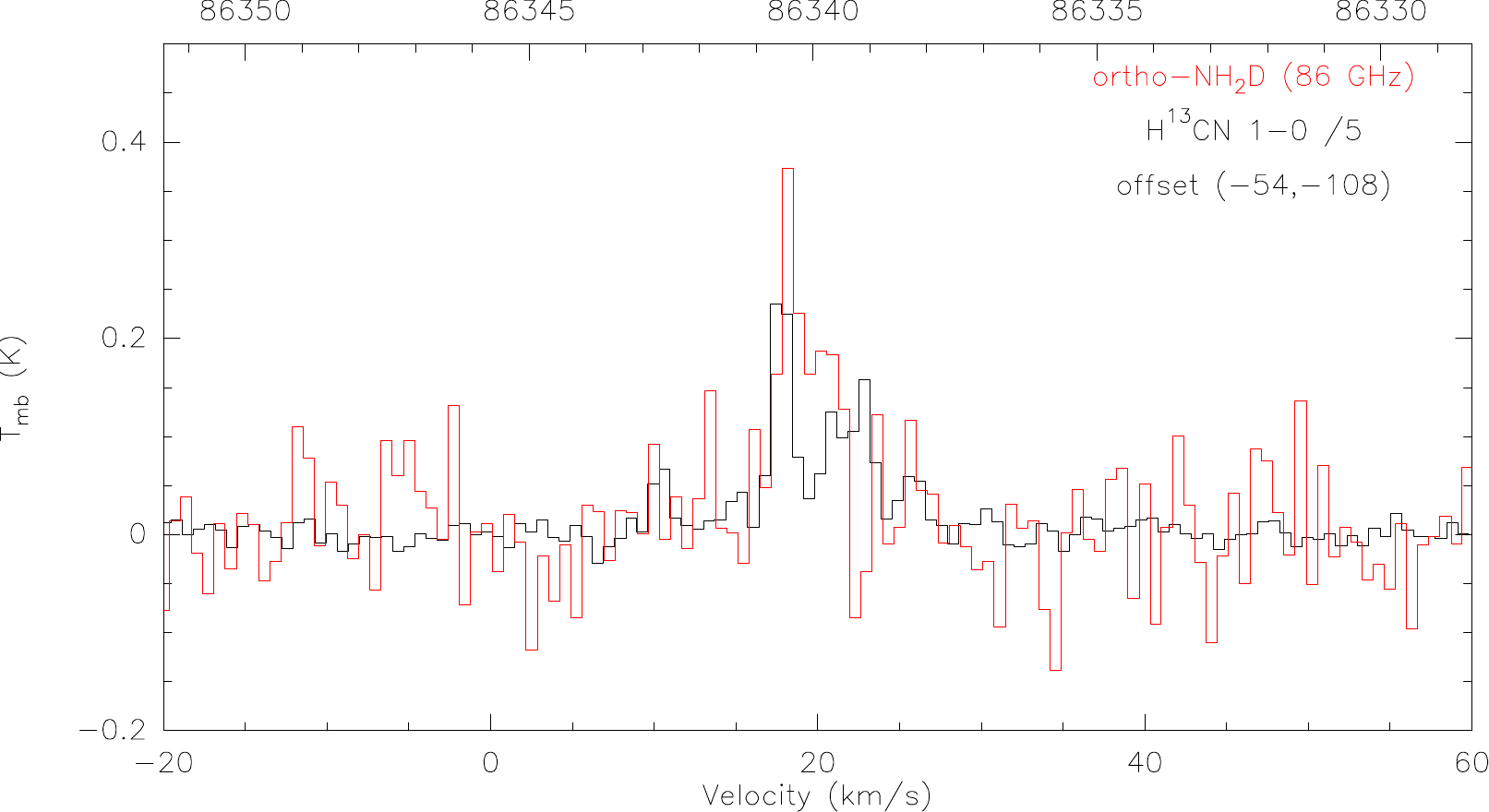}}
\caption{(a): NH$_2$D at 85.9263 GHz velocity integrated intensity contour (red contour) overlaid on H$^{13}$CN 1-0 velocity integrated intensity image (Gray scale and black contour) in G015.03-00.67. The contour levels start at 3$\sigma$ in steps of 2$\sigma$ for NH$_2$D, while the contour levels start at 18$\sigma$ in steps of 12$\sigma$ for H$^{13}$CN 1-0. The gray scale starts at 3$\sigma$. (b): NH$_2$D at 85.9263 GHz velocity integrated intensity contour (red contour) overlaid on H42$\alpha$ velocity integrated intensity image (Gray scale and black contour) in G015.03-00.67. The contour levels start at 3$\sigma$ in steps of 2$\sigma$ for NH$_2$D, while the contour levels start at 6$\sigma$ in steps of 15$\sigma$ for H42$\alpha$. The gray scale starts at 3$\sigma$. (c): Spectra of ortho-NH$_2$D at 85.9263 GHz and para-NH$_2$D at 110.1535 GHz in G015.03-00.67. The para-NH$_2$D was observed by IRAM-30m with position-switching mode. The red spectra is ortho-NH$_2$D and the black is para-NH$_2$D. The offset for two spectra is (45$^{\prime \prime}$,36$^{\prime \prime}$). The black box in the picture indicates the range of flux integration of ortho-NH$_2$D at 85.9263 GHz. (d): Spectra of ortho-NH$_2$D at 85.9263 GHz and H$^{13}$CN 1-0 in G015.03-00.67. The red spectra is ortho-NH$_2$D and the black is H$^{13}$CN 1-0. The offset for two spectra is (-9$^{\prime \prime}$,-18$^{\prime \prime}$). (e): Spectra of ortho-NH$_2$D at 85.9263 GHz and H$^{13}$CN 1-0 in G015.03-00.67.  The red spectra is ortho-NH$_2$D and the black is H$^{13}$CN 1-0. The offset for two spectra is (-54$^{\prime \prime}$,-108$^{\prime \prime}$).}
\label{app3}
\end{figure}

\begin{figure}
\centering
\subfigure[]{\includegraphics[width=0.5\textwidth]{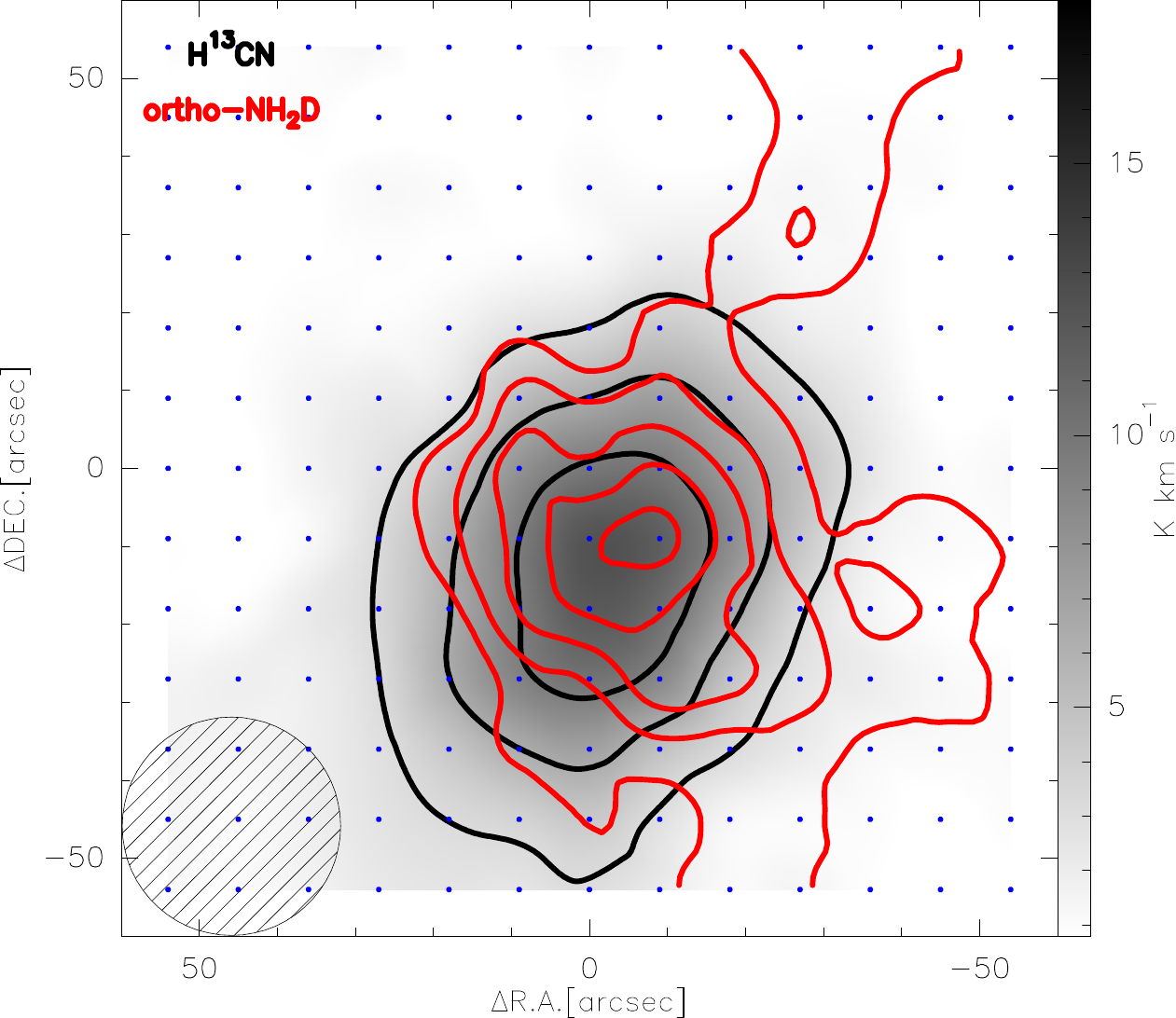}}\subfigure[]{\includegraphics[width=0.5\textwidth]{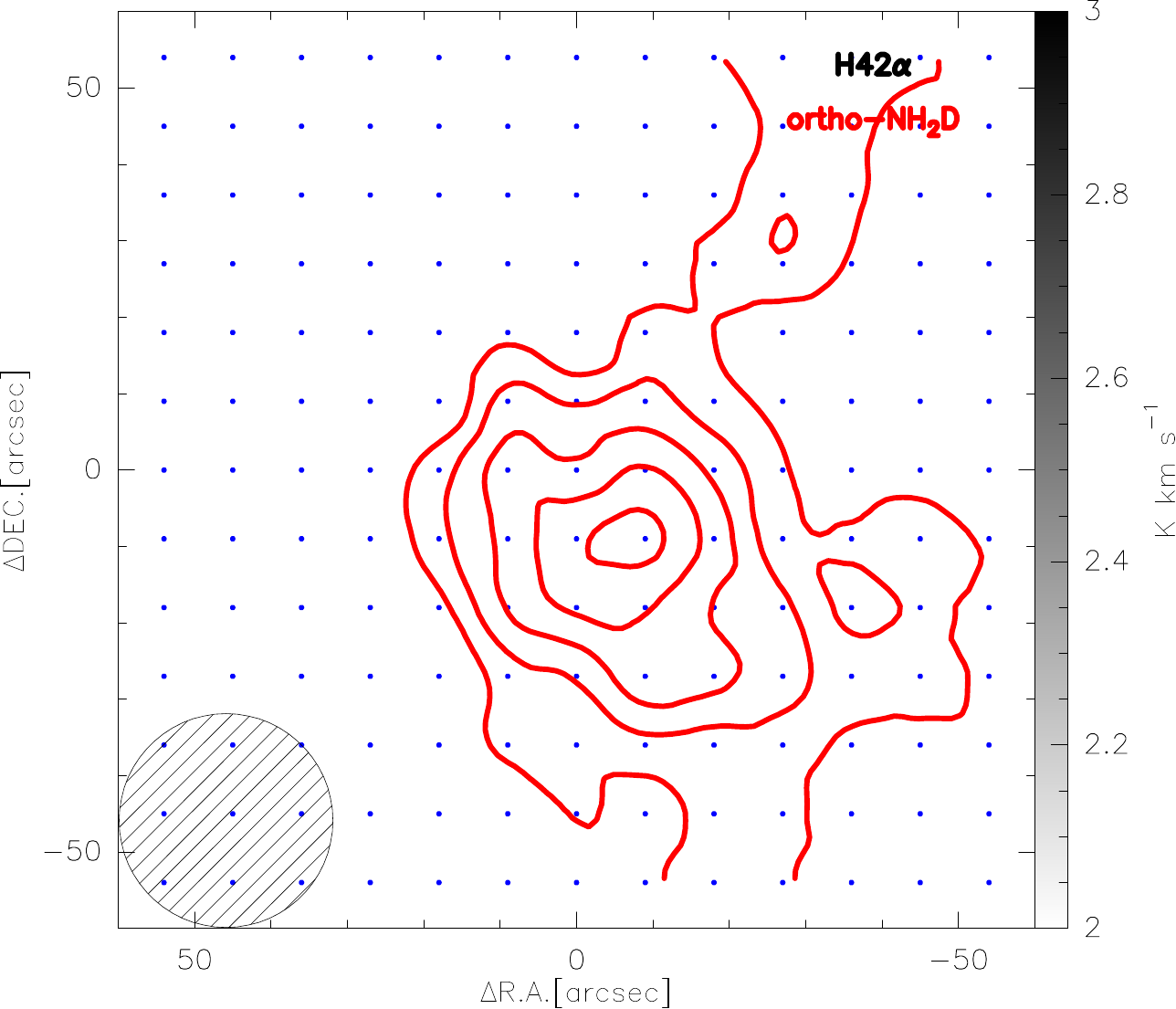}}
\subfigure[]{\includegraphics[width=0.5\textwidth]{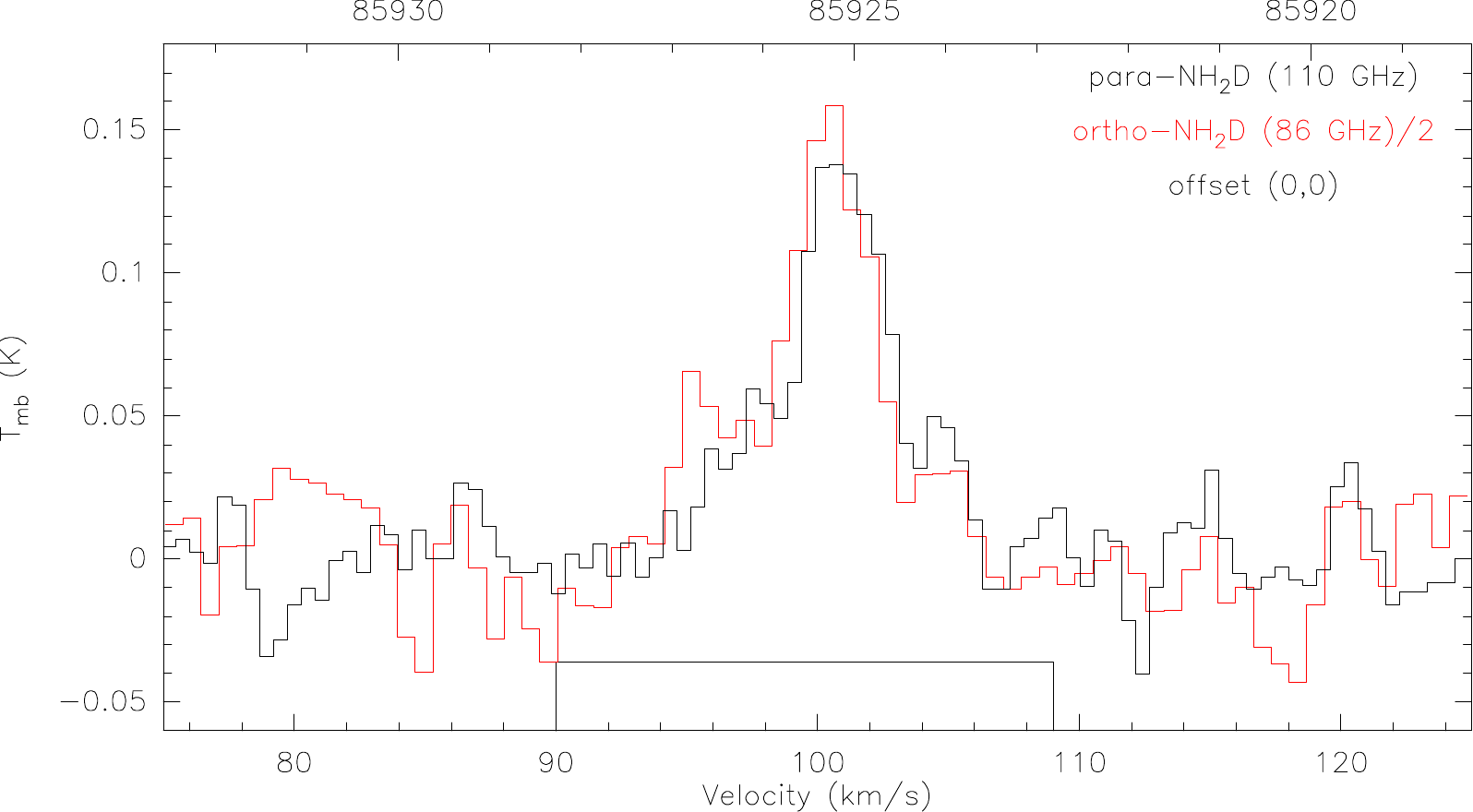}}\subfigure[]{\includegraphics[width=0.5\textwidth]{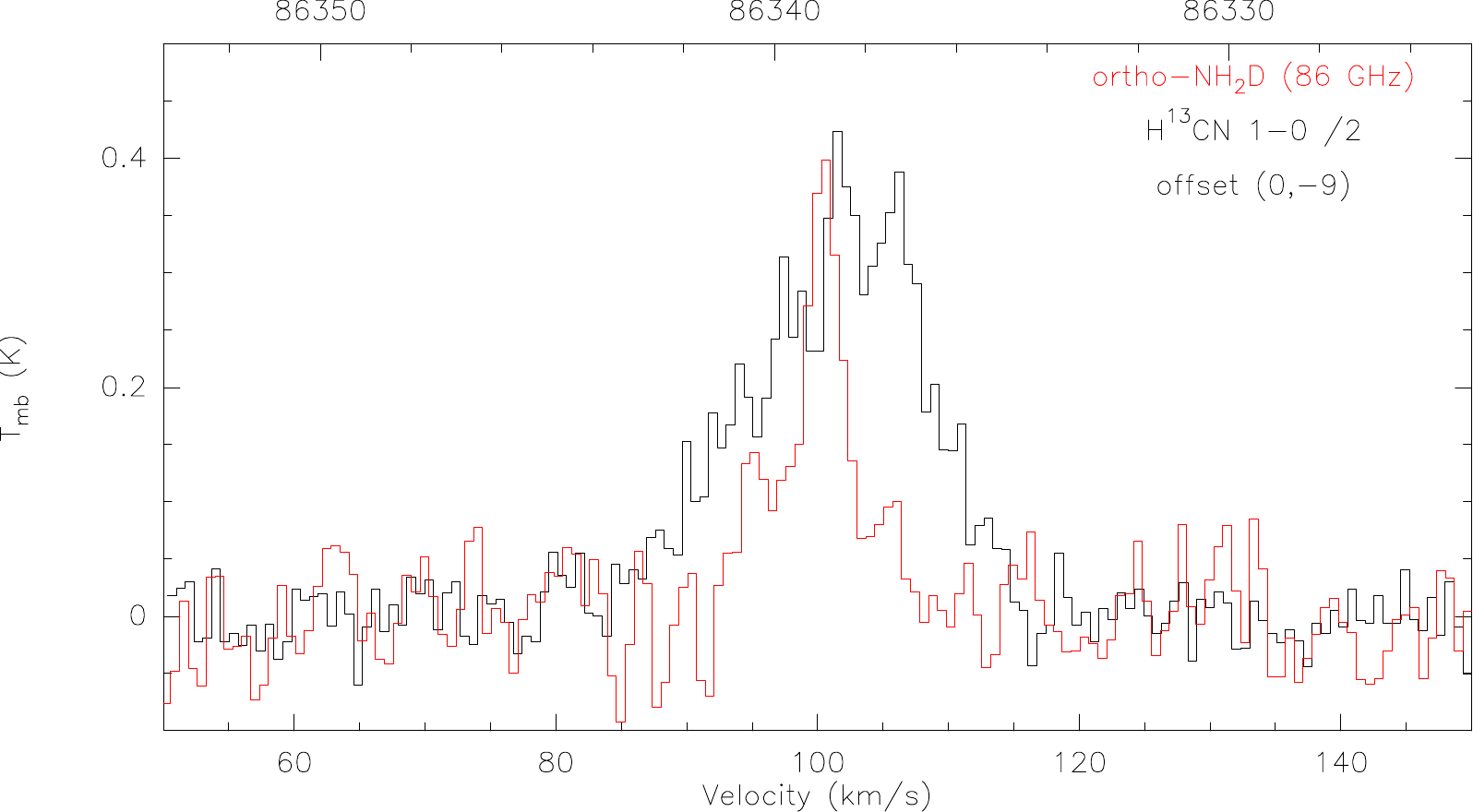}}
\caption{(a): NH$_2$D at 85.9263 GHz velocity integrated intensity contour (red contour) overlaid on H$^{13}$CN 1-0 velocity integrated intensity image (Gray scale and black contour) in G023.44-00.18. The contour levels start at 5$\sigma$ in steps of 3$\sigma$ for NH$_2$D, while the contour levels start at 12$\sigma$ in steps of 12$\sigma$ for H$^{13}$CN 1-0. The gray scale starts at 3$\sigma$. (b): NH$_2$D at 85.9263 GHz velocity integrated intensity contour (red contour) overlaid on H42$\alpha$ velocity integrated intensity image (Gray scale and black contour, while the H42$\alpha$ is not detected) in G023.44-00.18. The contour levels start at 5$\sigma$ in steps of 3$\sigma$ for NH$_2$D. (c): Spectra of ortho-NH$_2$D at 85.9263 GHz and para-NH$_2$D at 110.1535 GHz in G023.44-00.18. The para-NH$_2$D was observed by IRAM-30m with position-switching mode. The red spectra is ortho-NH$_2$D and the black is para-NH$_2$D. The offset for two spectra is (0$^{\prime \prime}$,0$^{\prime \prime}$). The black box in the picture indicates the range of flux integration of ortho-NH$_2$D at 85.9263 GHz. (d): Spectra of ortho-NH$_2$D at 85.9263 GHz and H$^{13}$CN 1-0 in G023.44-00.18. The red spectra is ortho-NH$_2$D and the black is H$^{13}$CN 1-0. The offset for two spectra is (0$^{\prime \prime}$,-9$^{\prime \prime}$).}
\label{app4}
\end{figure}

\begin{figure}
\centering
\subfigure[]{\includegraphics[width=0.5\textwidth]{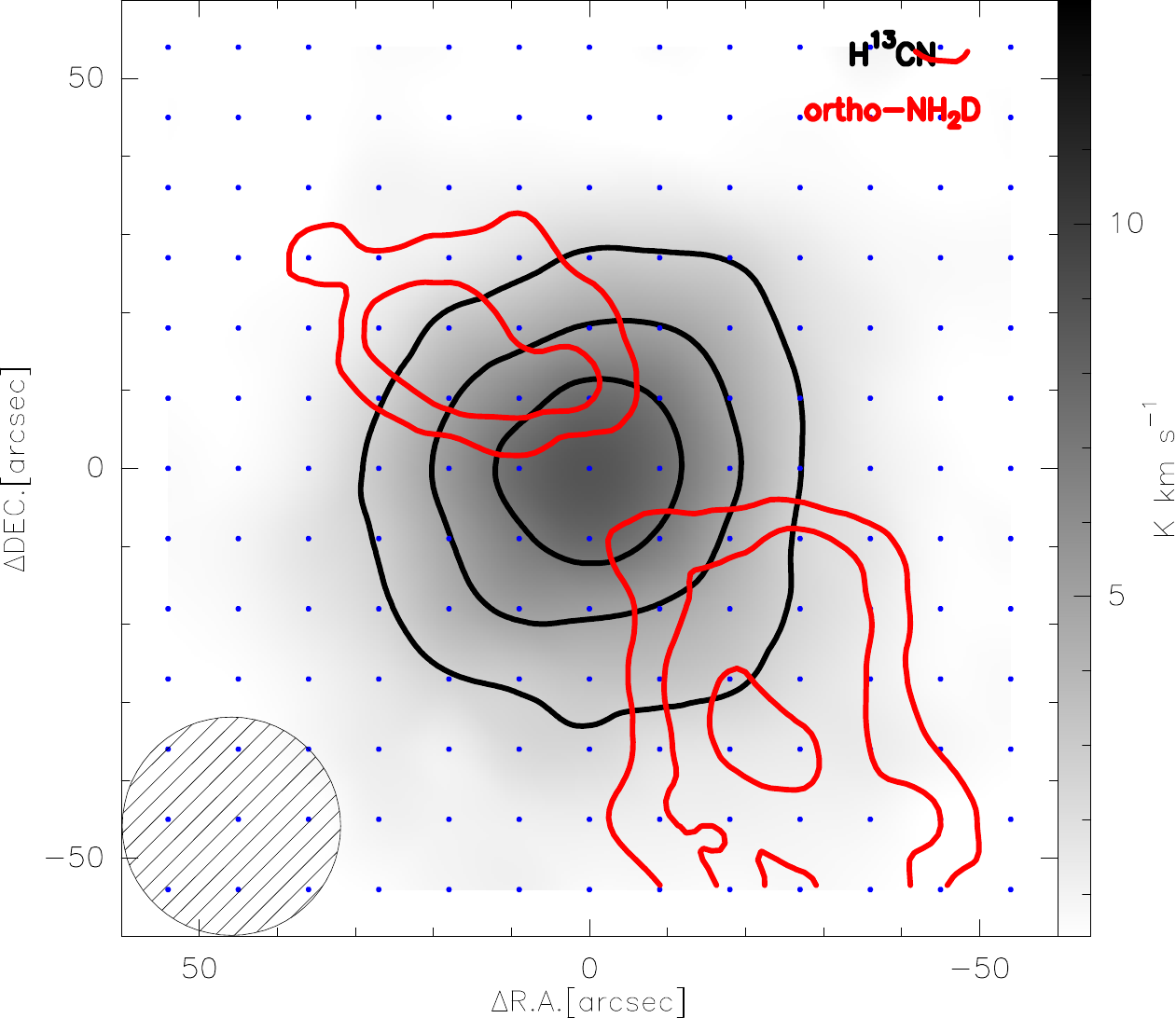}}\subfigure[]{\includegraphics[width=0.5\textwidth]{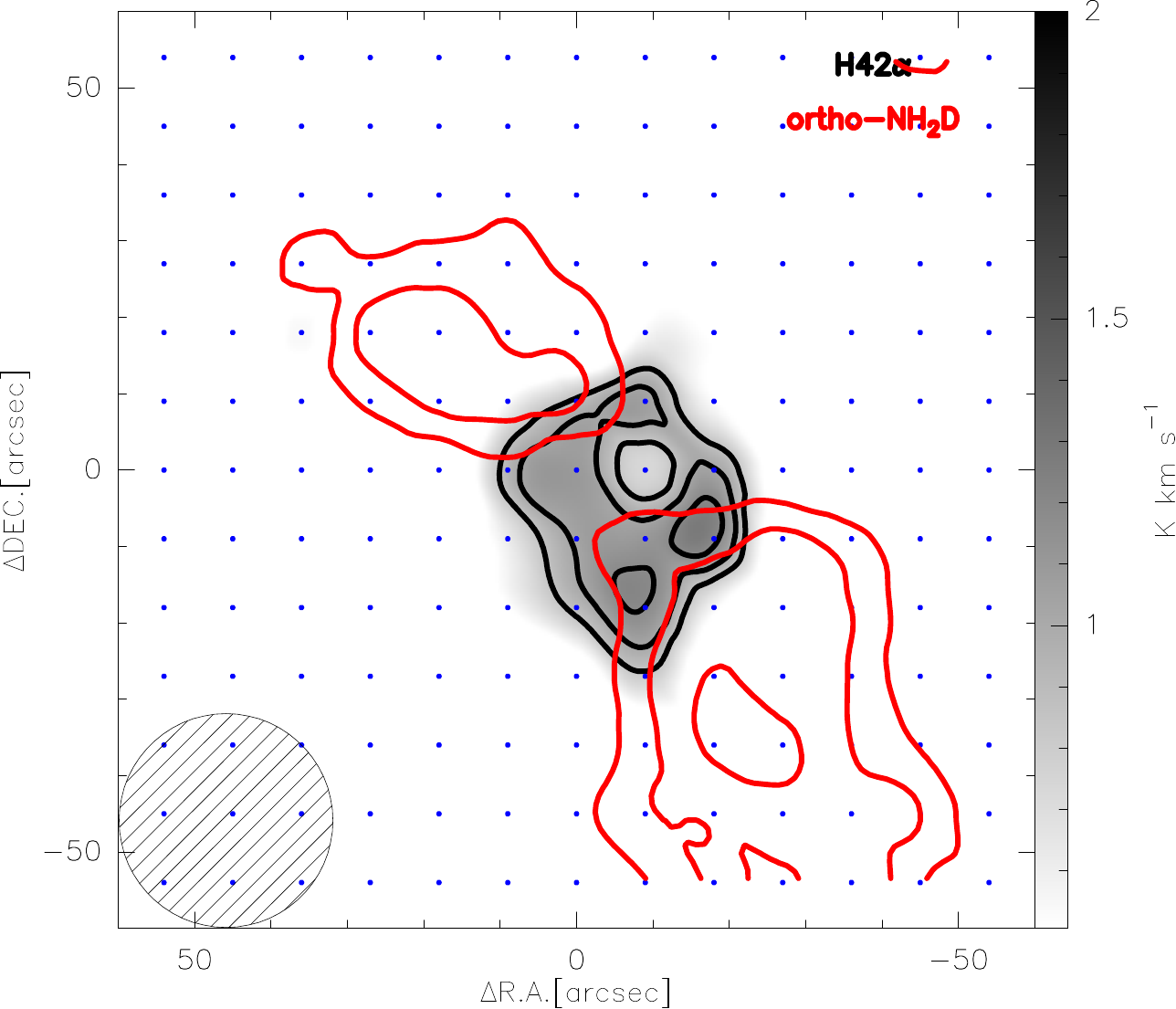}}
\subfigure[]{\includegraphics[width=0.5\textwidth]{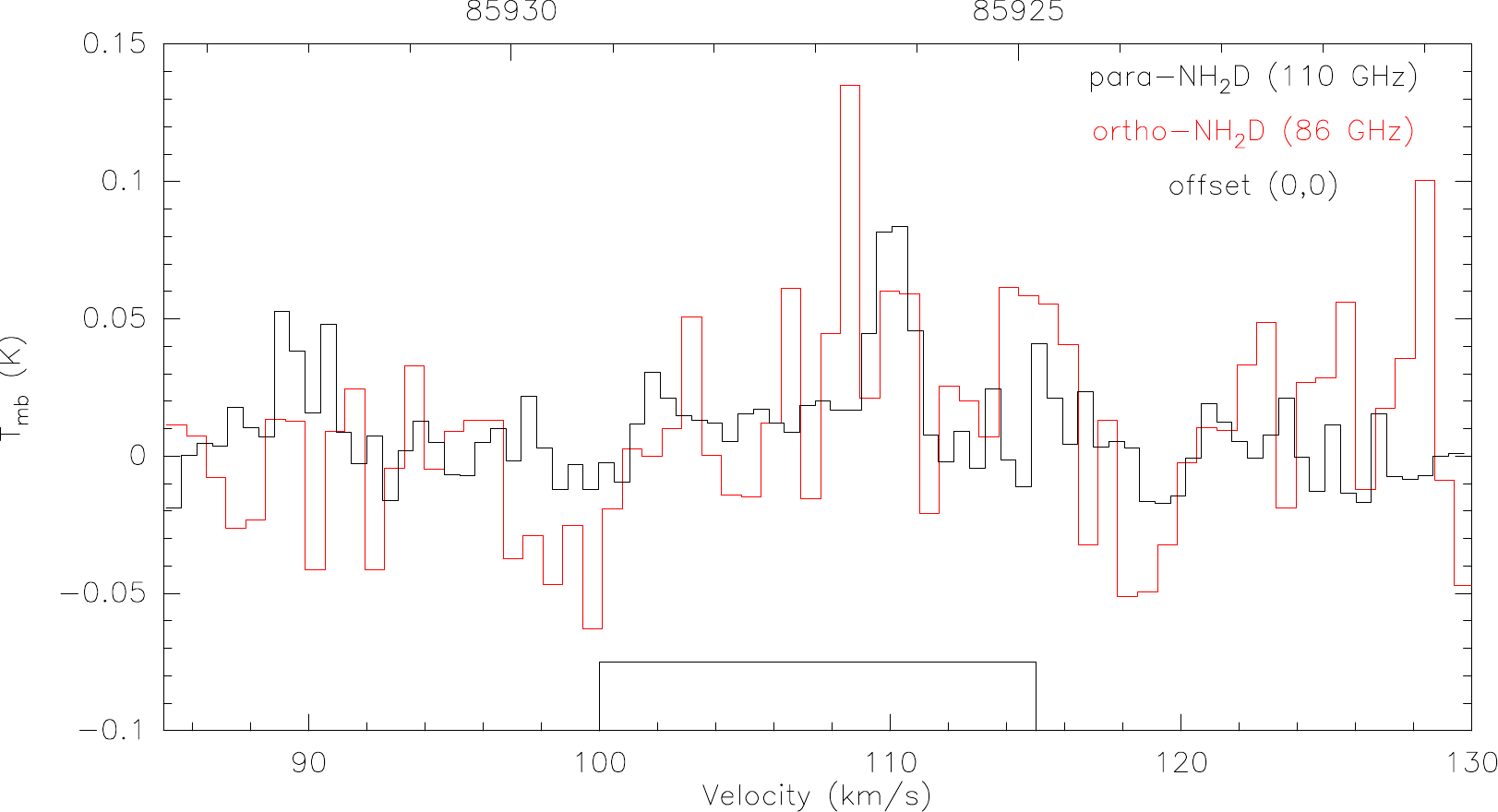}}\subfigure[]{\includegraphics[width=0.5\textwidth]{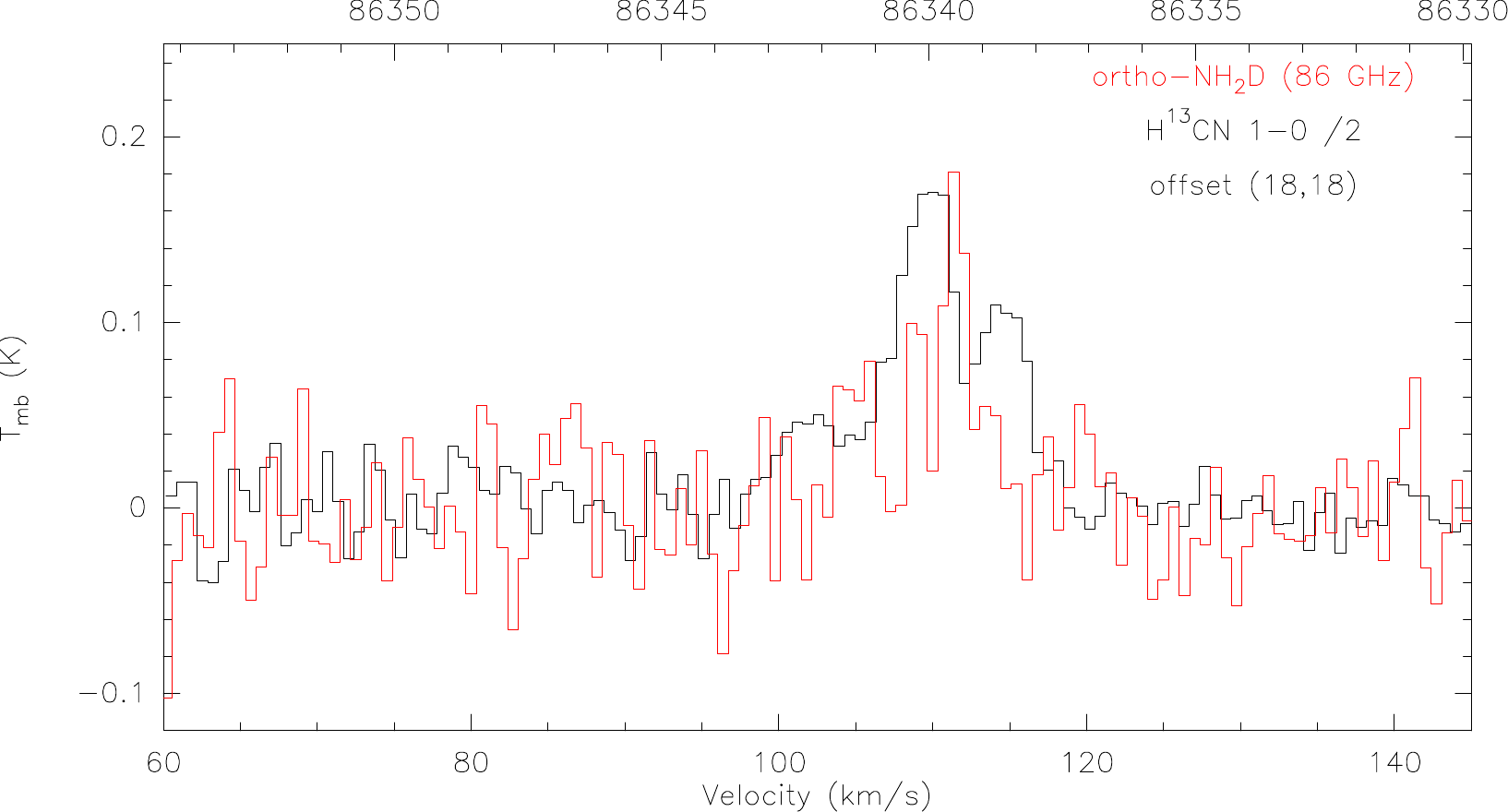}}
\subfigure[]{\includegraphics[width=0.5\textwidth]{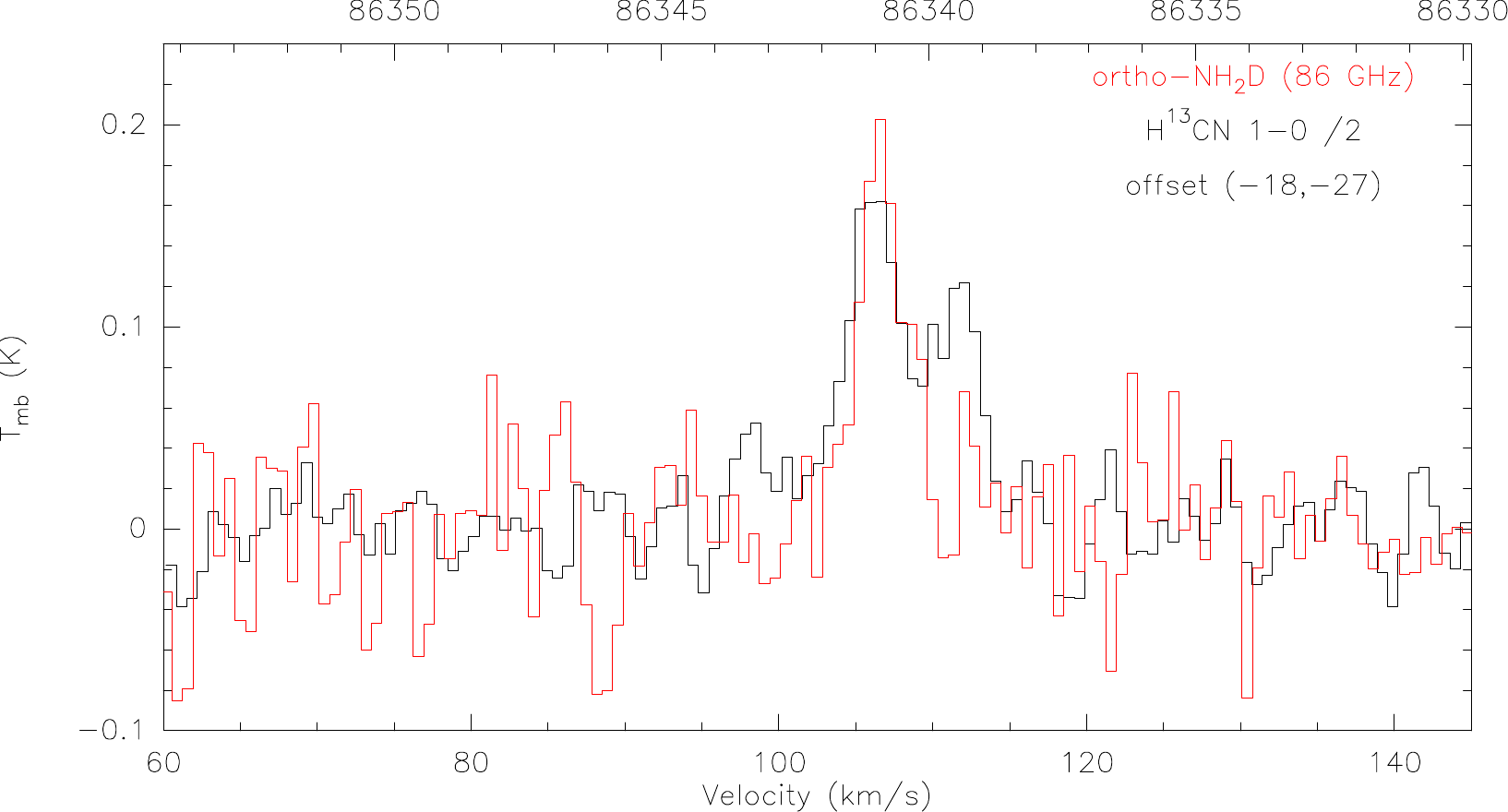}}
\caption{(a): NH$_2$D at 85.9263 GHz velocity integrated intensity contour (red contour) overlaid on H$^{13}$CN 1-0 velocity integrated intensity image (Gray scale and black contour) in G031.28+00.06. The contour levels start at 3$\sigma$ in steps of 2$\sigma$ for NH$_2$D, while the contour levels start at 18$\sigma$ in steps of 15$\sigma$ for H$^{13}$CN 1-0. The gray scale starts at 3$\sigma$. (b): NH$_2$D at 85.9263 GHz velocity integrated intensity contour (red contour) overlaid on H42$\alpha$ velocity integrated intensity image (Gray scale and black contour) in G031.28+00.06. The contour levels start at 3$\sigma$ in steps of 3$\sigma$ for NH$_2$D, while the contour levels start at 5$\sigma$ in steps of 1$\sigma$ for H42$\alpha$. The gray scale starts at 3$\sigma$. (c): Spectra of ortho-NH$_2$D at 85.9263 GHz and para-NH$_2$D at 110.1535 GHz in G031.28+00.06. The para-NH$_2$D was observed by IRAM-30m with position-switching mode. The red spectra is ortho-NH$_2$D and the black is para-NH$_2$D. The offset for two spectra is (0$^{\prime \prime}$,0$^{\prime \prime}$). The black box in the picture indicates the range of flux integration of ortho-NH$_2$D at 85.9263 GHz. (d): Spectra of ortho-NH$_2$D at 85.9263 GHz and H$^{13}$CN 1-0 in G031.28+00.06. The red spectra is ortho-NH$_2$D and the black is H$^{13}$CN 1-0. The offset for two spectra is (+18$^{\prime \prime}$,+18$^{\prime \prime}$). (e): Spectra of ortho-NH$_2$D at 85.9263 GHz and H$^{13}$CN 1-0 in G031.28+00.06.  The red spectra is ortho-NH$_2$D and the black is H$^{13}$CN 1-0. The offset for two spectra is (-18$^{\prime \prime}$,-27$^{\prime \prime}$).}
\label{app5}
\end{figure}

\begin{figure}
\centering
\subfigure[]{\includegraphics[width=0.5\textwidth]{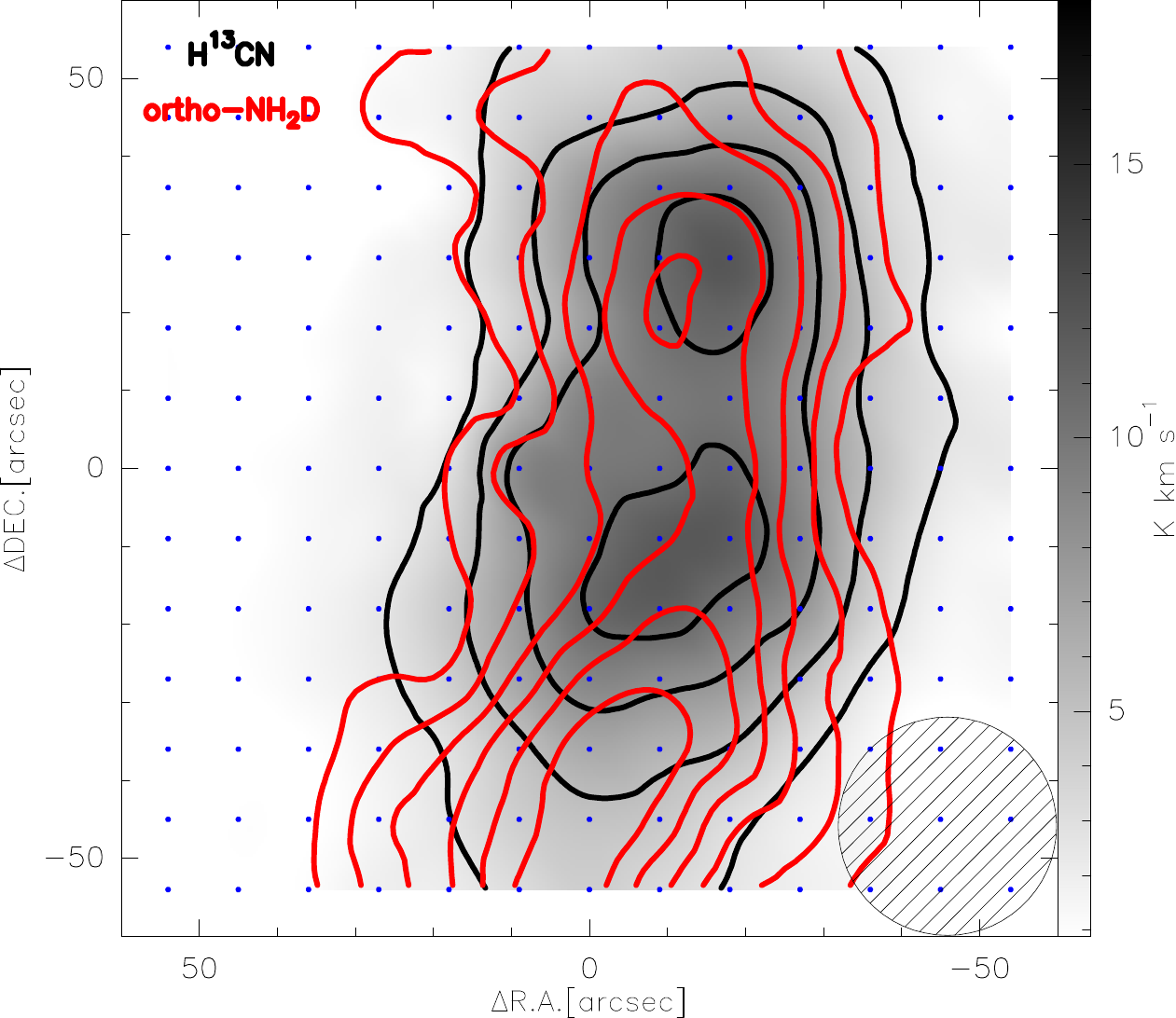}}\subfigure[]{\includegraphics[width=0.5\textwidth]{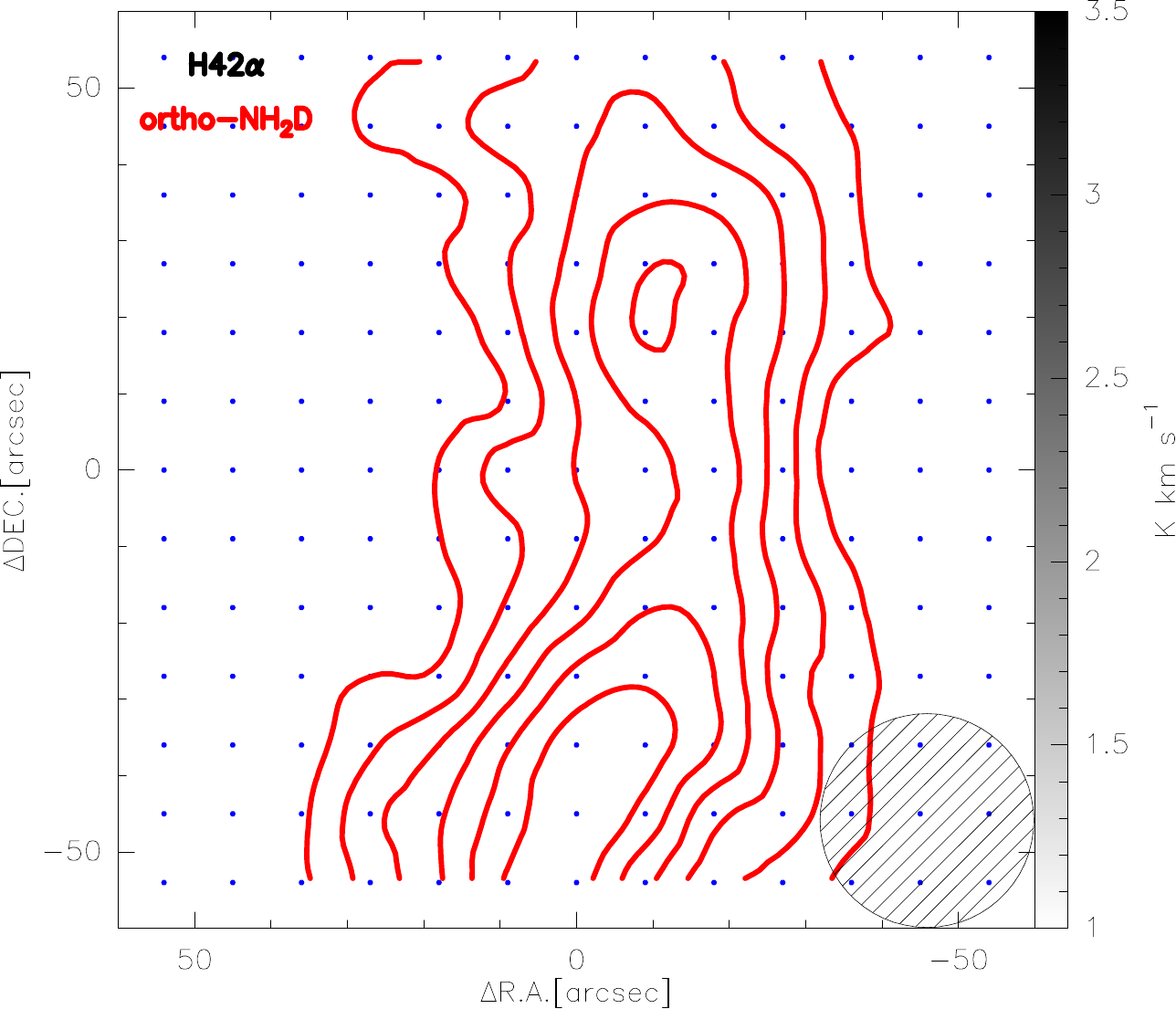}}
\subfigure[]{\includegraphics[width=0.5\textwidth]{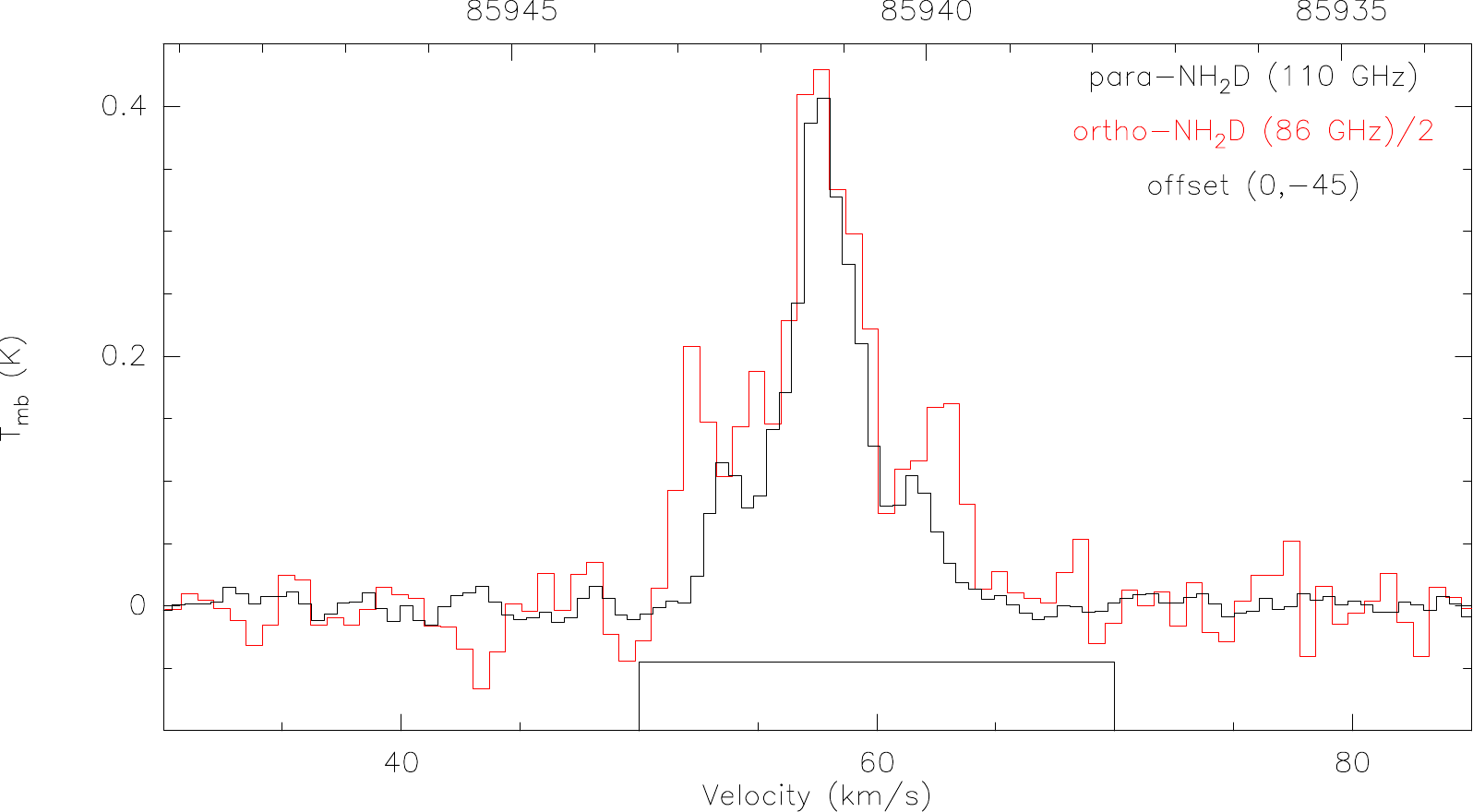}}\subfigure[]{\includegraphics[width=0.5\textwidth]{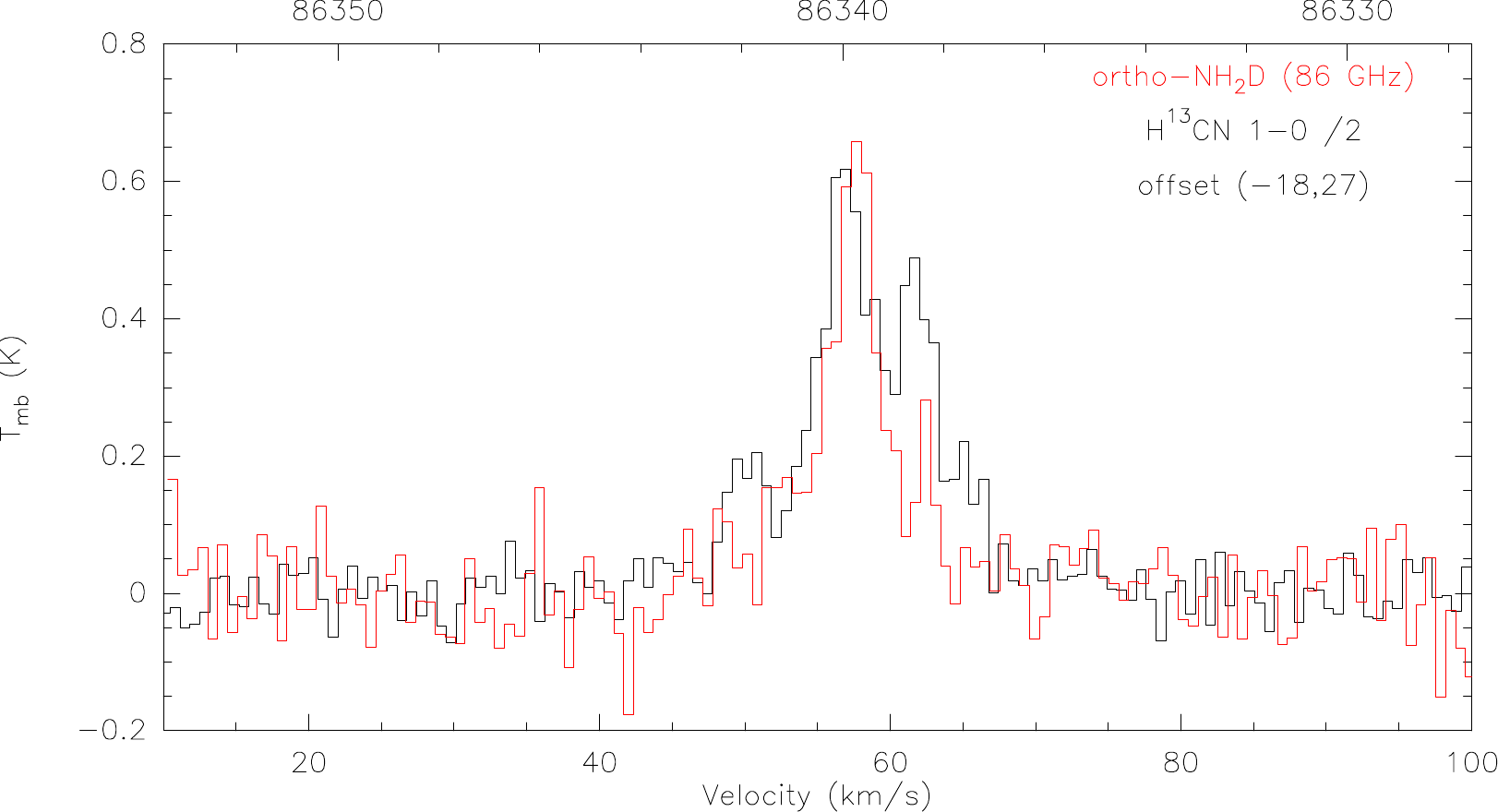}}
\subfigure[]{\includegraphics[width=0.5\textwidth]{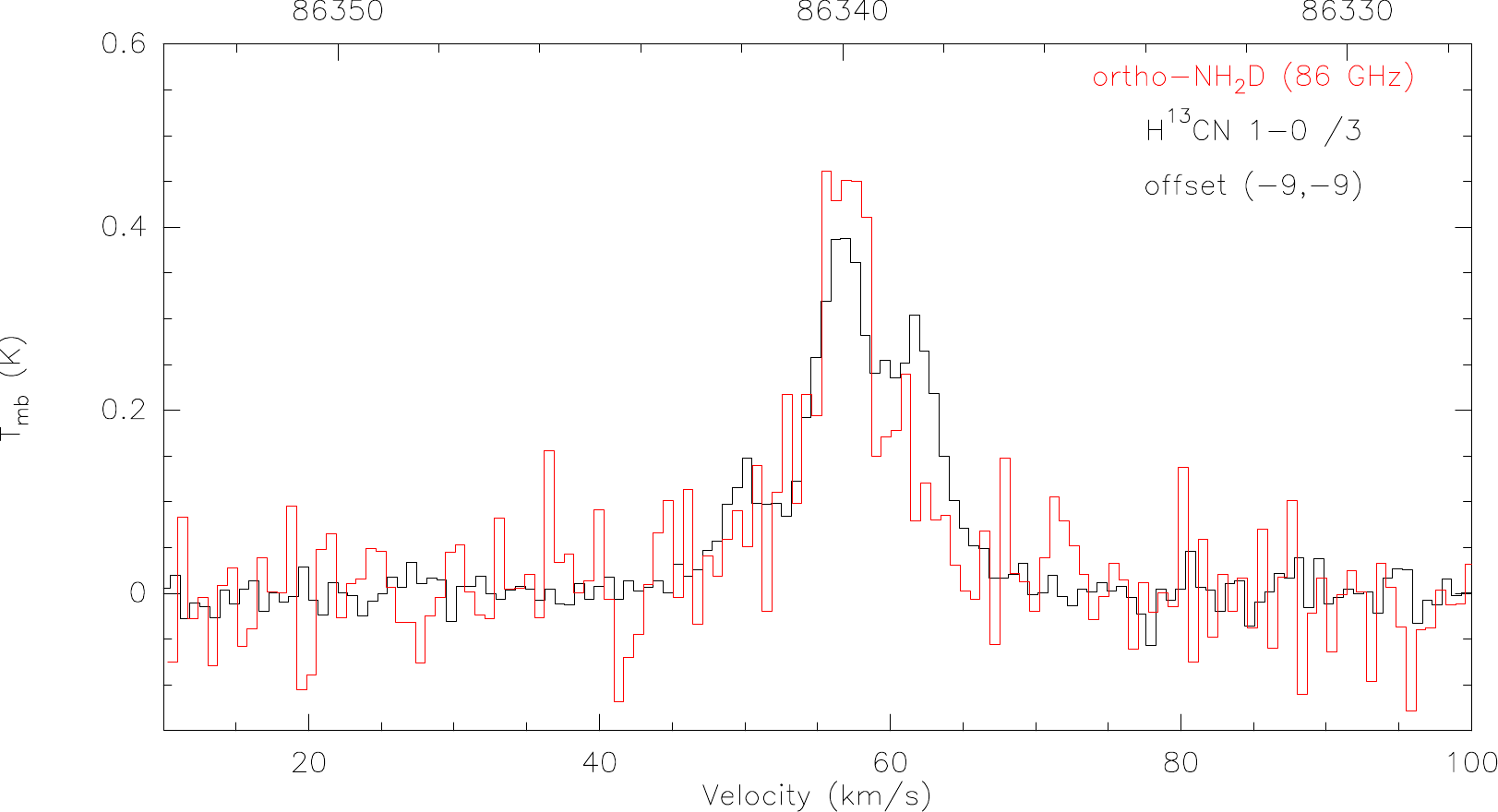}}
\caption{(a): NH$_2$D at 85.9263 GHz velocity integrated intensity contour (red contour) overlaid on H$^{13}$CN 1-0 velocity integrated intensity image (Gray scale and black contour) in G034.39+00.22. The contour levels start at 5$\sigma$ in steps of 3$\sigma$ for NH$_2$D, while the contour levels start at 9$\sigma$ in steps of 9$\sigma$ for H$^{13}$CN 1-0. The gray scale starts at 3$\sigma$. (b): NH$_2$D at 85.9263 GHz velocity integrated intensity contour (red contour) overlaid on H42$\alpha$ velocity integrated intensity image (Gray scale and black contour, while the H42$\alpha$ is not detected) in G034.39+00.22. The contour levels start at 5$\sigma$ in steps of 3$\sigma$ for NH$_2$D. (c): Spectra of ortho-NH$_2$D at 85.9263 GHz and para-NH$_2$D at 110.1535 GHz in G034.39+00.22. The para-NH$_2$D was observed by IRAM-30m with position-switching mode. The red spectra is ortho-NH$_2$D and the black is para-NH$_2$D. The offset for two spectra is (0$^{\prime \prime}$,-45$^{\prime \prime}$). The black box in the picture indicates the range of flux integration of ortho-NH$_2$D at 85.9263 GHz. (d): Spectra of ortho-NH$_2$D at 85.9263 GHz and H$^{13}$CN 1-0 in G034.39+00.22. The red spectra is ortho-NH$_2$D and the black is H$^{13}$CN 1-0. The offset for two spectra is (-18$^{\prime \prime}$,-27$^{\prime \prime}$). (e): Spectra of ortho-NH$_2$D at 85.9263 GHz and H$^{13}$CN 1-0 in G034.39+00.22.  The red spectra is ortho-NH$_2$D and the black is H$^{13}$CN 1-0. The offset for two spectra is (-9$^{\prime \prime}$,-9$^{\prime \prime}$).}
\label{app6}
\end{figure}

\begin{figure}
\centering
\subfigure[]{\includegraphics[width=0.5\textwidth]{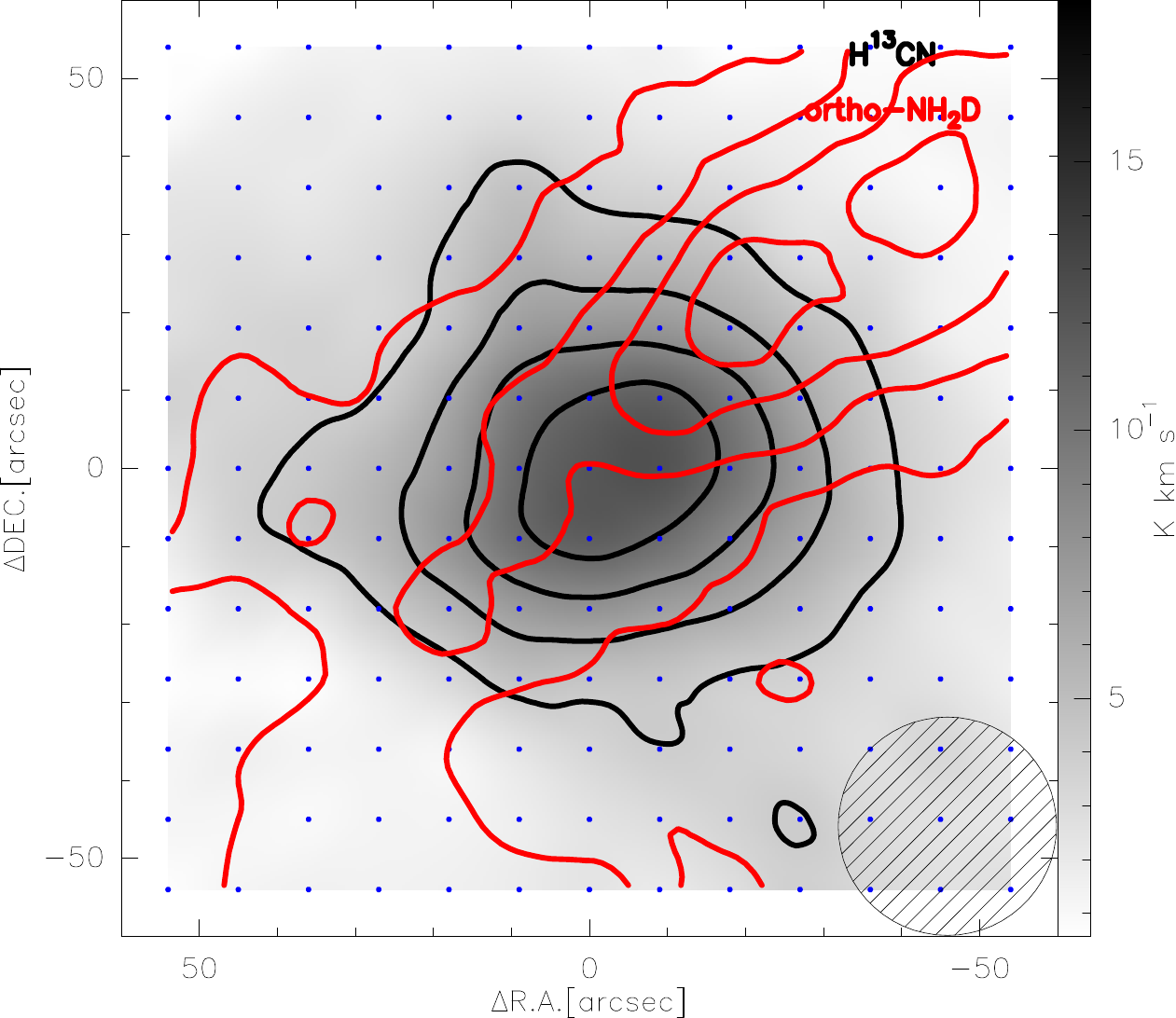}}\subfigure[]{\includegraphics[width=0.5\textwidth]{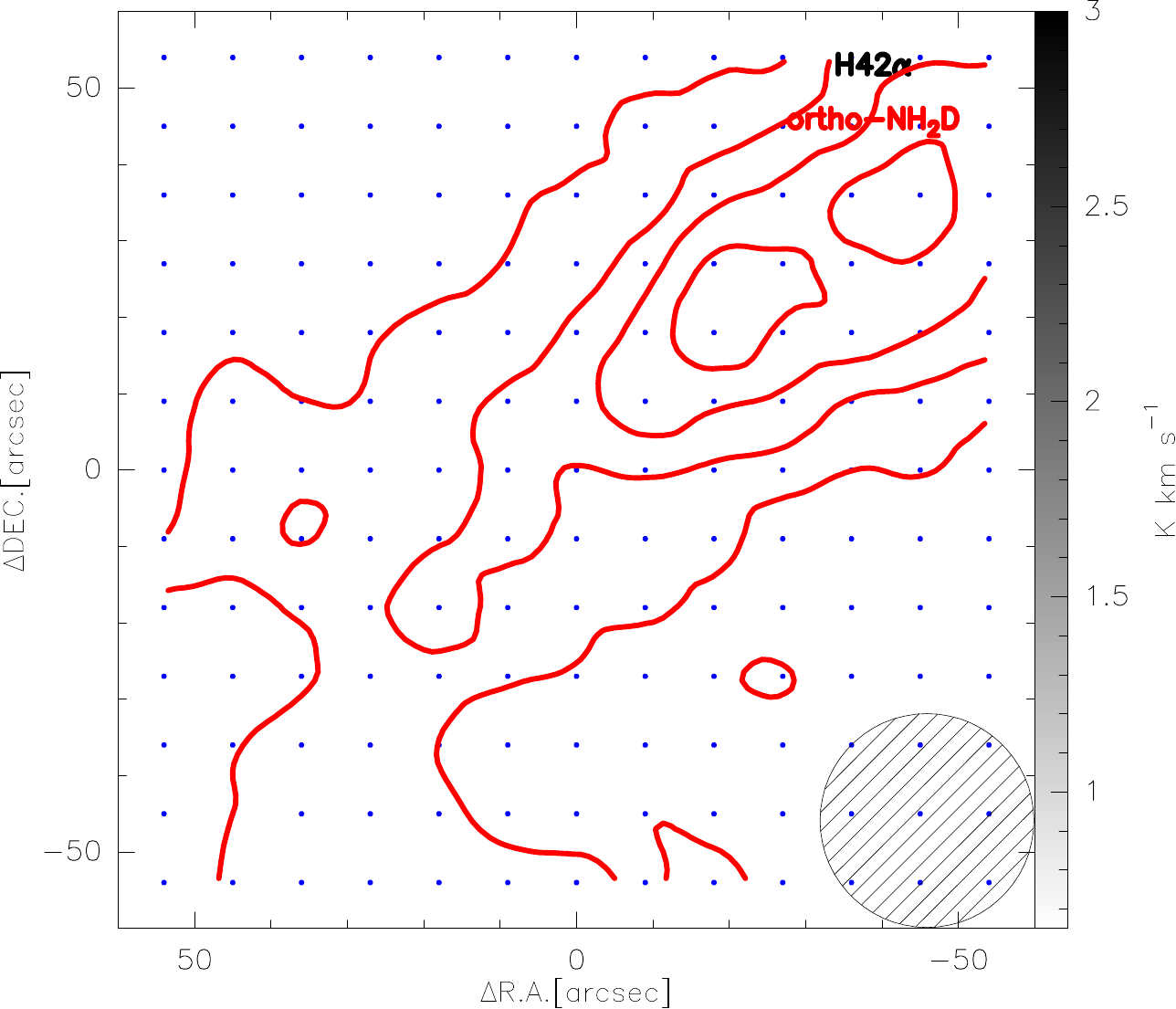}}
\subfigure[]{\includegraphics[width=0.5\textwidth]{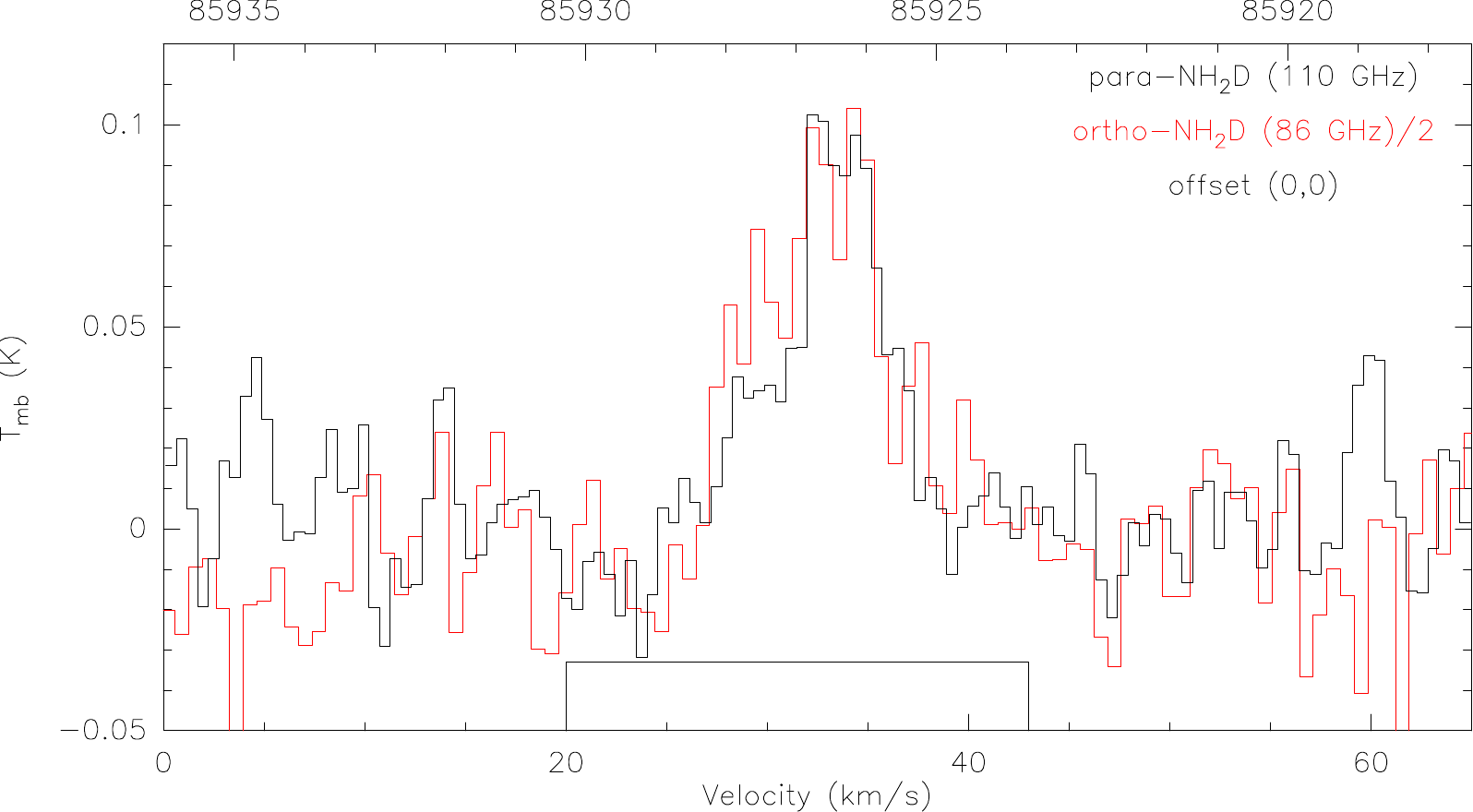}}\subfigure[]{\includegraphics[width=0.5\textwidth]{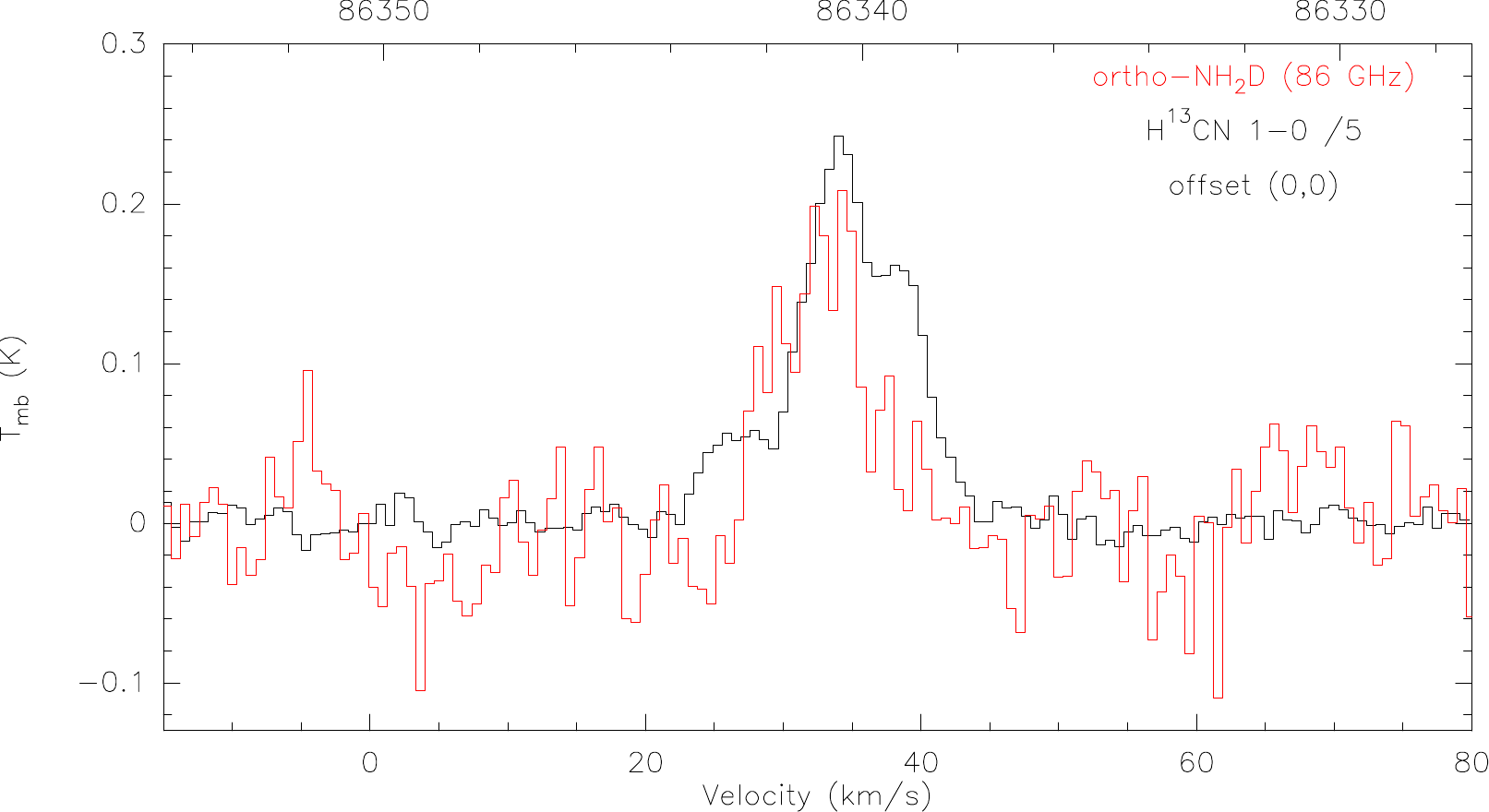}}
\subfigure[]{\includegraphics[width=0.5\textwidth]{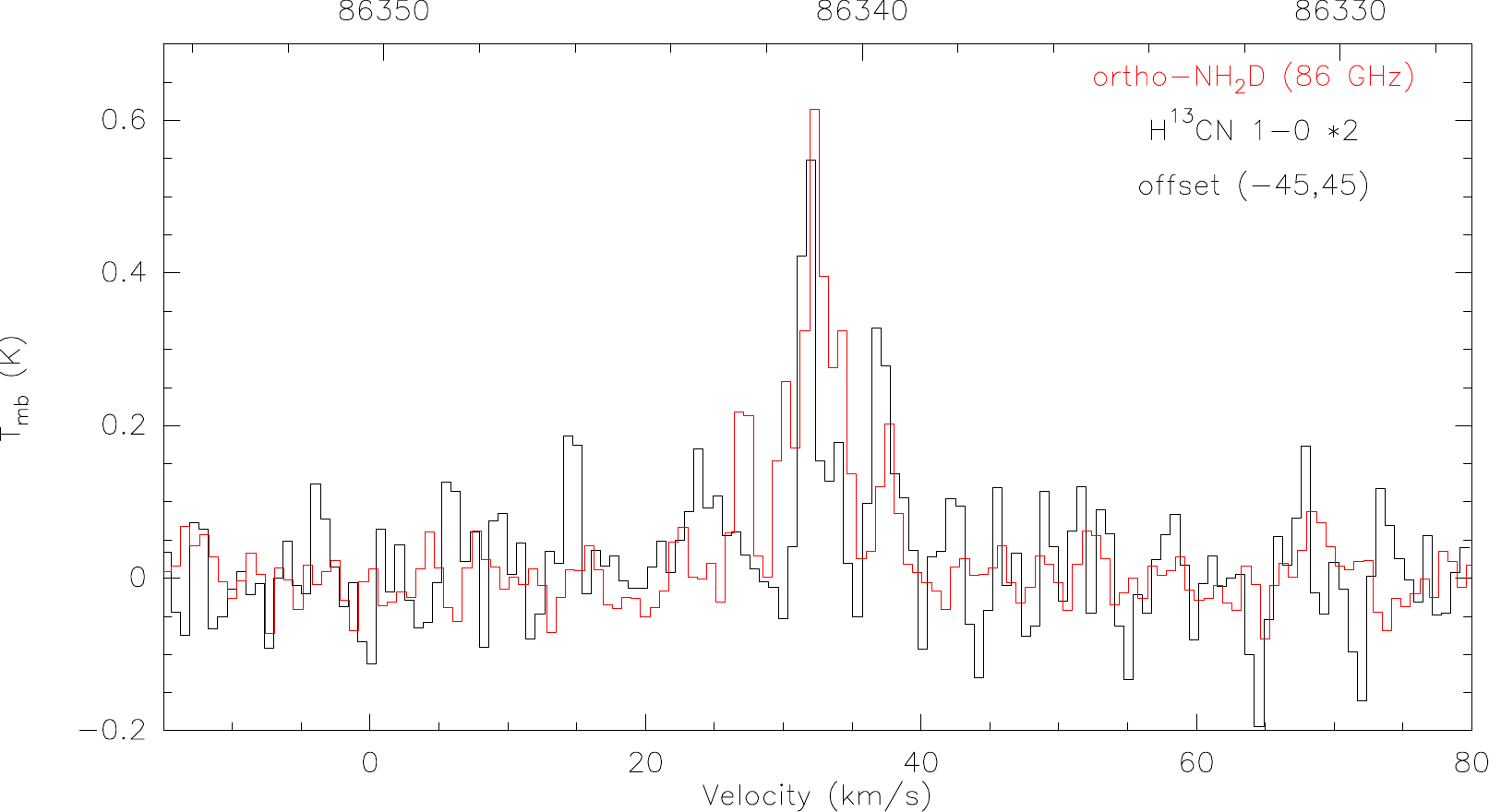}}\subfigure[]{\includegraphics[width=0.5\textwidth]{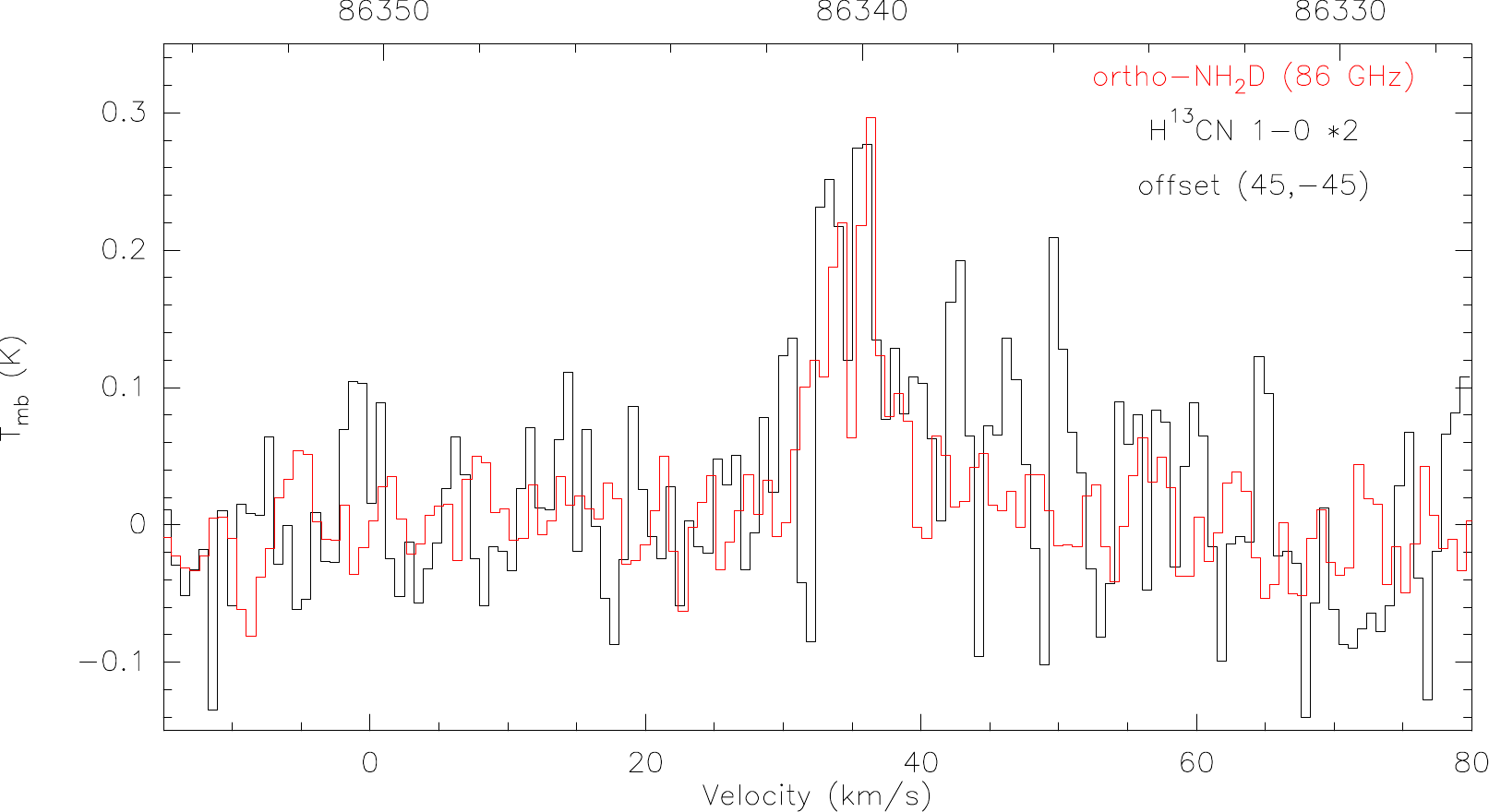}}
\caption{(a): NH$_2$D at 85.9263 GHz velocity integrated intensity contour (red contour) overlaid on H$^{13}$CN 1-0 velocity integrated intensity image (Gray scale and black contour) in G035.19-00.74. The contour levels start at 5$\sigma$ in steps of 5$\sigma$ for NH$_2$D, while the contour levels start at 21$\sigma$ in steps of 12$\sigma$ for H$^{13}$CN 1-0. The gray scale starts at 3$\sigma$. (b): NH$_2$D at 85.9263 GHz velocity integrated intensity contour (red contour) overlaid on H42$\alpha$ velocity integrated intensity image (Gray scale and black contour, while the H42$\alpha$ is not detected) in G035.19-00.74. The contour levels start at 5$\sigma$ in steps of 5$\sigma$ for NH$_2$D. (c): Spectra of ortho-NH$_2$D at 85.9263 GHz and para-NH$_2$D at 110.1535 GHz in G035.19-00.74. The para-NH$_2$D was observed by IRAM-30m with position-switching mode. The red spectra is ortho-NH$_2$D and the black is para-NH$_2$D. The offset for two spectra is (0$^{\prime \prime}$,0$^{\prime \prime}$). The black box in the picture indicates the range of flux integration of ortho-NH$_2$D at 85.9263 GHz. (d): Spectra of ortho-NH$_2$D at 85.9263 GHz and H$^{13}$CN 1-0 in G035.19-00.74. The red spectra is ortho-NH$_2$D and the black is H$^{13}$CN 1-0. The offset for two spectra is (0$^{\prime \prime}$,0$^{\prime \prime}$). (e): Spectra of ortho-NH$_2$D at 85.9263 GHz and H$^{13}$CN 1-0 in G035.19-00.74. The red spectra is ortho-NH$_2$D and the black is H$^{13}$CN 1-0. The offset for two spectra is (-45$^{\prime \prime}$,45$^{\prime \prime}$). (f): Spectra of ortho-NH$_2$D at 85.9263 GHz and H$^{13}$CN 1-0 in G035.19-00.74. The red spectra is ortho-NH$_2$D and the black is H$^{13}$CN 1-0. The offset for two spectra is (45$^{\prime \prime}$,-45$^{\prime \prime}$).}
\label{app7}
\end{figure}

\begin{figure}
\centering
\subfigure[]{\includegraphics[width=0.5\textwidth]{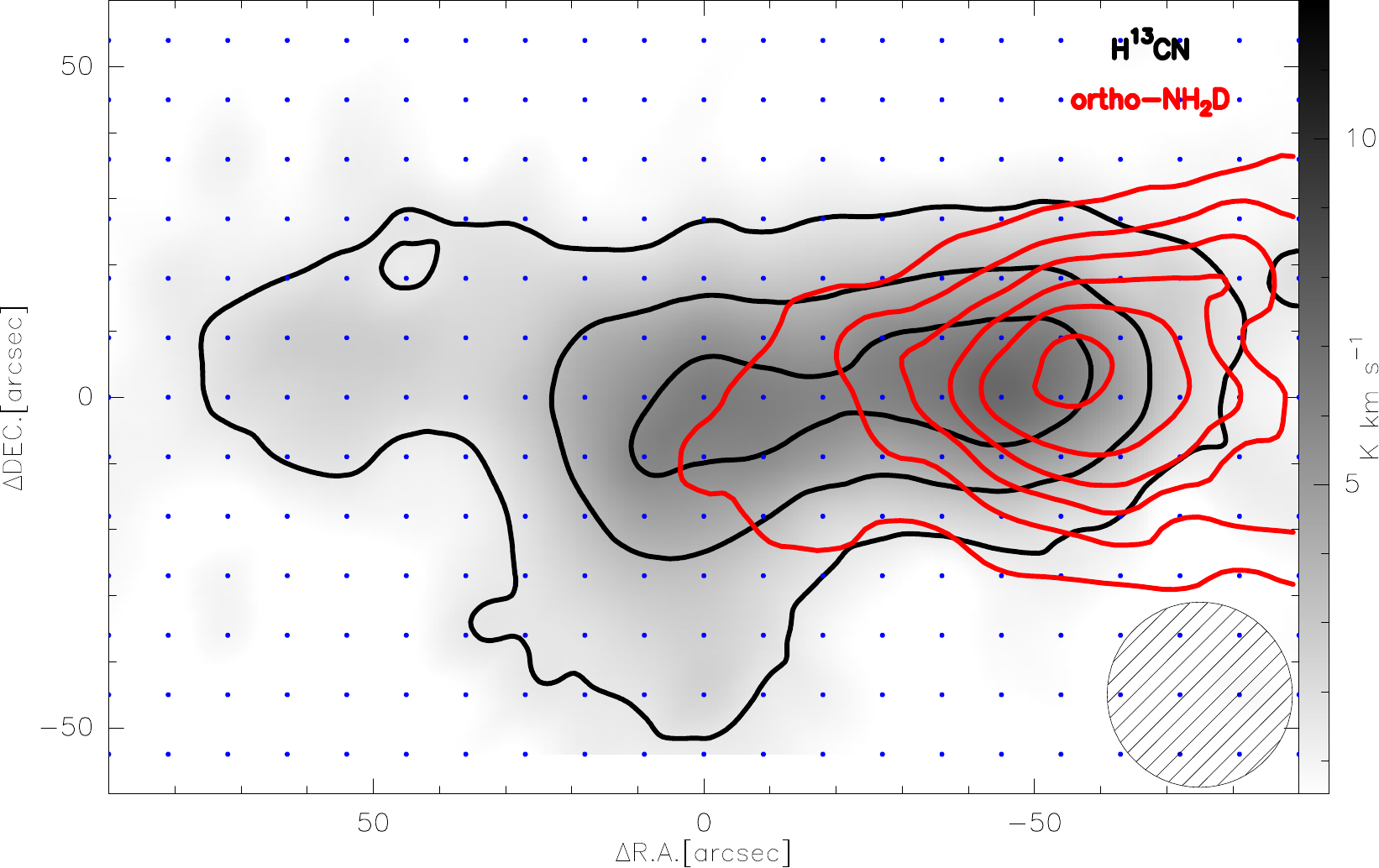}}\subfigure[]{\includegraphics[width=0.5\textwidth]{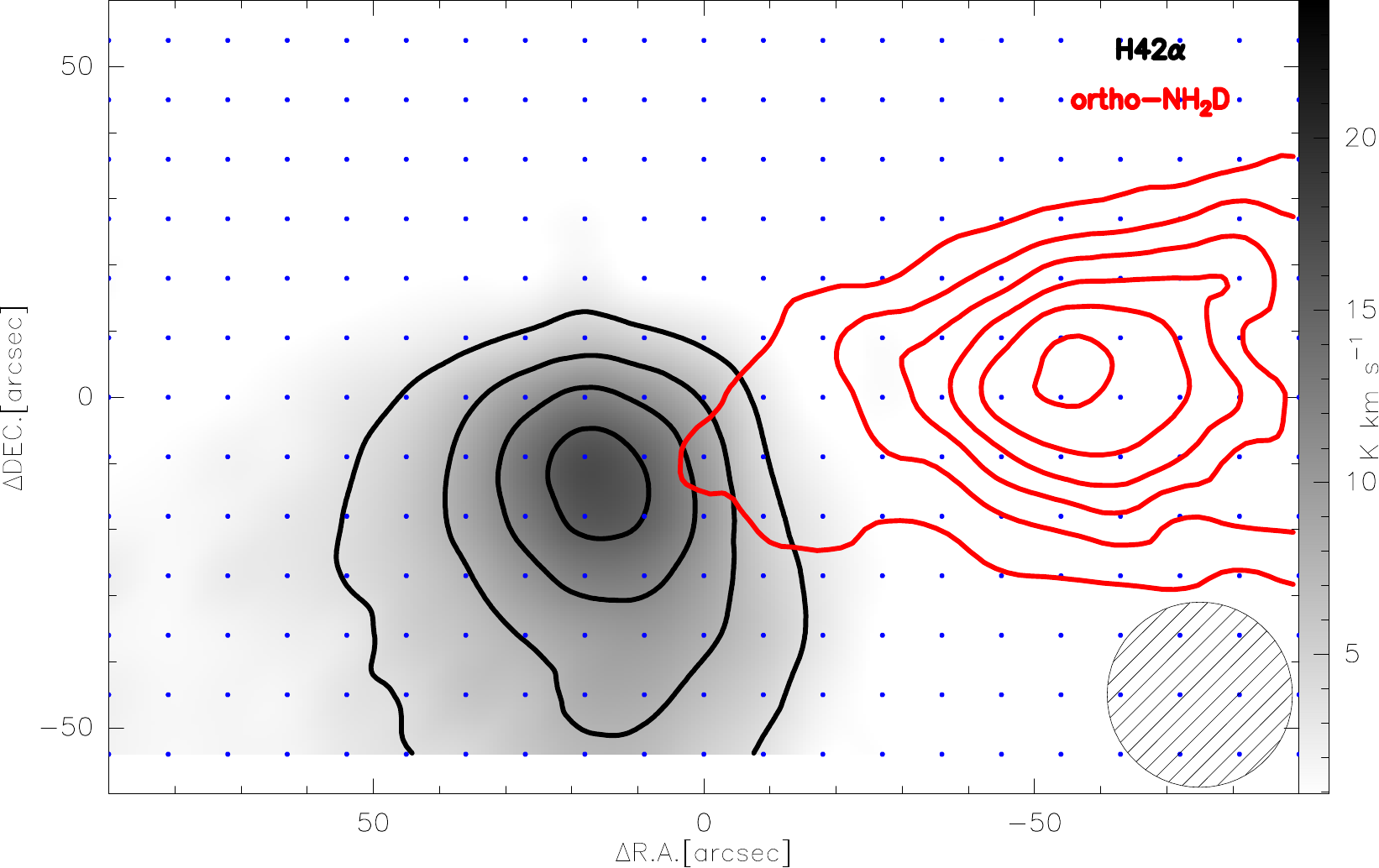}}
\subfigure[]{\includegraphics[width=0.5\textwidth]{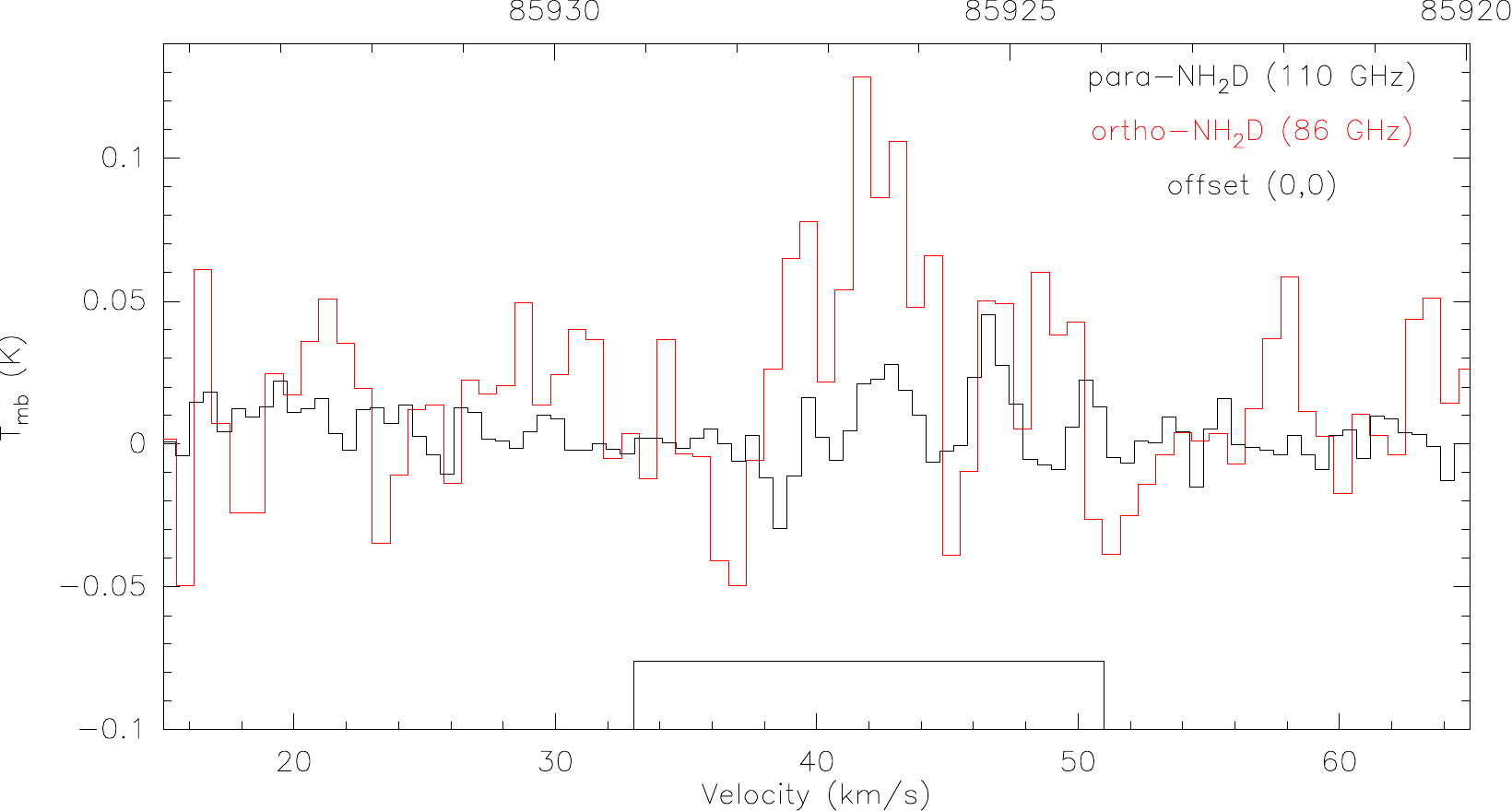}}\subfigure[]{\includegraphics[width=0.5\textwidth]{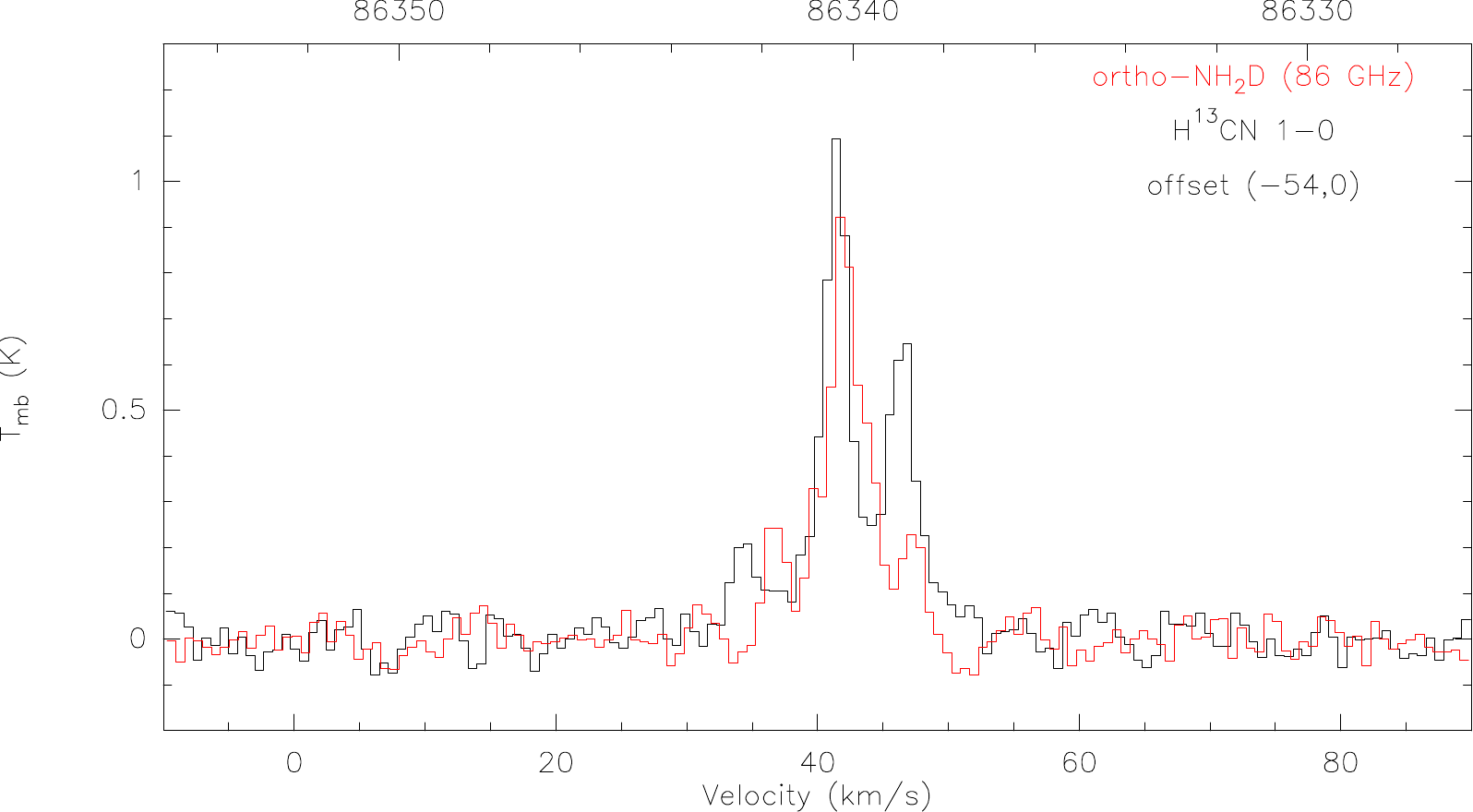}}
\caption{(a): NH$_2$D at 85.9263 GHz velocity integrated intensity contour (red contour) overlaid on H$^{13}$CN 1-0 velocity integrated intensity image (Gray scale and black contour) in G035.20-01.73. The contour levels start at 5$\sigma$ in steps of 5$\sigma$ for NH$_2$D, while the contour levels start at 9$\sigma$ in steps of 12$\sigma$ for H$^{13}$CN 1-0. The gray scale starts at 3$\sigma$. (b): NH$_2$D at 85.9263 GHz velocity integrated intensity contour (red contour) overlaid on H42$\alpha$ velocity integrated intensity image (Gray scale and black contour) in G035.20-01.73. The contour levels start at 5$\sigma$ in steps of 5$\sigma$ for NH$_2$D, while the contour levels start at 12$\sigma$ in steps of 12$\sigma$ for H42$\alpha$. The gray scale starts at 3$\sigma$. (c): Spectra of ortho-NH$_2$D at 85.9263 GHz and para-NH$_2$D at 110.1535 GHz in G035.20-01.73. The para-NH$_2$D was observed by IRAM-30m with position-switching mode. The red spectra is ortho-NH$_2$D and the black is para-NH$_2$D. The offset for two spectra is (0$^{\prime \prime}$,0$^{\prime \prime}$). The black box in the picture indicates the range of flux integration of ortho-NH$_2$D at 85.9263 GHz. (d): Spectra of ortho-NH$_2$D at 85.9263 GHz and H$^{13}$CN 1-0 in G035.20-01.73. The red spectra is ortho-NH$_2$D and the black is H$^{13}$CN 1-0. The offset for two spectra is (-54$^{\prime \prime}$,0$^{\prime \prime}$).}
\label{app8}
\end{figure}

\begin{figure}
\centering
\subfigure[]{\includegraphics[width=0.5\textwidth]{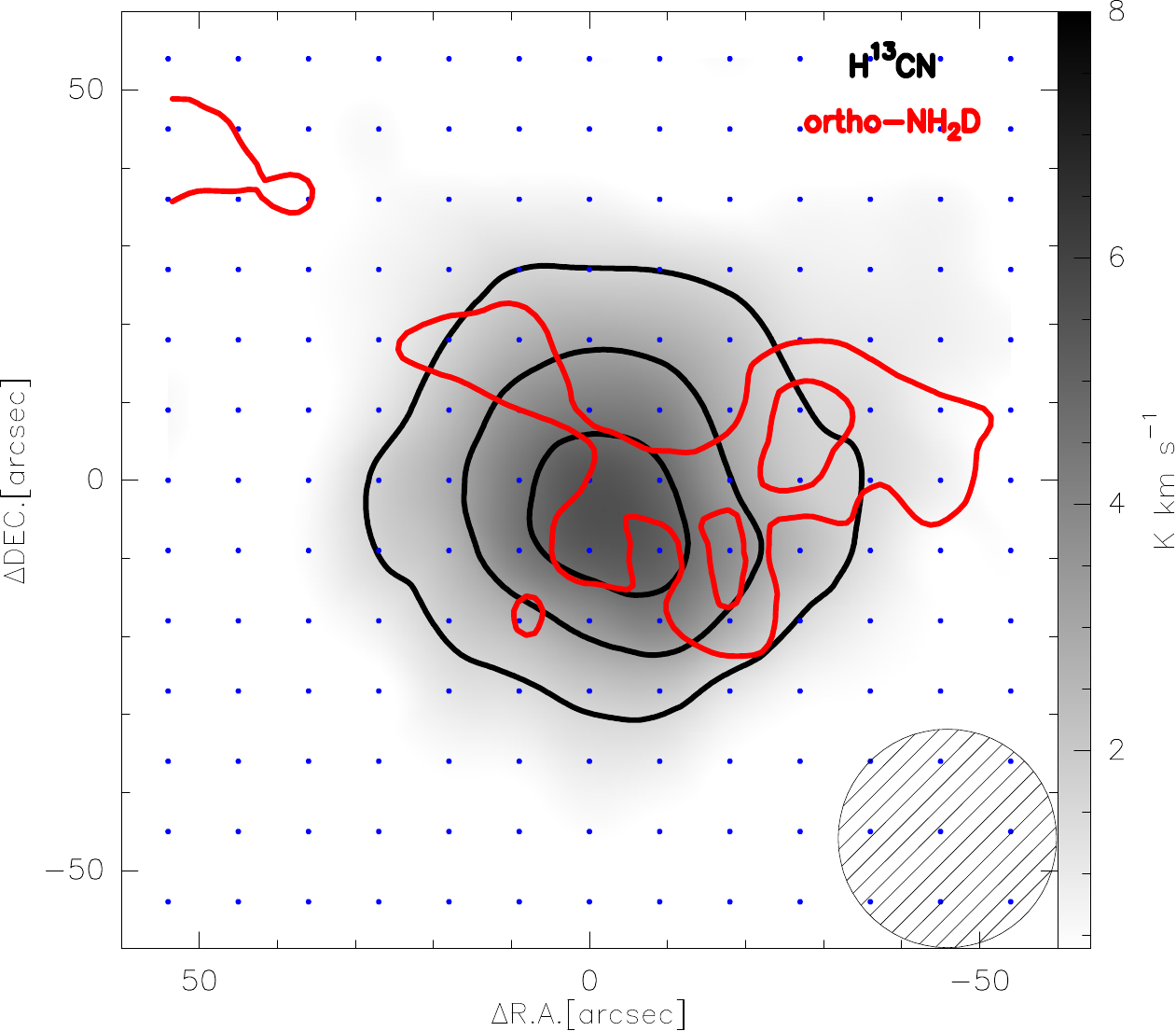}}\subfigure[]{\includegraphics[width=0.5\textwidth]{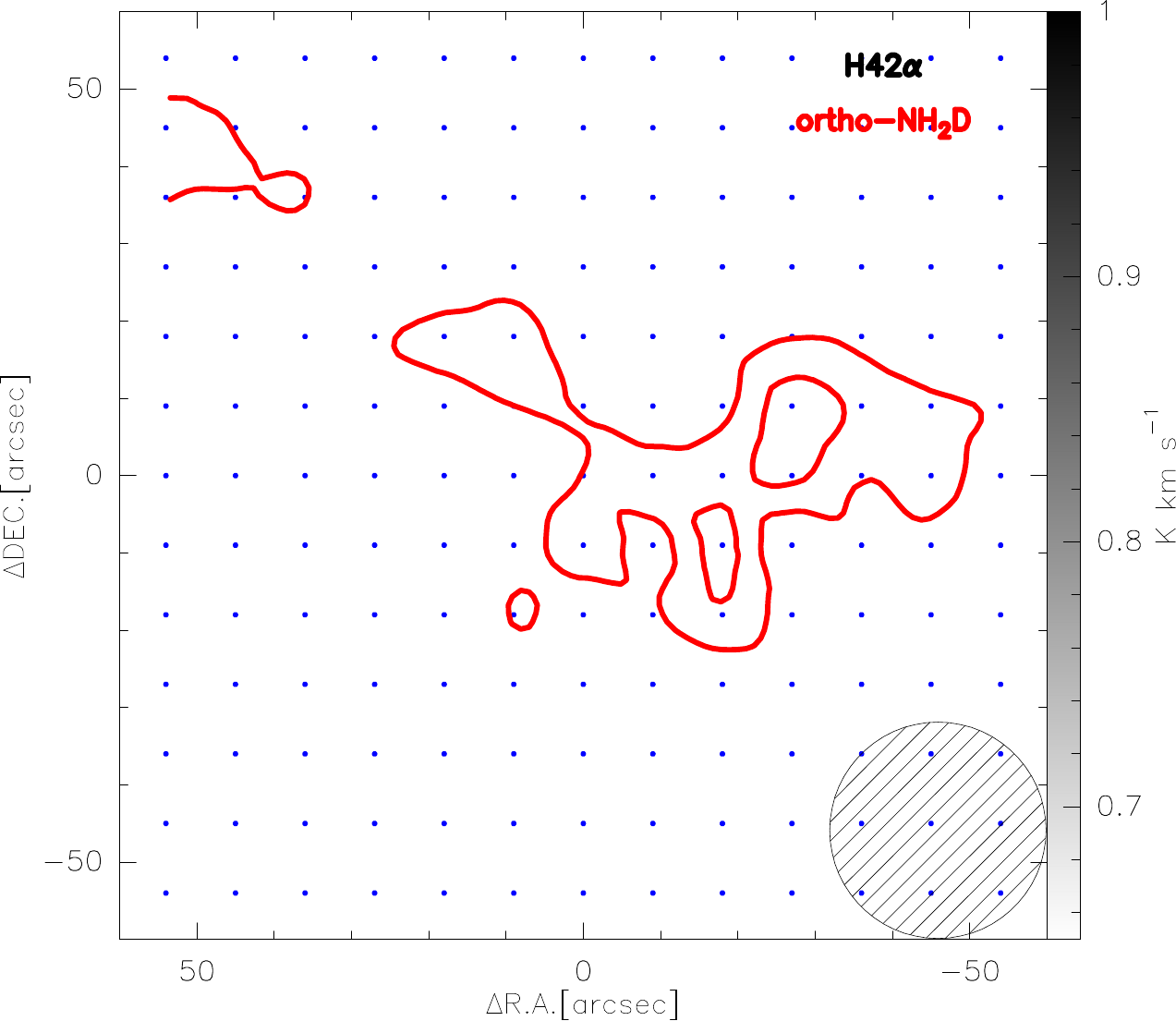}}
\subfigure[]{\includegraphics[width=0.5\textwidth]{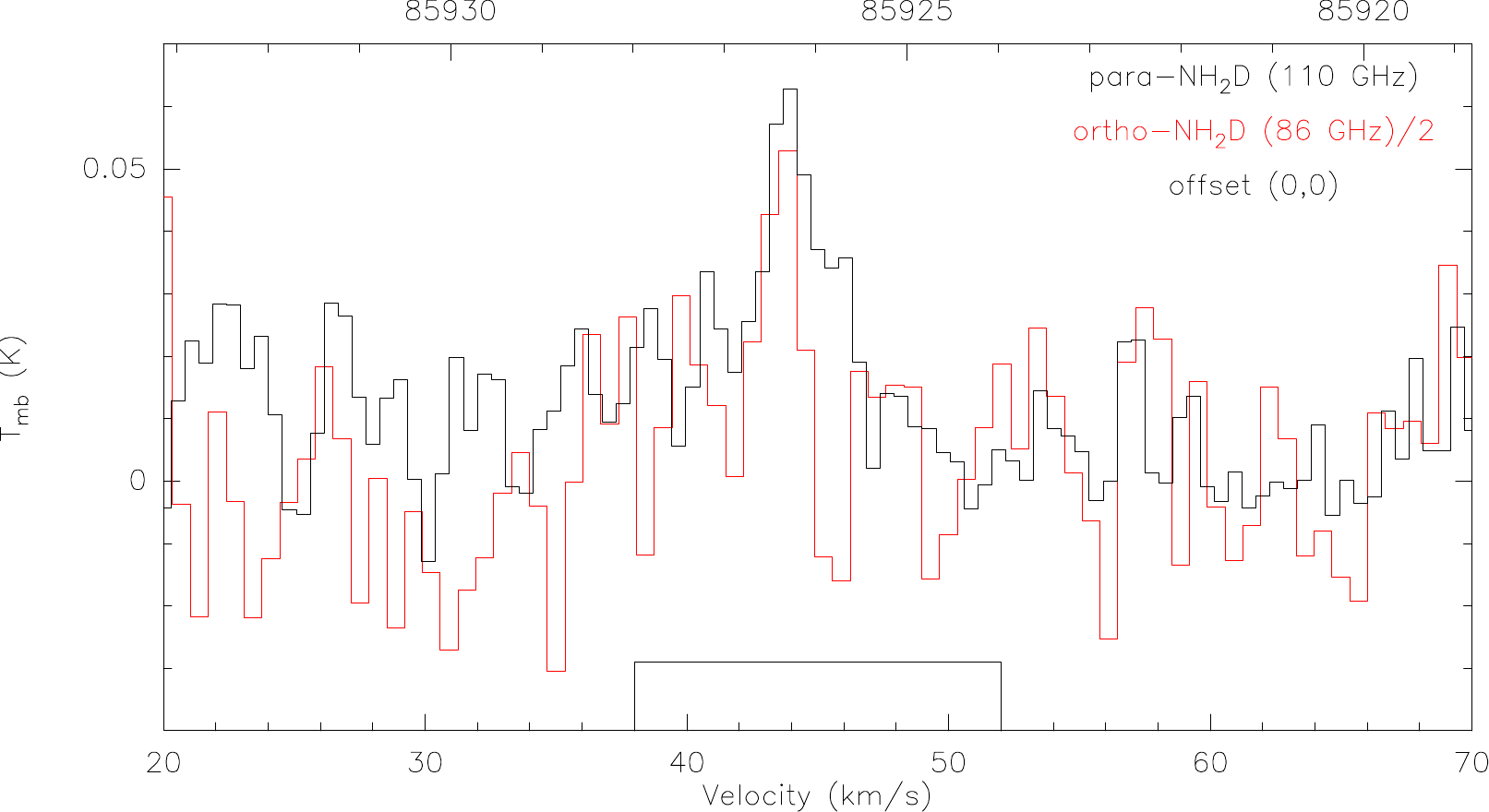}}\subfigure[]{\includegraphics[width=0.5\textwidth]{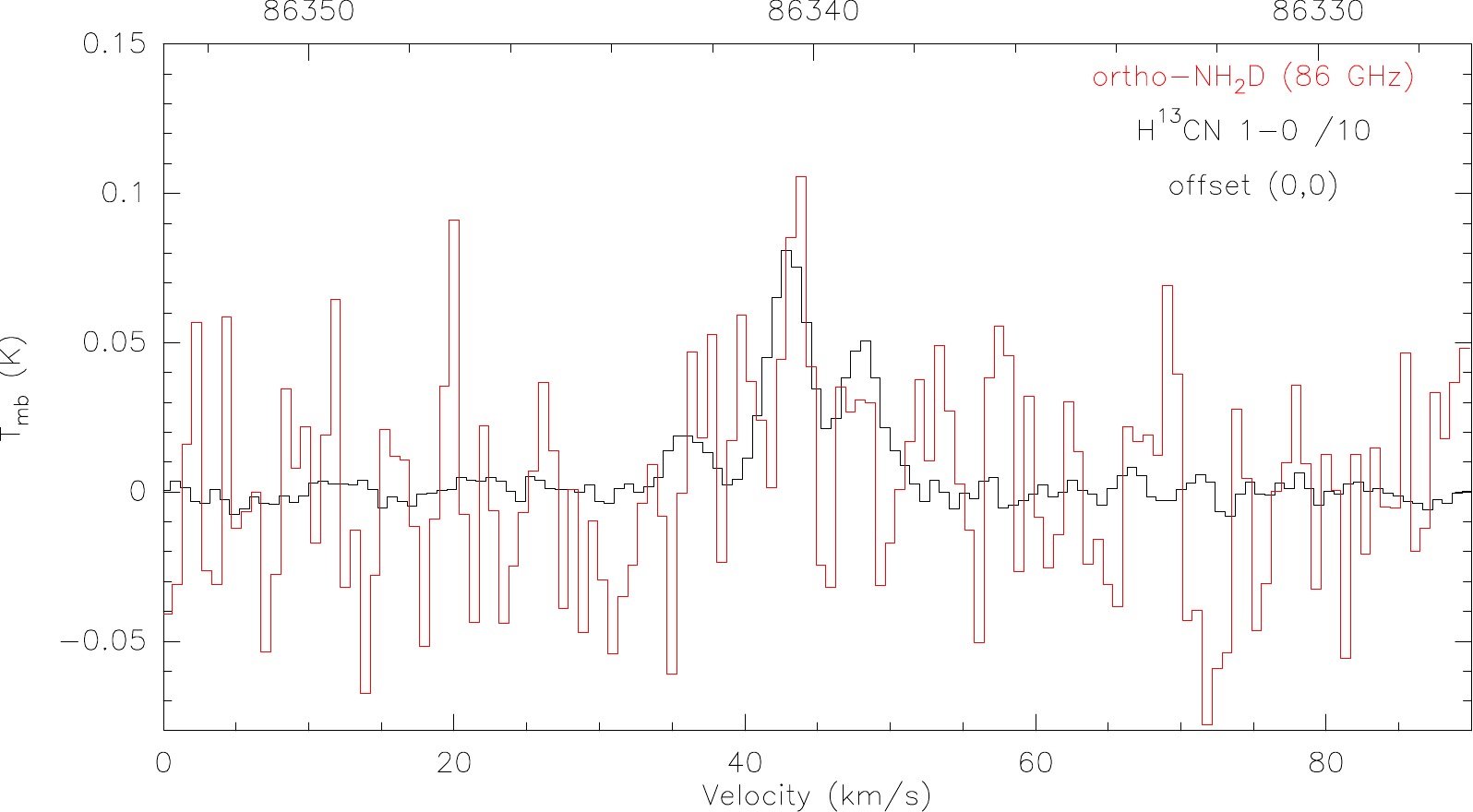}}
\caption{(a): NH$_2$D at 85.9263 GHz velocity integrated intensity contour (red contour) overlaid on H$^{13}$CN 1-0 velocity integrated intensity image (Gray scale and black contour) in G037.43+01.51. The contour levels start at 3$\sigma$ in steps of 2$\sigma$ for NH$_2$D, while the contour levels start at 12$\sigma$ in steps of 12$\sigma$ for H$^{13}$CN 1-0. The gray scale starts at 3$\sigma$. (b): NH$_2$D at 85.9263 GHz velocity integrated intensity contour (red contour) overlaid on H42$\alpha$ velocity integrated intensity image (Gray scale and black contour, while the H42$\alpha$ is not detected) in G037.43+01.51. The contour levels start at 3$\sigma$ in steps of 2$\sigma$ for NH$_2$D. (c): Spectra of ortho-NH$_2$D at 85.9263 GHz and para-NH$_2$D at 110.1535 GHz in G037.43+01.51. The para-NH$_2$D was observed by IRAM-30m with position-switching mode. The red spectra is ortho-NH$_2$D and the black is para-NH$_2$D. The offset for two spectra is (0$^{\prime \prime}$,0$^{\prime \prime}$). The black box in the picture indicates the range of flux integration of ortho-NH$_2$D at 85.9263 GHz. (d): Spectra of ortho-NH$_2$D at 85.9263 GHz and H$^{13}$CN 1-0 in G037.43+01.51. The red spectra is ortho-NH$_2$D and the black is H$^{13}$CN 1-0. The offset for two spectra is (0$^{\prime \prime}$,0$^{\prime \prime}$).}
\label{app9}
\end{figure}

\begin{figure}
\centering
\subfigure[]{\includegraphics[width=0.5\textwidth]{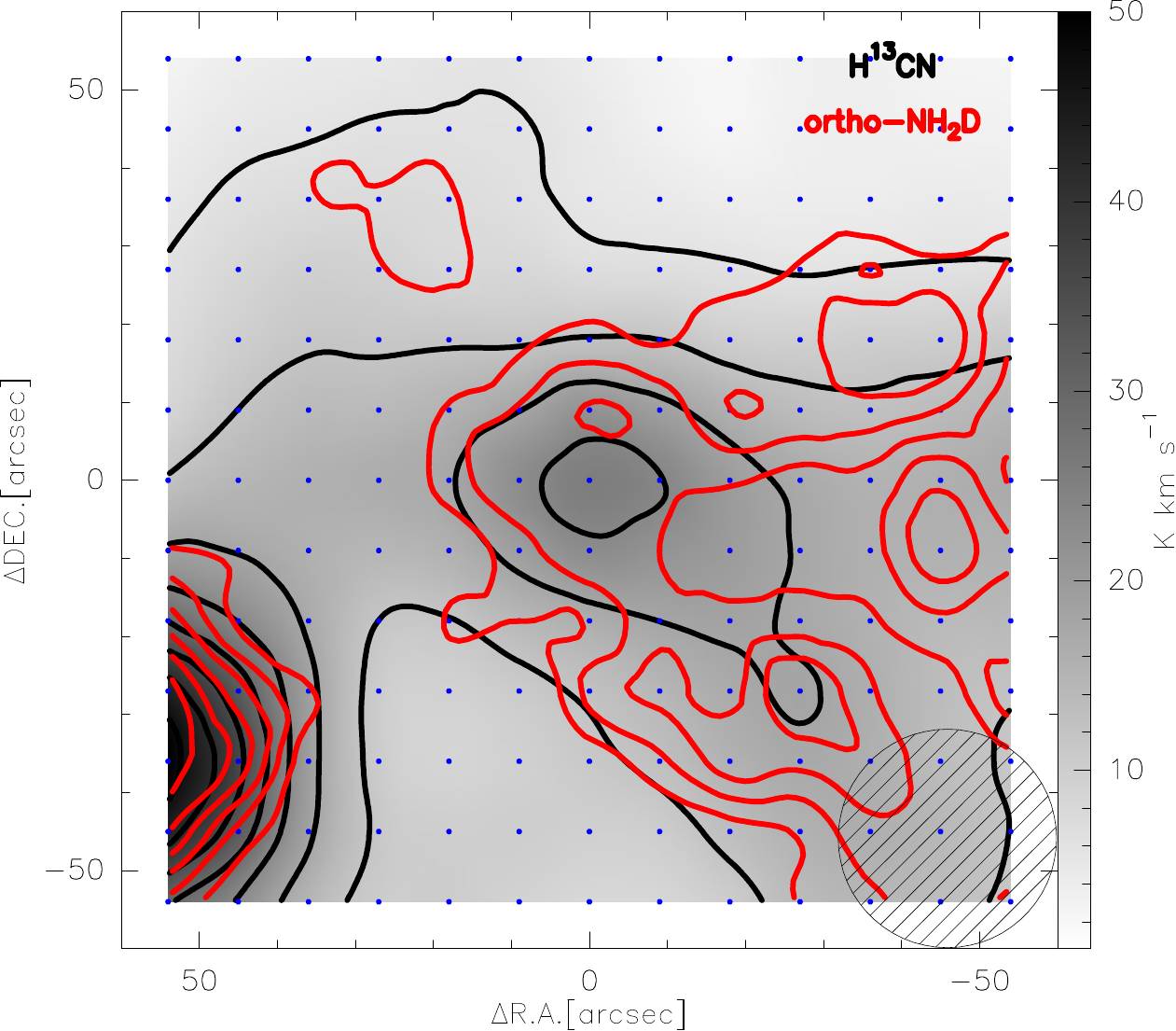}}\subfigure[]{\includegraphics[width=0.5\textwidth]{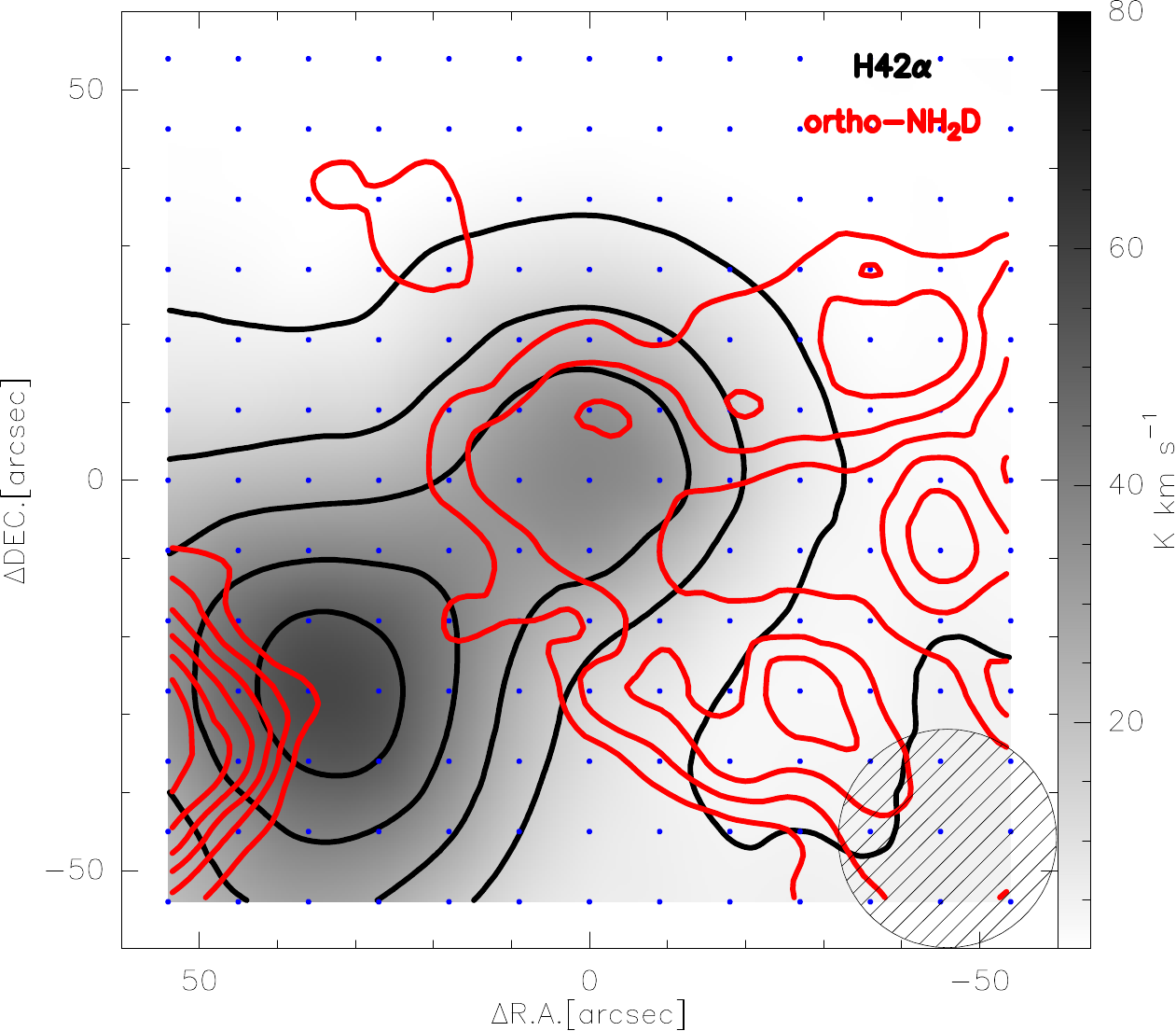}}
\subfigure[]{\includegraphics[width=0.5\textwidth]{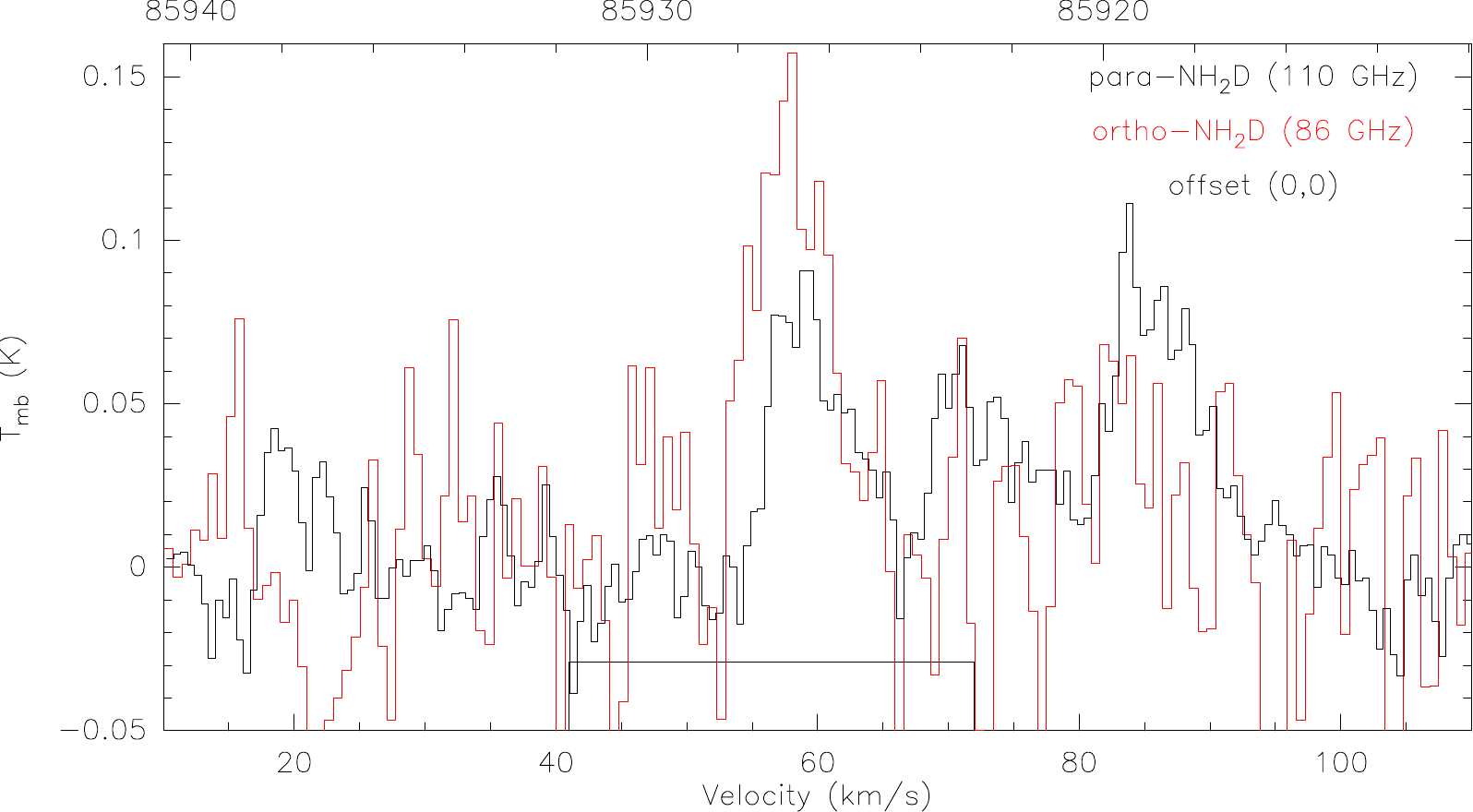}}\subfigure[]{\includegraphics[width=0.5\textwidth]{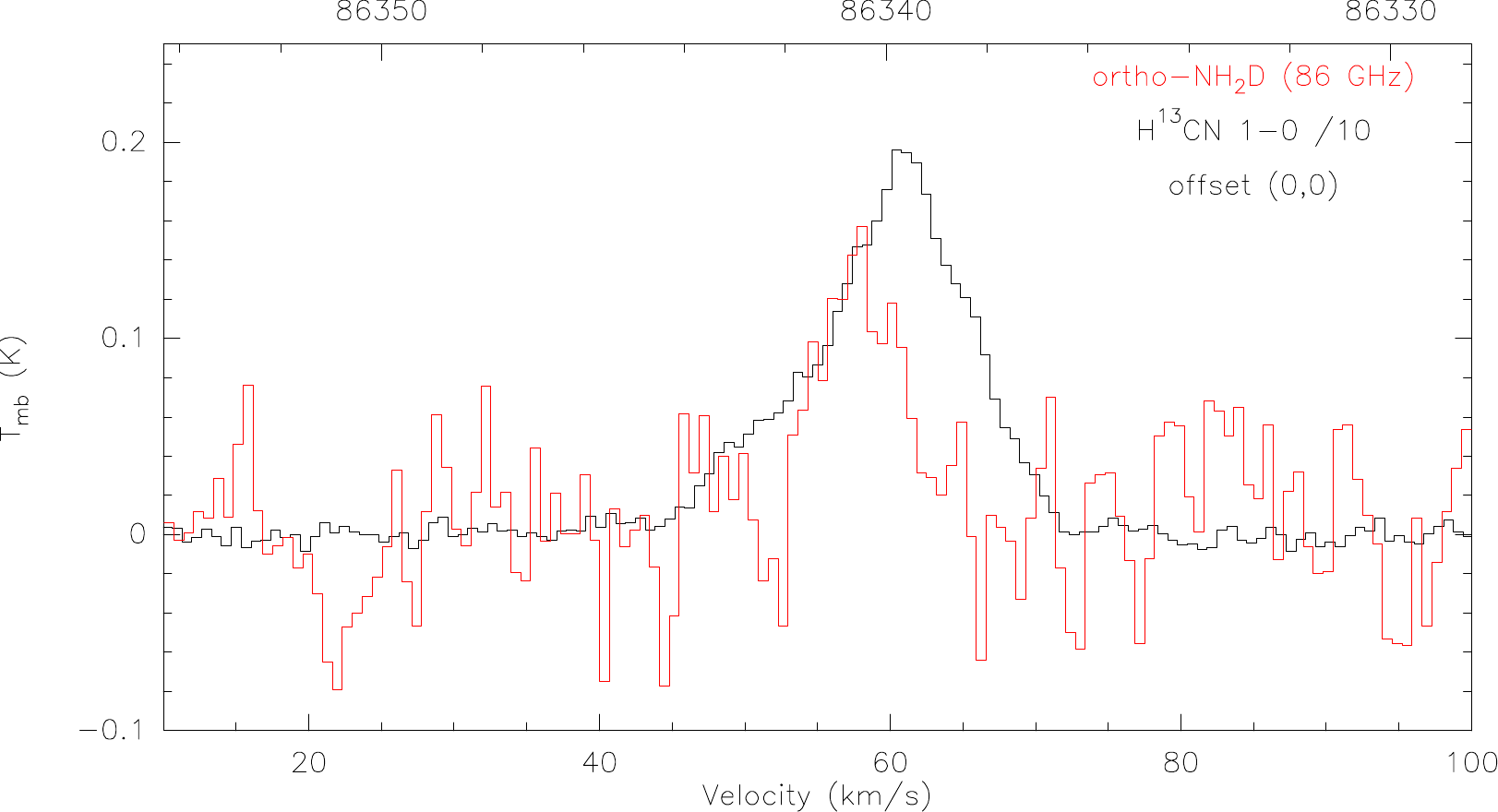}}
\caption{(a): NH$_2$D at 85.9263 GHz velocity integrated intensity contour (red contour) overlaid on H$^{13}$CN 1-0 velocity integrated intensity image (Gray scale and black contour) in G049.48-00.36. The contour levels start at 5$\sigma$ in steps of 2$\sigma$ for NH$_2$D, while the contour levels start at 30$\sigma$ in steps of 30$\sigma$ for H$^{13}$CN 1-0. The gray scale starts at 3$\sigma$. (b): NH$_2$D at 85.9263 GHz velocity integrated intensity contour (red contour) overlaid on H42$\alpha$ velocity integrated intensity image (Gray scale and black contour) in G049.48-00.36. The contour levels start at 5$\sigma$ in steps of 2$\sigma$ for NH$_2$D, while the contour levels start at 15$\sigma$ in steps of 40$\sigma$ for H42$\alpha$. The gray scale starts at 3$\sigma$. (c): Spectra of ortho-NH$_2$D at 85.9263 GHz and para-NH$_2$D at 110.1535 GHz in G049.48-00.36. The para-NH$_2$D was observed by IRAM-30m with position-switching mode. The red spectra is ortho-NH$_2$D and the black is para-NH$_2$D. The offset for two spectra is (0$^{\prime \prime}$,0$^{\prime \prime}$). The black box in the picture indicates the range of flux integration of ortho-NH$_2$D at 85.9263 GHz. (d): Spectra of ortho-NH$_2$D at 85.9263 GHz and H$^{13}$CN 1-0 in G049.48-00.36. The red spectra is ortho-NH$_2$D and the black is H$^{13}$CN 1-0. The offset for two spectra is (0$^{\prime \prime}$,0$^{\prime \prime}$).}
\label{app10}
\end{figure}

\begin{figure}
\centering
\subfigure[]{\includegraphics[width=0.5\textwidth]{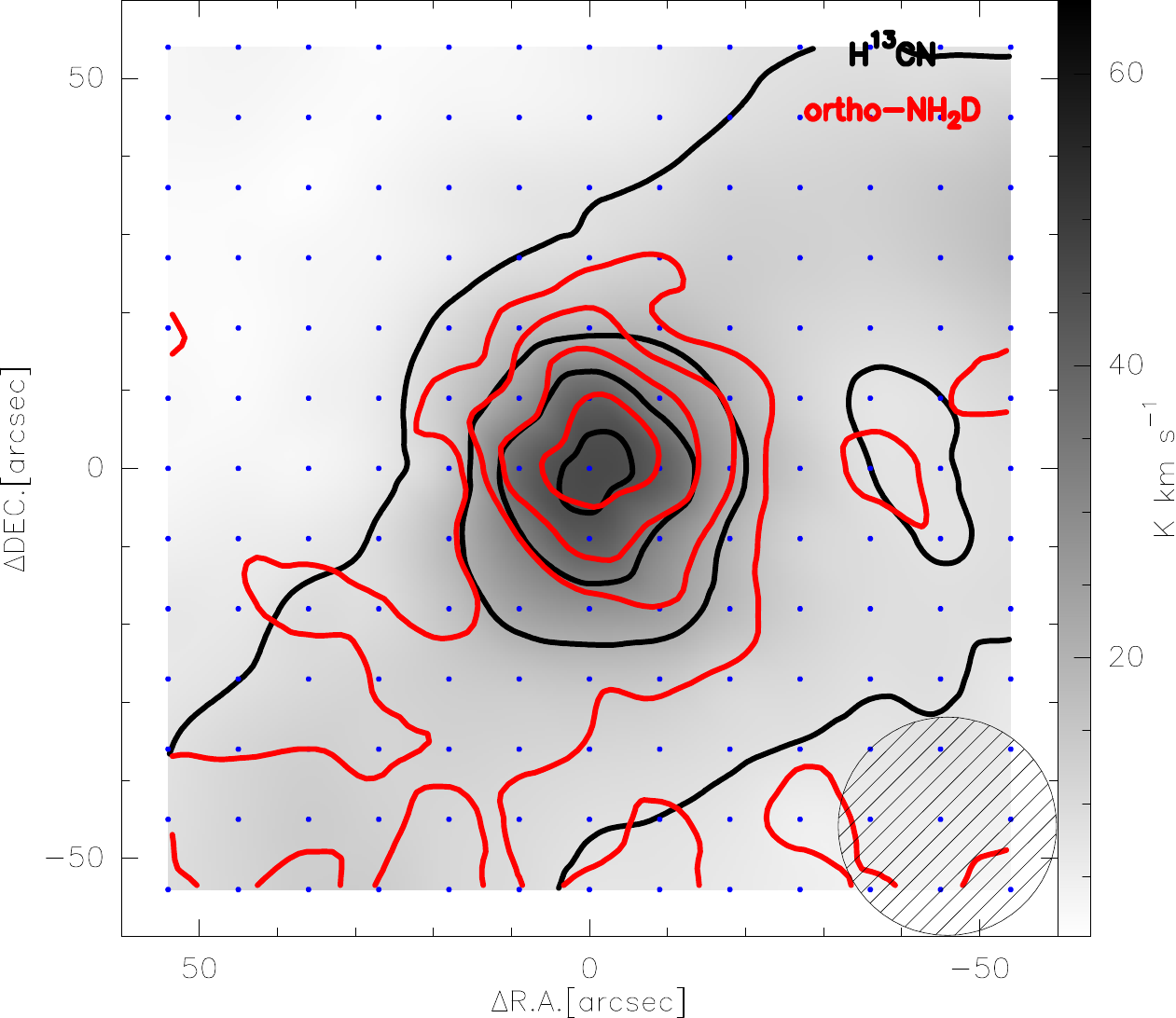}}\subfigure[]{\includegraphics[width=0.5\textwidth]{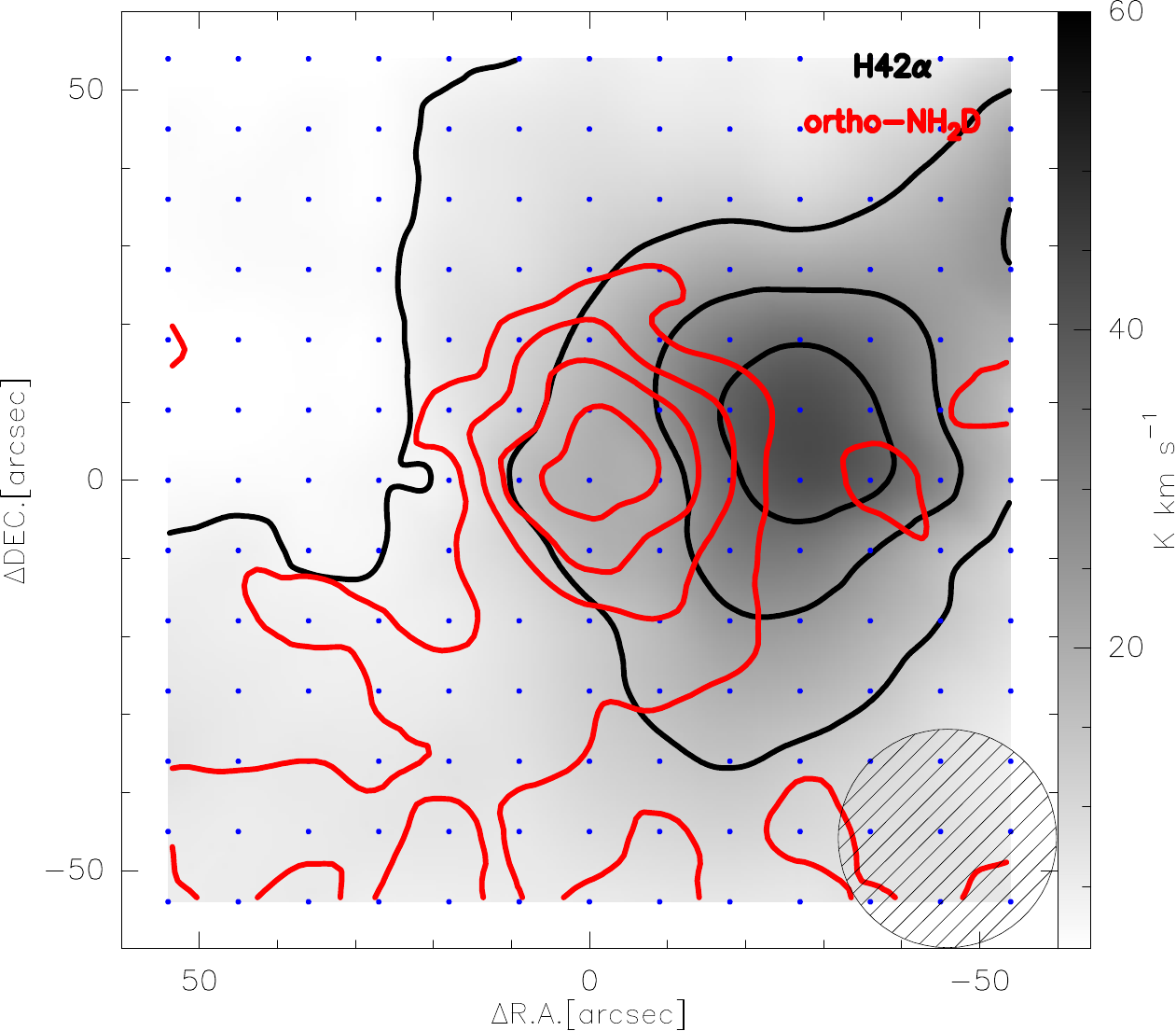}}
\subfigure[]{\includegraphics[width=0.5\textwidth]{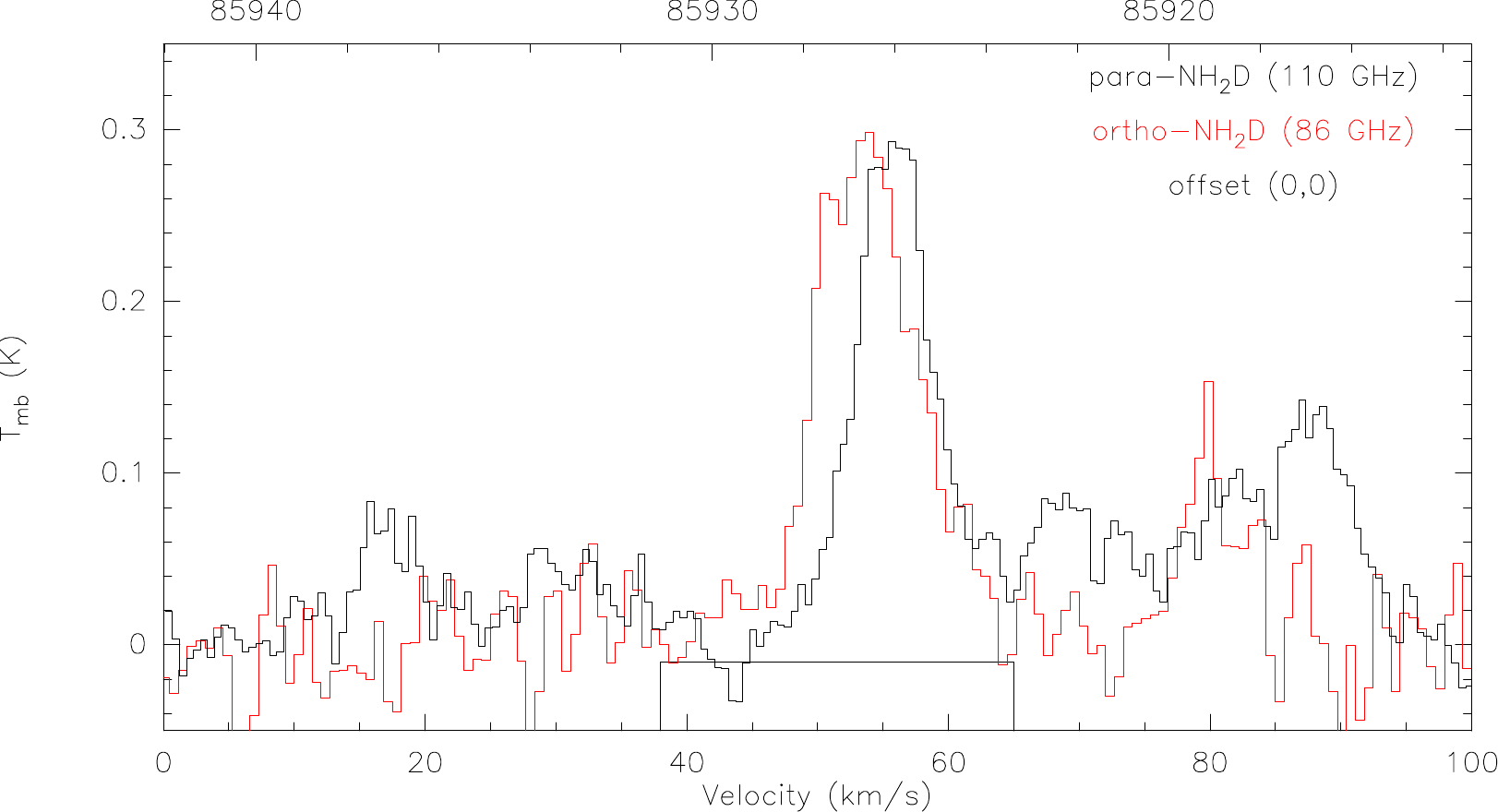}}\subfigure[]{\includegraphics[width=0.5\textwidth]{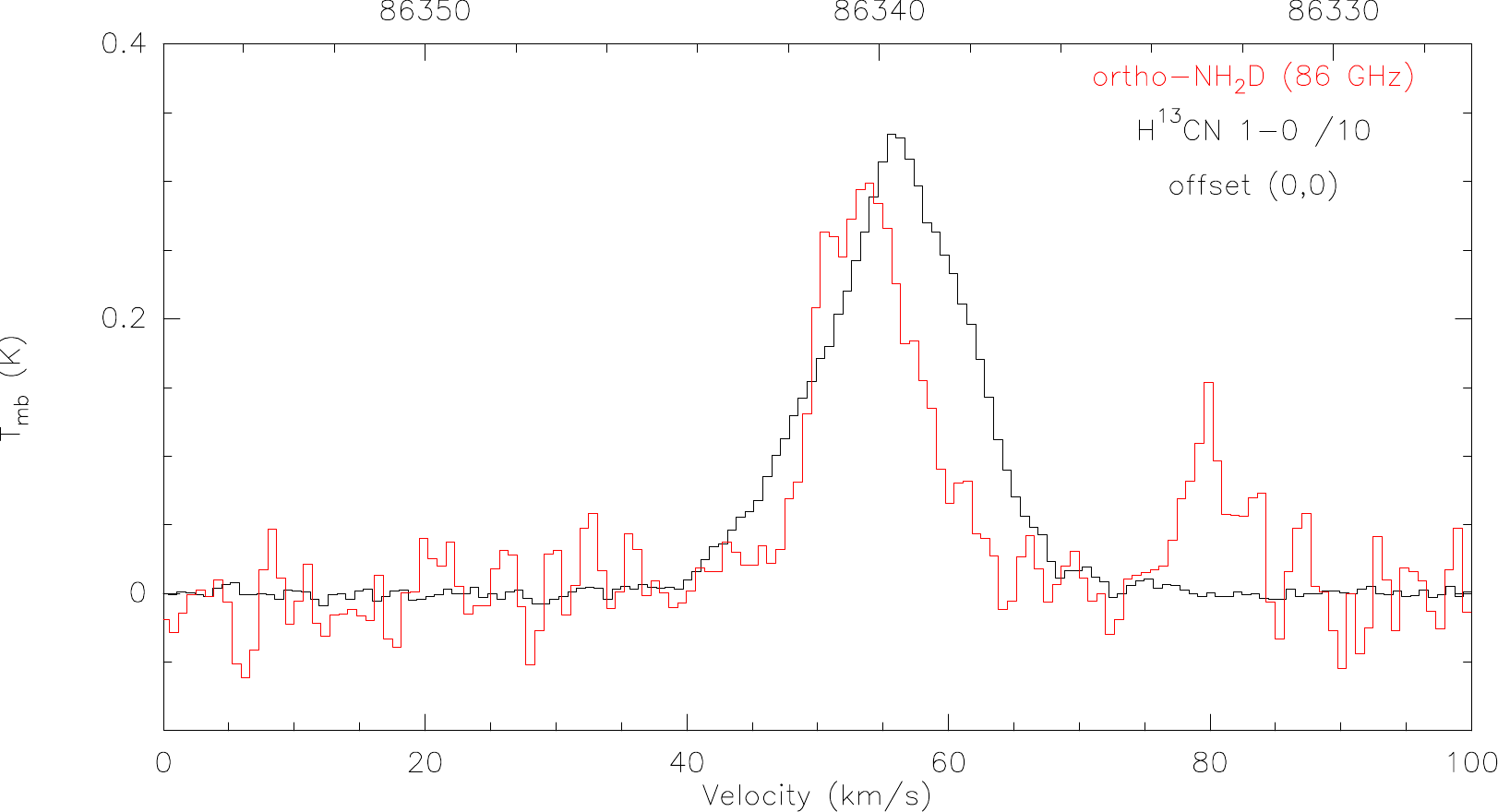}}
\subfigure[]{\includegraphics[width=0.5\textwidth]{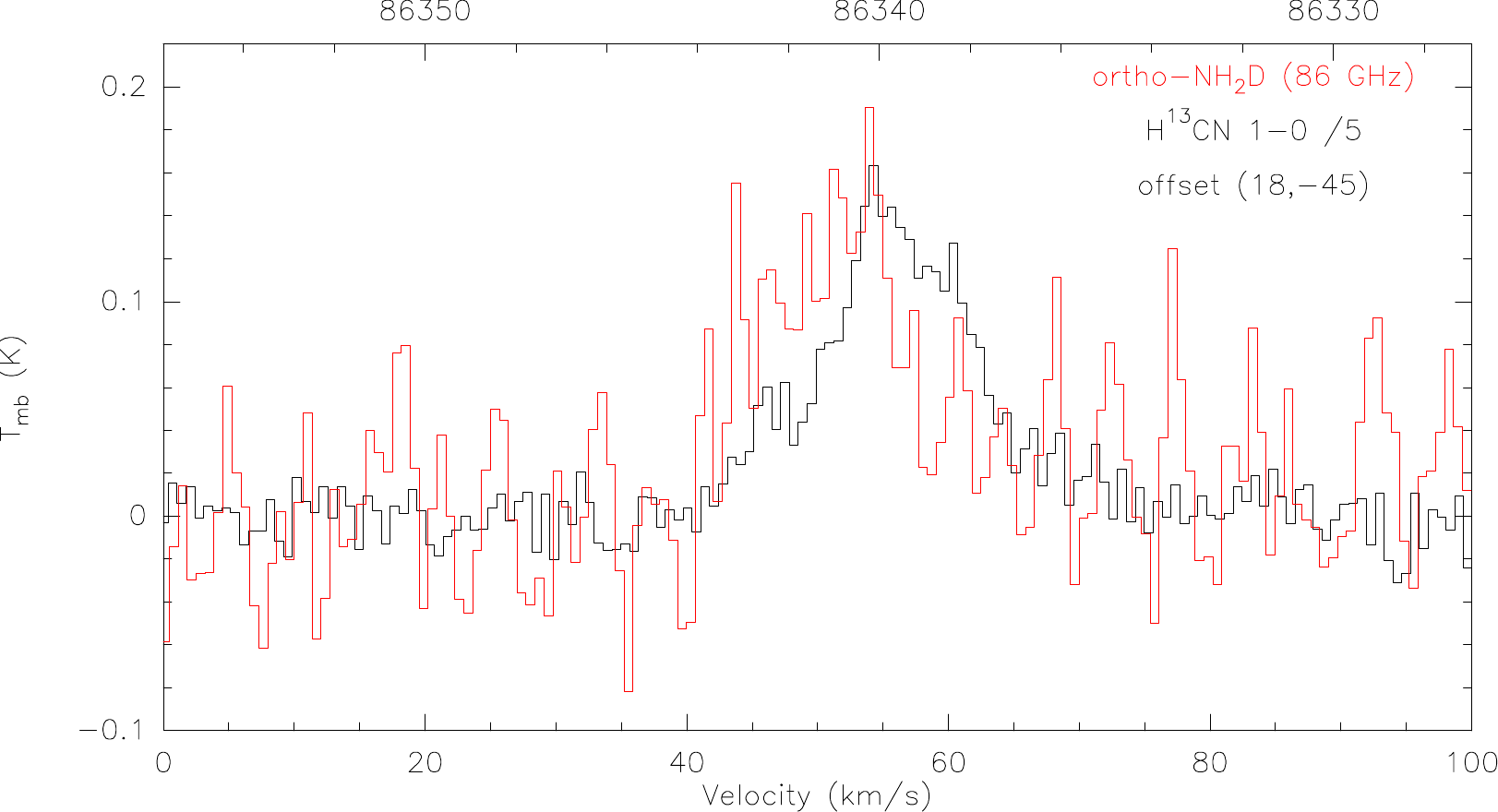}}
\caption{(a): NH$_2$D at 85.9263 GHz velocity integrated intensity contour (red contour) overlaid on H$^{13}$CN 1-0 velocity integrated intensity image (Gray scale and black contour) in G049.48-00.38. The contour levels start at 4$\sigma$ in steps of 3$\sigma$ for NH$_2$D, while the contour levels start at 27$\sigma$ in steps of 40$\sigma$ for H$^{13}$CN 1-0. The gray scale starts at 3$\sigma$. (b): NH$_2$D at 85.9263 GHz velocity integrated intensity contour (red contour) overlaid on H42$\alpha$ velocity integrated intensity image (Gray scale and black contour) in G049.48-00.38. The contour levels start at 4$\sigma$ in steps of 3$\sigma$ for NH$_2$D, while the contour levels start at 9$\sigma$ in steps of 30$\sigma$ for H42$\alpha$. The gray scale starts at 3$\sigma$. (c): Spectra of ortho-NH$_2$D at 85.9263 GHz and para-NH$_2$D at 110.1535 GHz in G049.48-00.38. The para-NH$_2$D was observed by IRAM-30m with position-switching mode. The red spectra is ortho-NH$_2$D and the black is para-NH$_2$D. The offset for two spectra is (0$^{\prime \prime}$,0$^{\prime \prime}$). (d): Spectra of ortho-NH$_2$D at 85.9263 GHz and H$^{13}$CN 1-0 in G049.48-00.38. The red spectra is ortho-NH$_2$D and the black is H$^{13}$CN 1-0. The offset for two spectra is (0$^{\prime \prime}$,0$^{\prime \prime}$). The black box in the picture indicates the range of flux integration of ortho-NH$_2$D at 85.9263 GHz. (e): Spectra of ortho-NH$_2$D at 85.9263 GHz and H$^{13}$CN 1-0 in G049.48-00.38.  The red spectra is ortho-NH$_2$D and the black is H$^{13}$CN 1-0. The offset for two spectra is (+18$^{\prime \prime}$,-45$^{\prime \prime}$).}
\label{app11}
\end{figure}

\begin{figure}
\centering
\subfigure[]{\includegraphics[width=0.5\textwidth]{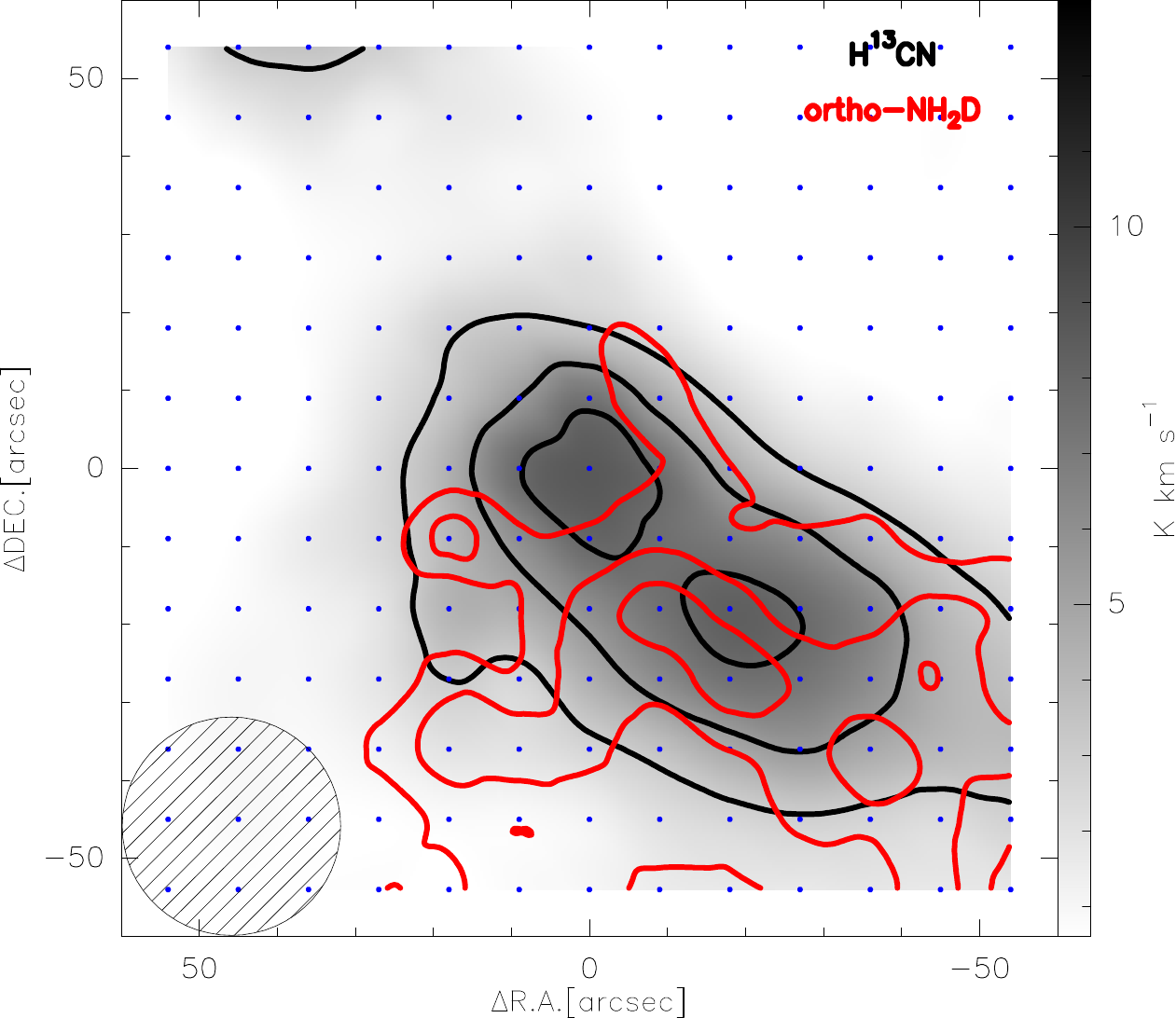}}\subfigure[]{\includegraphics[width=0.5\textwidth]{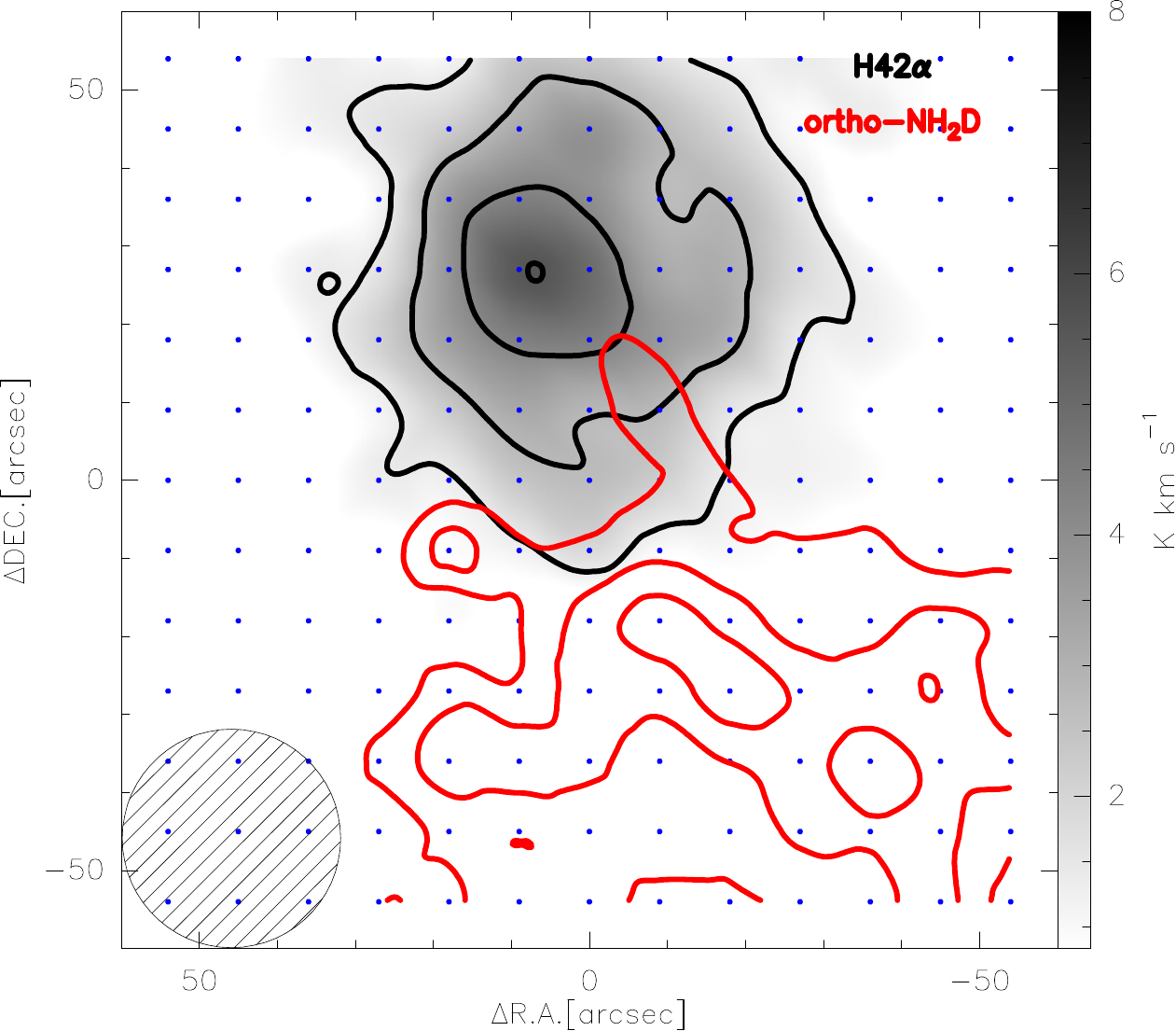}}
\subfigure[]{\includegraphics[width=0.5\textwidth]{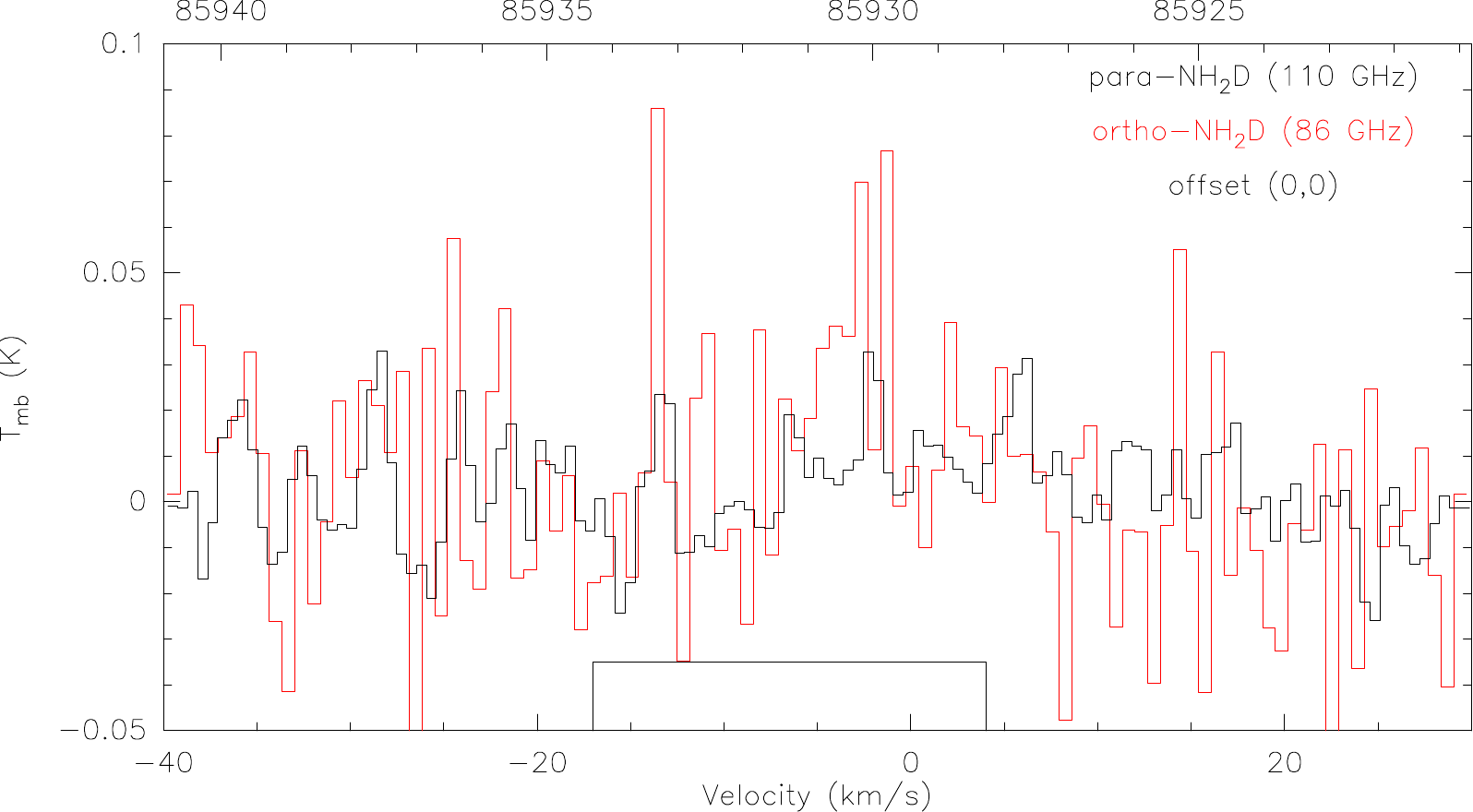}}\subfigure[]{\includegraphics[width=0.5\textwidth]{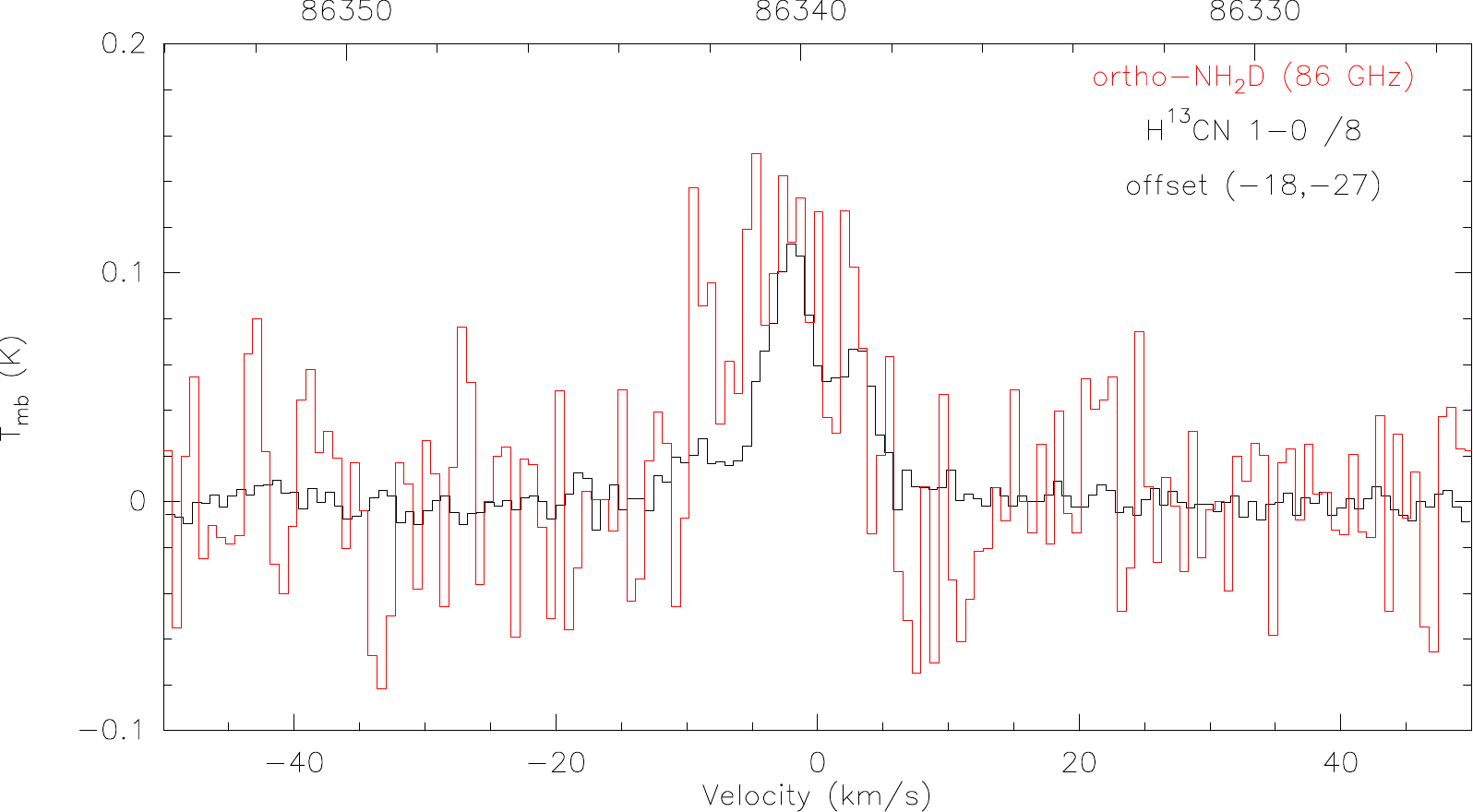}}
\caption{(a): NH$_2$D at 85.9263 GHz velocity integrated intensity contour (red contour) overlaid on H$^{13}$CN 1-0 velocity integrated intensity image (Gray scale and black contour) in G075.76+00.33. The contour levels start at 3$\sigma$ in steps of 2$\sigma$ for NH$_2$D, while the contour levels start at 15$\sigma$ in steps of 12$\sigma$ for H$^{13}$CN 1-0. The gray scale starts at 3$\sigma$. (b): NH$_2$D at 85.9263 GHz velocity integrated intensity contour (red contour) overlaid on H42$\alpha$ velocity integrated intensity image (Gray scale and black contour) in G075.76+00.33. The contour levels start at 3$\sigma$ in steps of 2$\sigma$ for NH$_2$D, while the contour levels start at 5$\sigma$ in steps of 5$\sigma$ for H42$\alpha$. The gray scale starts at 3$\sigma$. (c): Spectra of ortho-NH$_2$D at 85.9263 GHz and para-NH$_2$D at 110.1535 GHz in G075.76+00.33. The para-NH$_2$D was observed by IRAM-30m with position-switching mode. The red spectra is ortho-NH$_2$D and the black is para-NH$_2$D. The offset for two spectra is (0$^{\prime \prime}$,0$^{\prime \prime}$). The black box in the picture indicates the range of flux integration of ortho-NH$_2$D at 85.9263 GHz. (d): Spectra of ortho-NH$_2$D at 85.9263 GHz and H$^{13}$CN 1-0 in G075.76+00.33. The red spectra is ortho-NH$_2$D and the black is H$^{13}$CN 1-0. The offset for two spectra is (-18$^{\prime \prime}$,-27$^{\prime \prime}$).}
\label{app12}
\end{figure}

\begin{figure}
\centering
\subfigure[]{\includegraphics[width=0.4\textwidth]{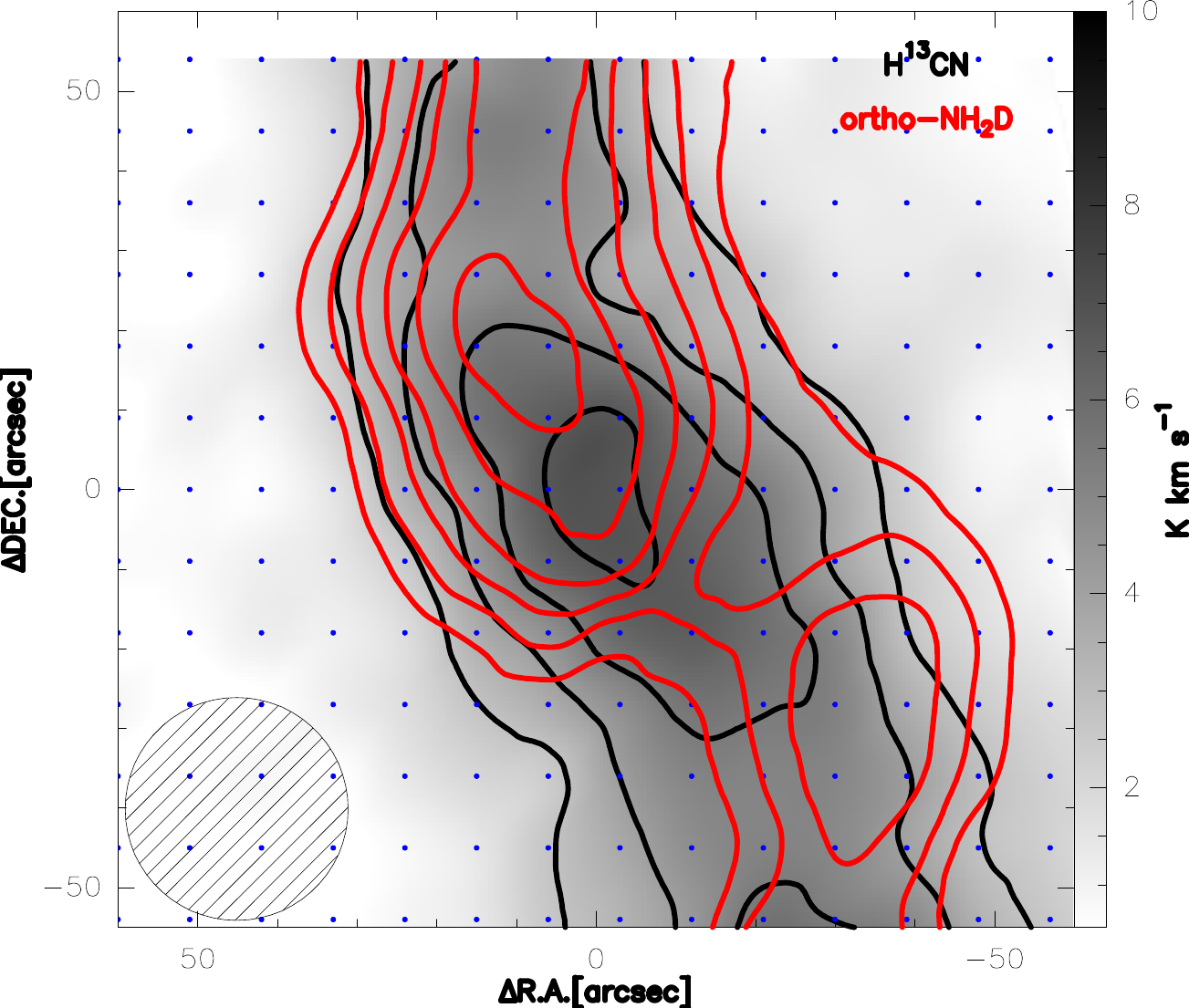}}\subfigure[]{\includegraphics[width=0.4\textwidth]{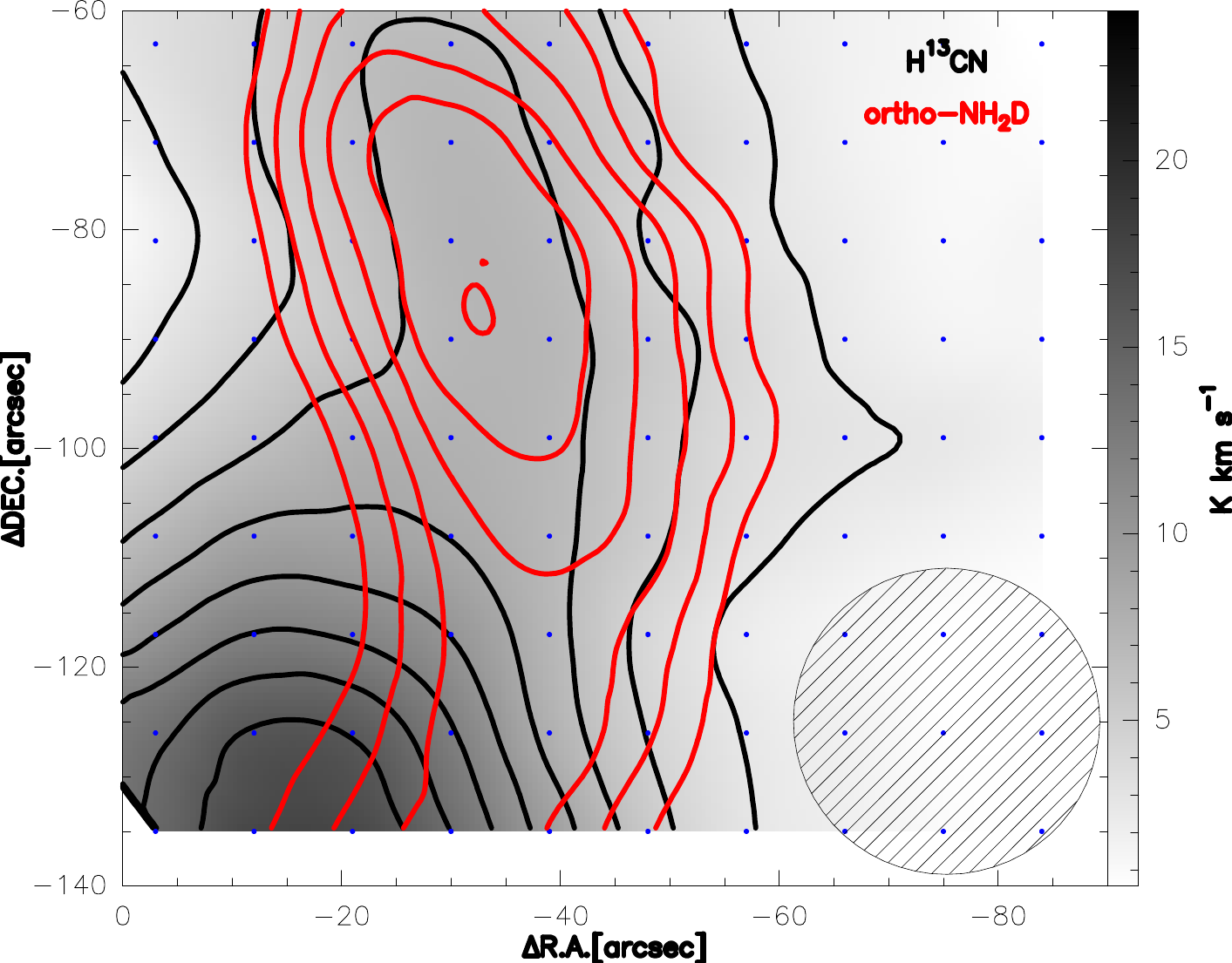}}
\subfigure[]{\includegraphics[width=0.3\textwidth]{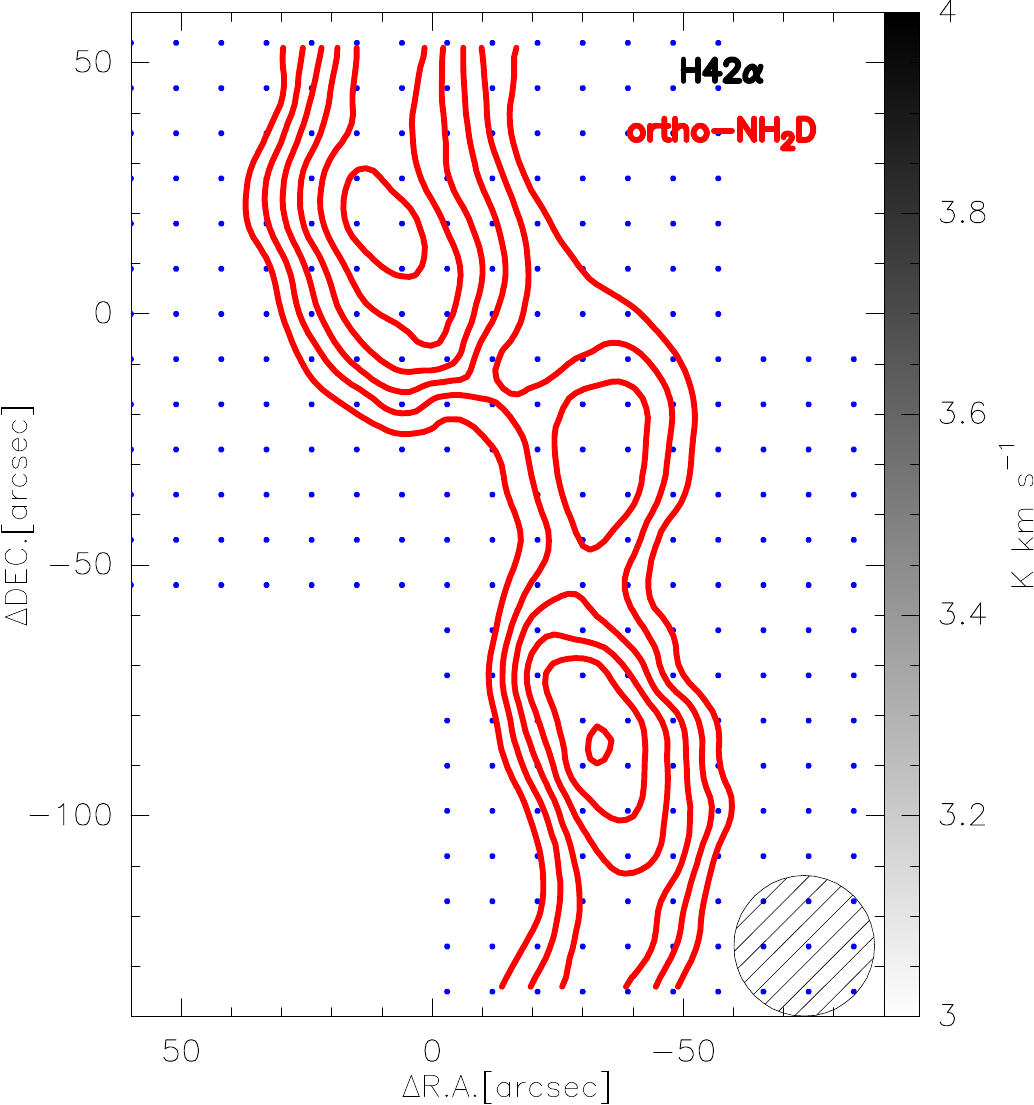}}\subfigure[]{\includegraphics[width=0.5\textwidth]{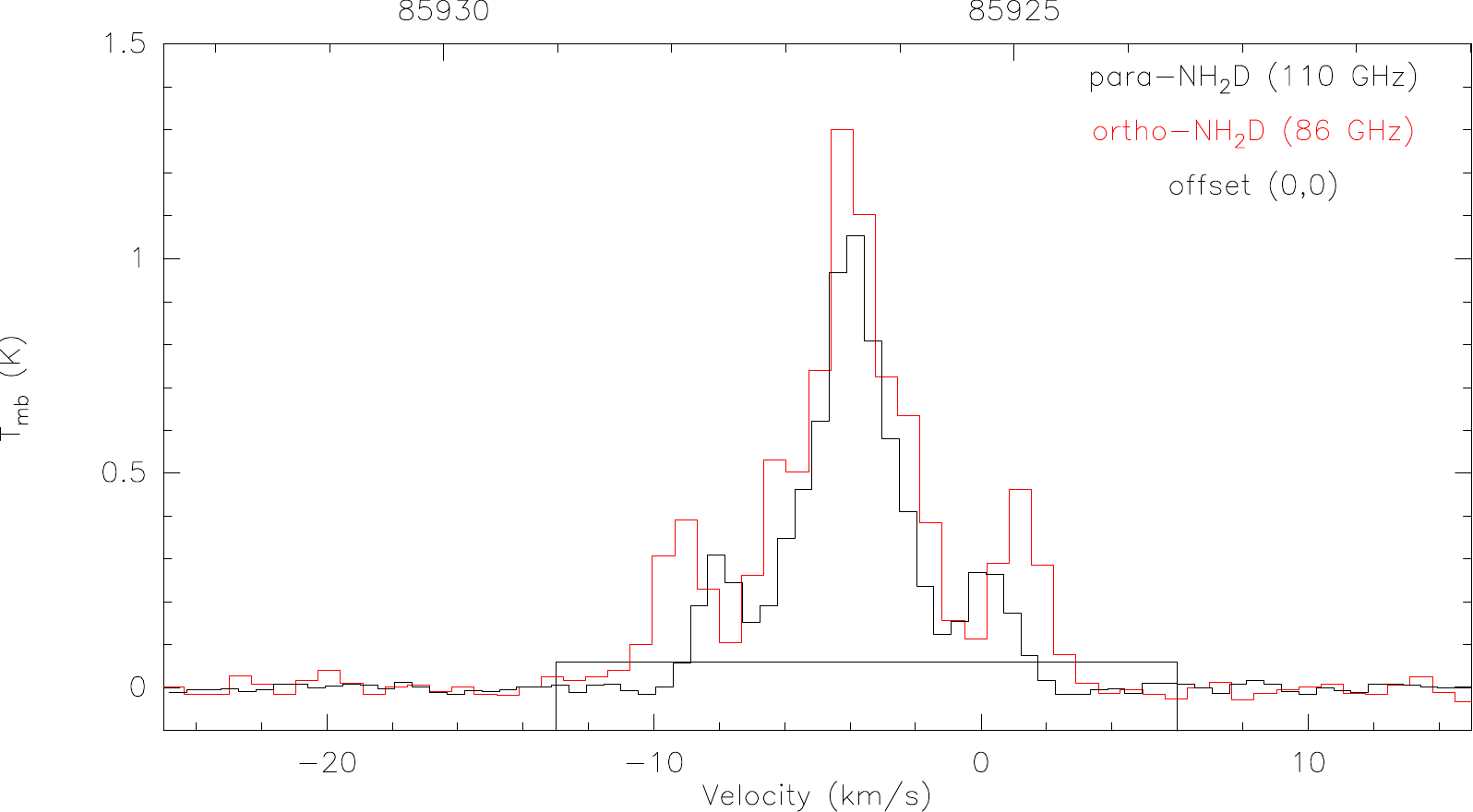}}
\subfigure[]{\includegraphics[width=0.5\textwidth]{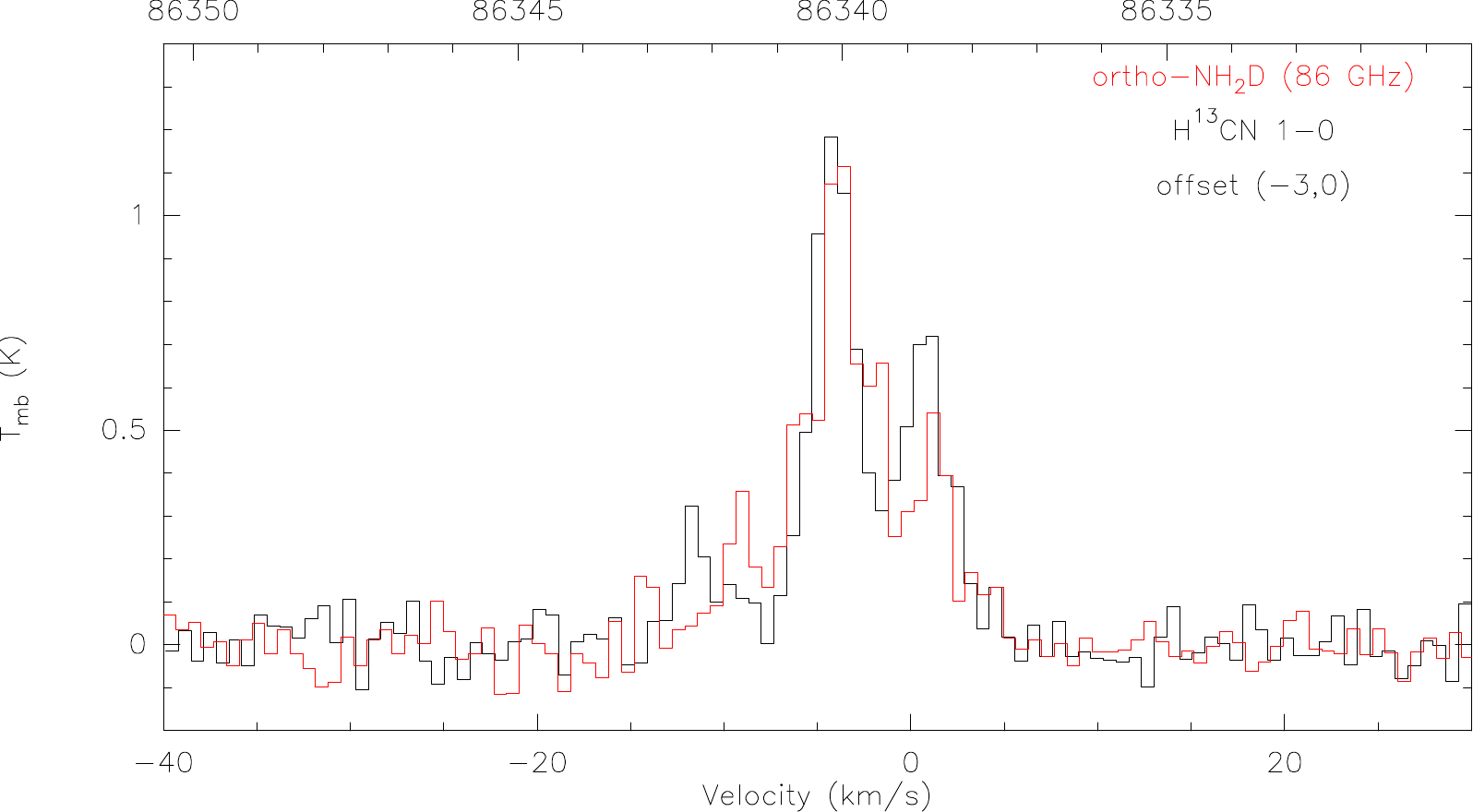}}\subfigure[]{\includegraphics[width=0.5\textwidth]{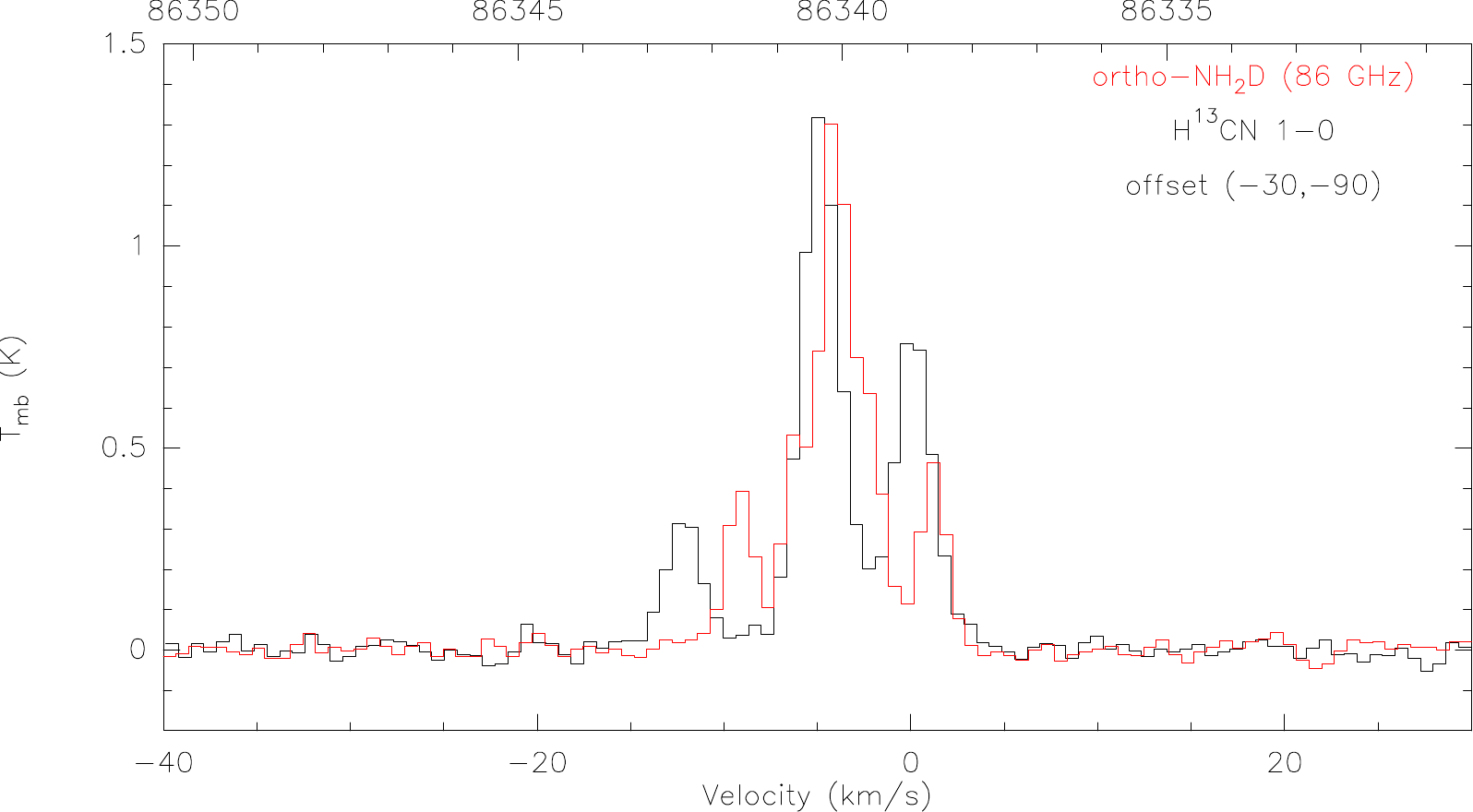}}
\caption{(a): NH$_2$D at 85.9263 GHz velocity integrated intensity contour (red contour) overlaid on H$^{13}$CN 1-0 velocity integrated intensity image (Gray scale and black contour) in G081.75+00.59 (W 75S). The contour levels start at 18$\sigma$ in steps of 5$\sigma$ for NH$_2$D, while the contour levels start at 15$\sigma$ in steps of 7$\sigma$ for H$^{13}$CN 1-0. The gray scale starts at 3$\sigma$. (b): NH$_2$D at 85.9263 GHz velocity integrated intensity contour (red contour) overlaid on H$^{13}$CN 1-0 velocity integrated intensity image (Gray scale and black contour) in G081.75+00.59 (DR 21(OH)). The contour levels start at 18$\sigma$ in steps of 5$\sigma$ for NH$_2$D, while the contour levels start at 15$\sigma$ in steps of 10$\sigma$ for H$^{13}$CN 1-0. The gray scale starts at 3$\sigma$. (c): NH$_2$D at 85.9263 GHz velocity integrated intensity contour (red contour) overlaid on H42$\alpha$ velocity integrated intensity image (Gray scale and black contour, while the H42$\alpha$ is not detected) in G081.75+00.59. The contour levels start at 18$\sigma$ in steps of 5$\sigma$ for NH$_2$D. (d): Spectra of ortho-NH$_2$D at 85.9263 GHz and para-NH$_2$D at 110.1535 GHz in G081.75+00.59. The para-NH$_2$D was observed by IRAM-30m with position-switching mode. The red spectra is ortho-NH$_2$D and the black is para-NH$_2$D. The offset for two spectra is (0$^{\prime \prime}$,-3$^{\prime \prime}$). The black box in the picture indicates the range of flux integration of ortho-NH$_2$D at 85.9263 GHz. (e): Spectra of ortho-NH$_2$D at 85.9263 GHz and H$^{13}$CN 1-0 in G081.75+00.59. The red spectra is ortho-NH$_2$D and the black is H$^{13}$CN 1-0. The offset for two spectra is (-3$^{\prime \prime}$,0$^{\prime \prime}$). (f): Spectra of ortho-NH$_2$D at 85.9263 GHz and H$^{13}$CN 1-0 in G081.75+00.59.  The red spectra is ortho-NH$_2$D and the black is H$^{13}$CN 1-0. The offset for two spectra is (-30$^{\prime \prime}$,-90$^{\prime \prime}$).}
\label{app13}
\end{figure}

\begin{figure}
\centering
\subfigure[]{\includegraphics[width=0.5\textwidth]{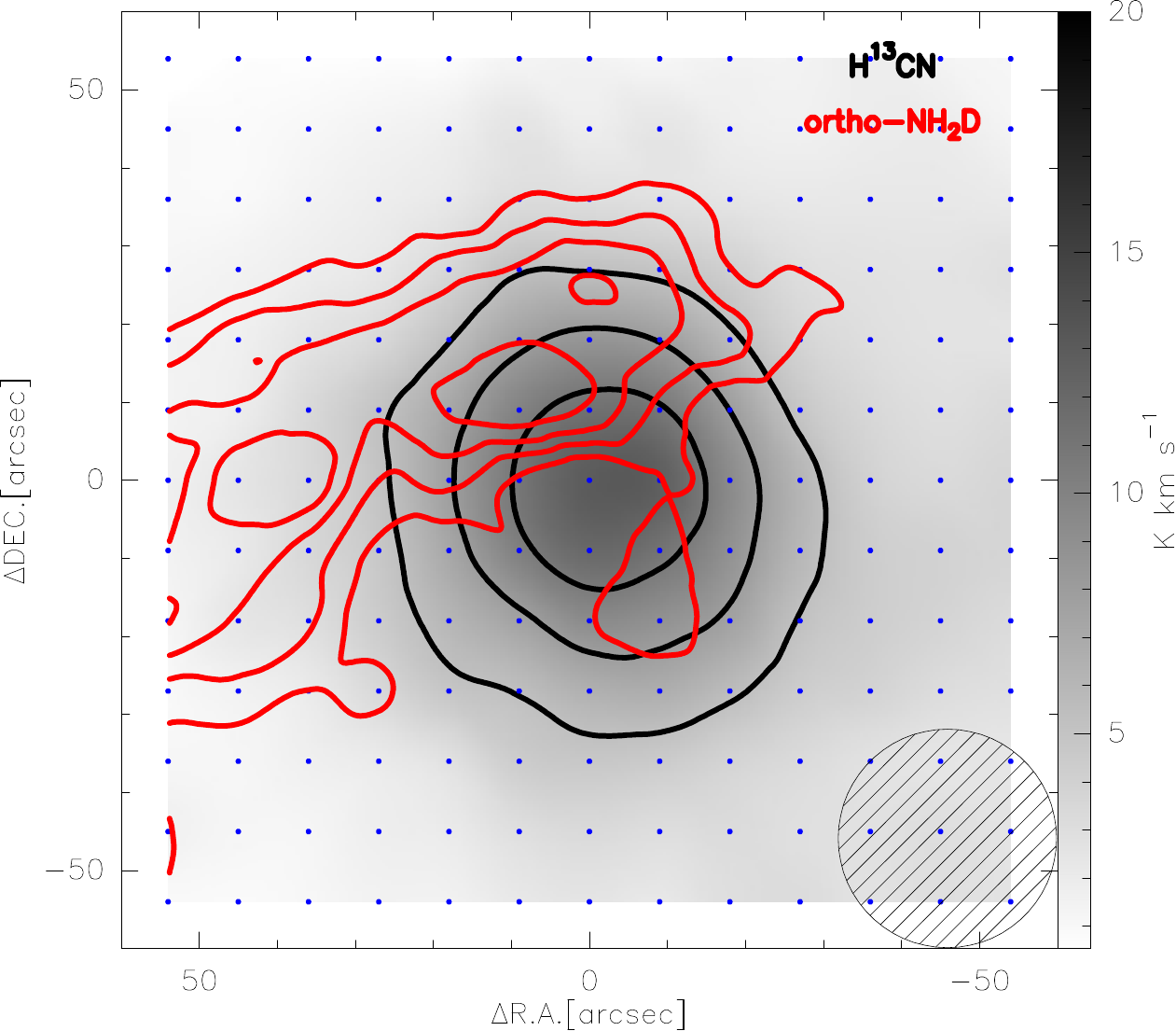}}\subfigure[]{\includegraphics[width=0.5\textwidth]{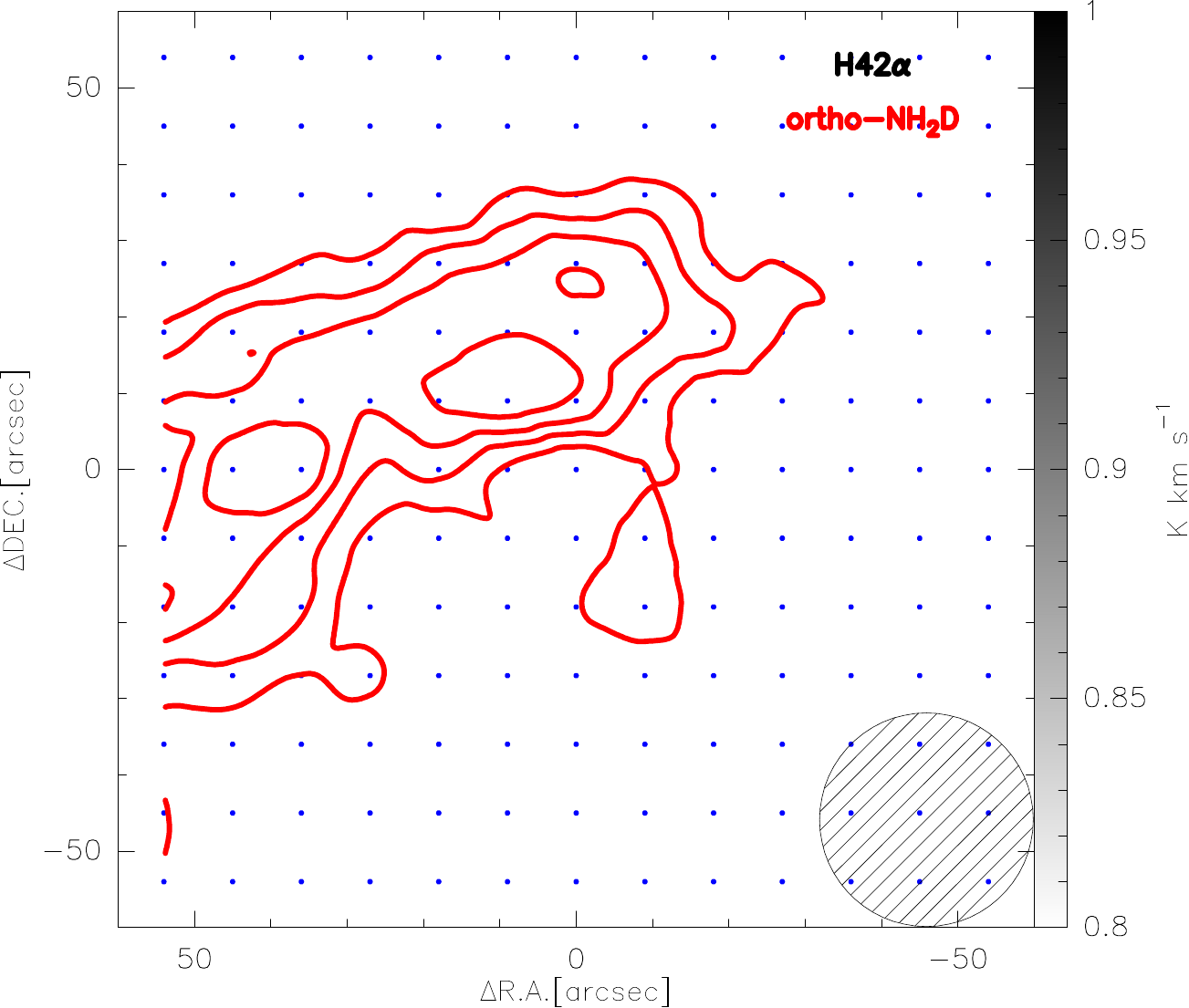}}
\subfigure[]{\includegraphics[width=0.5\textwidth]{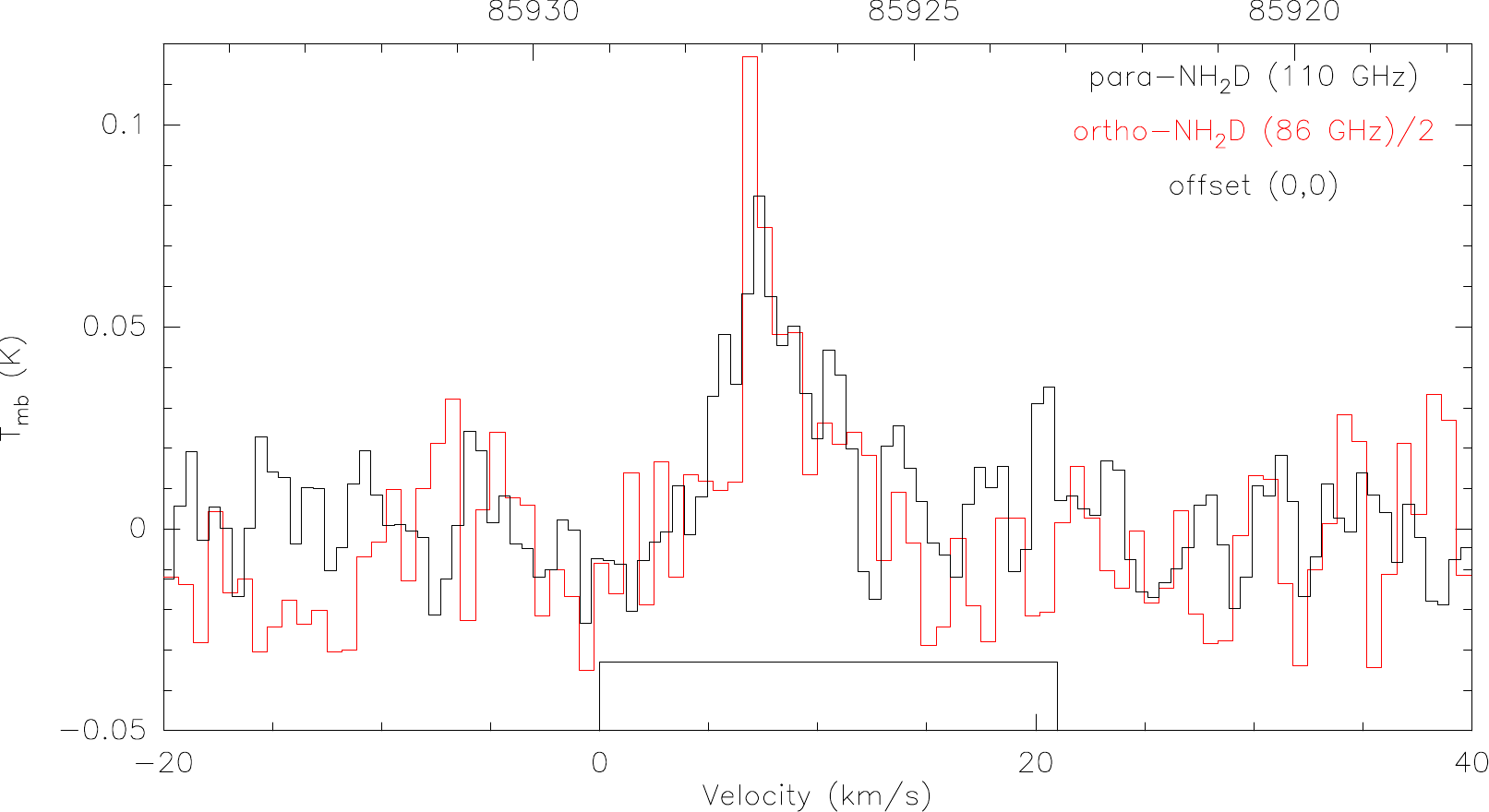}}\subfigure[]{\includegraphics[width=0.5\textwidth]{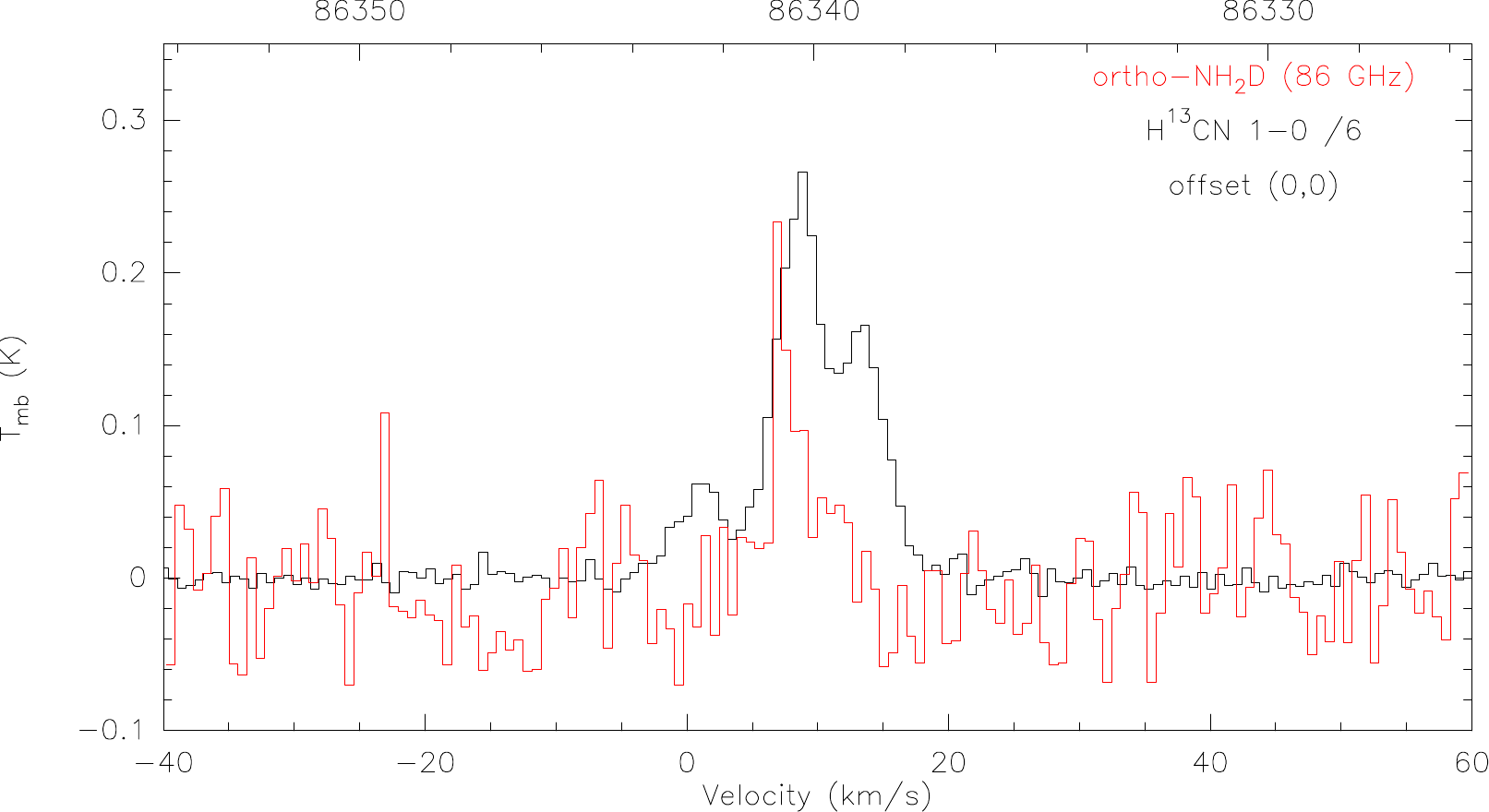}}
\subfigure[]{\includegraphics[width=0.5\textwidth]{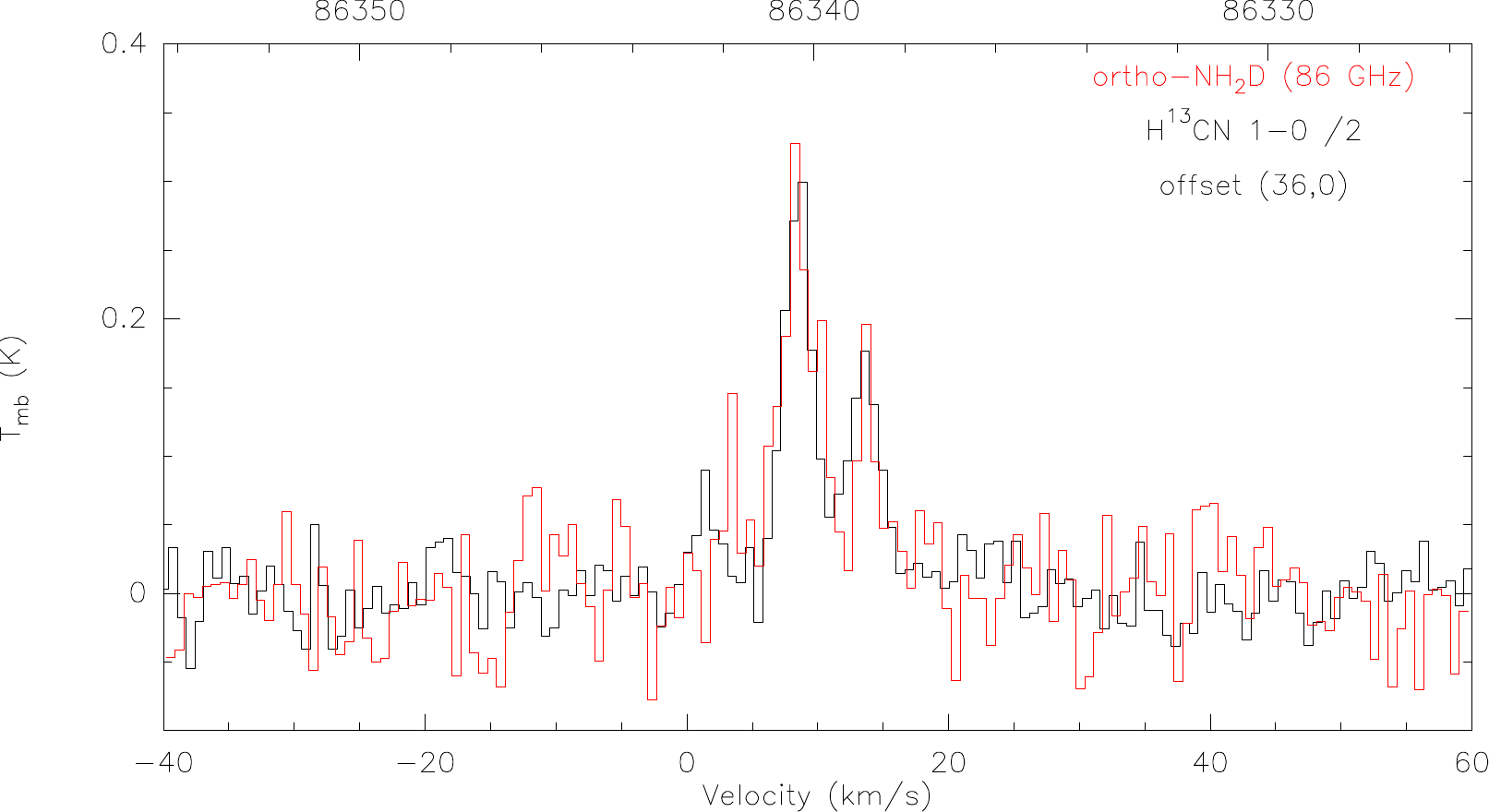}}
\caption{(a): NH$_2$D at 85.9263 GHz velocity integrated intensity contour (red contour) overlaid on H$^{13}$CN 1-0 velocity integrated intensity image (Gray scale and black contour) in G081.87+00.78. The contour levels start at 5$\sigma$ in steps of 2$\sigma$ for NH$_2$D, while the contour levels start at 30$\sigma$ in steps of 15$\sigma$ for H$^{13}$CN 1-0. The gray scale starts at 3$\sigma$. (b): NH$_2$D at 85.9263 GHz velocity integrated intensity contour (red contour) overlaid on H42$\alpha$ velocity integrated intensity image (Gray scale and black contour, while the H42$\alpha$ is not detected) in G081.87+00.78. The contour levels start at 5$\sigma$ in steps of 2$\sigma$ for NH$_2$D. (c): Spectra of ortho-NH$_2$D at 85.9263 GHz and para-NH$_2$D at 110.1535 GHz in G081.87+00.78. The para-NH$_2$D was observed by IRAM-30m with position-switching mode. The red spectra is ortho-NH$_2$D and the black is para-NH$_2$D. The offset for two spectra is (0$^{\prime \prime}$,0$^{\prime \prime}$). The black box in the picture indicates the range of flux integration of ortho-NH$_2$D at 85.9263 GHz. (d): Spectra of ortho-NH$_2$D at 85.9263 GHz and H$^{13}$CN 1-0 in G081.87+00.78. The red spectra is ortho-NH$_2$D and the black is H$^{13}$CN 1-0. The offset for two spectra is (0$^{\prime \prime}$,0$^{\prime \prime}$). (e): Spectra of ortho-NH$_2$D at 85.9263 GHz and H$^{13}$CN 1-0 in G081.87+00.78.  The red spectra is ortho-NH$_2$D and the black is H$^{13}$CN 1-0. The offset for two spectra is (+36$^{\prime \prime}$,0$^{\prime \prime}$).}
\label{app14}
\end{figure}

\begin{figure}
\centering
\subfigure[]{\includegraphics[width=0.3\textwidth]{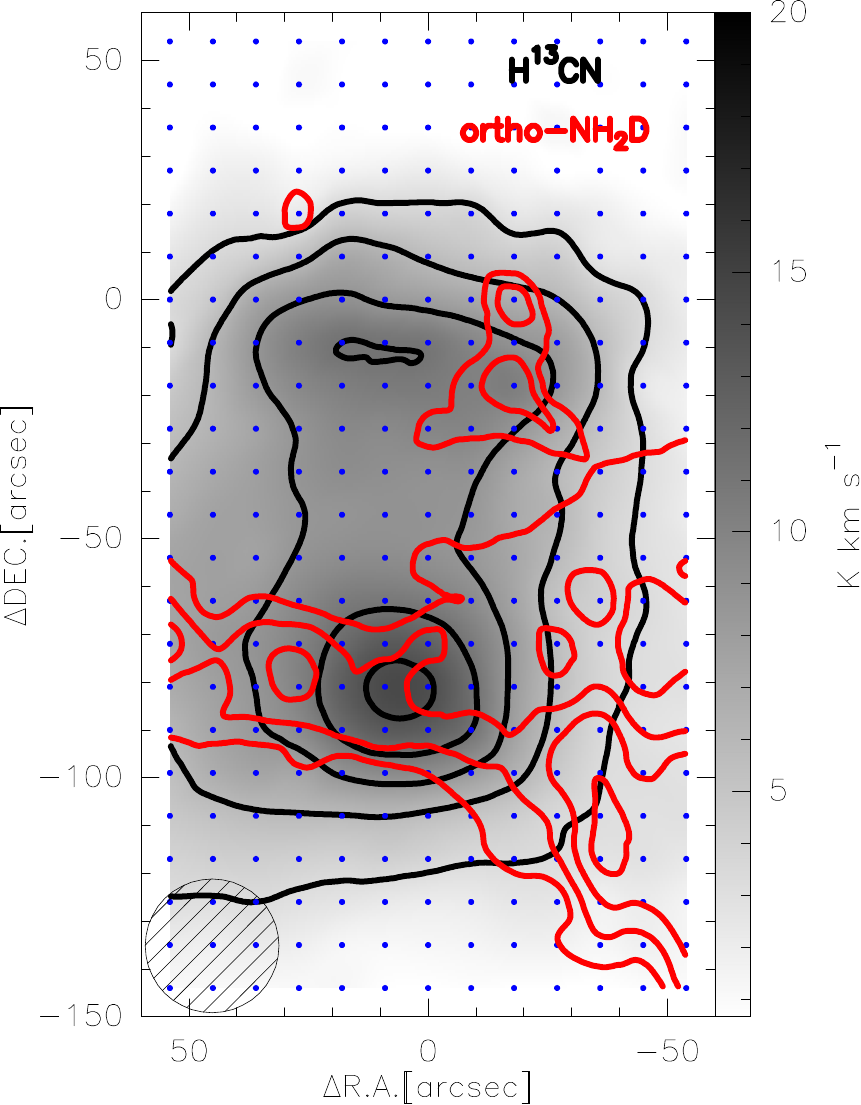}}\subfigure[]{\includegraphics[width=0.3\textwidth]{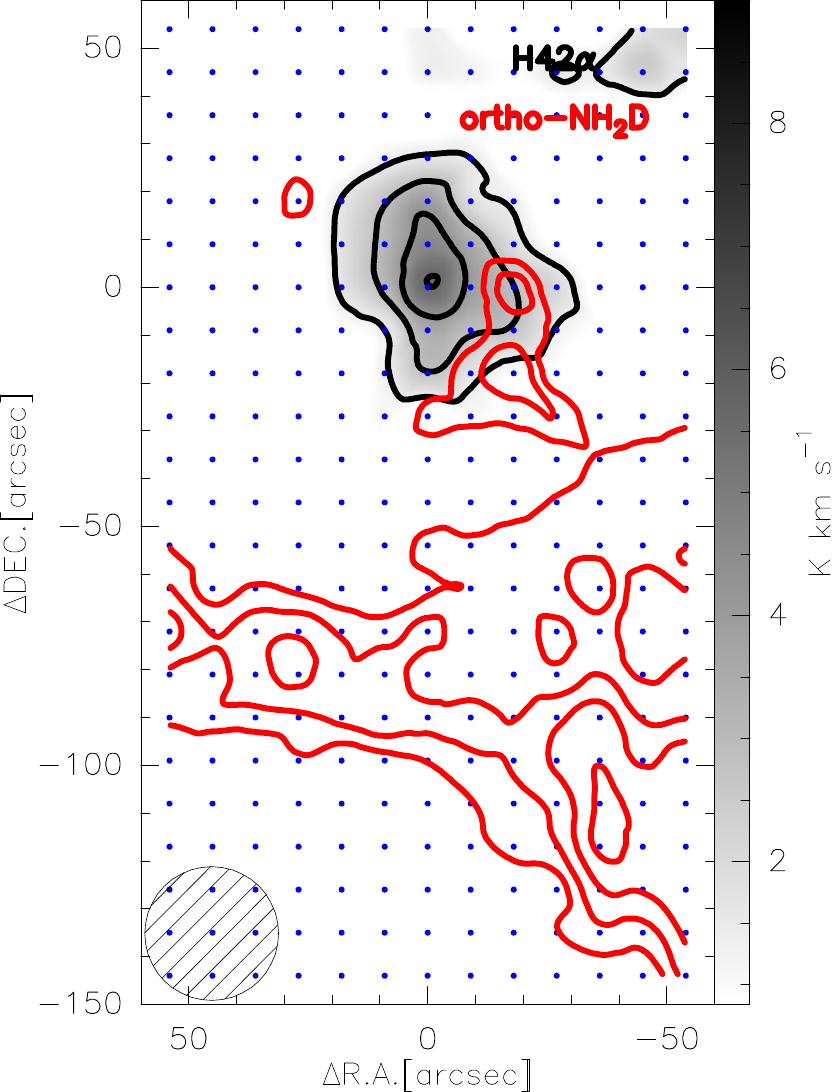}}
\subfigure[]{\includegraphics[width=0.5\textwidth]{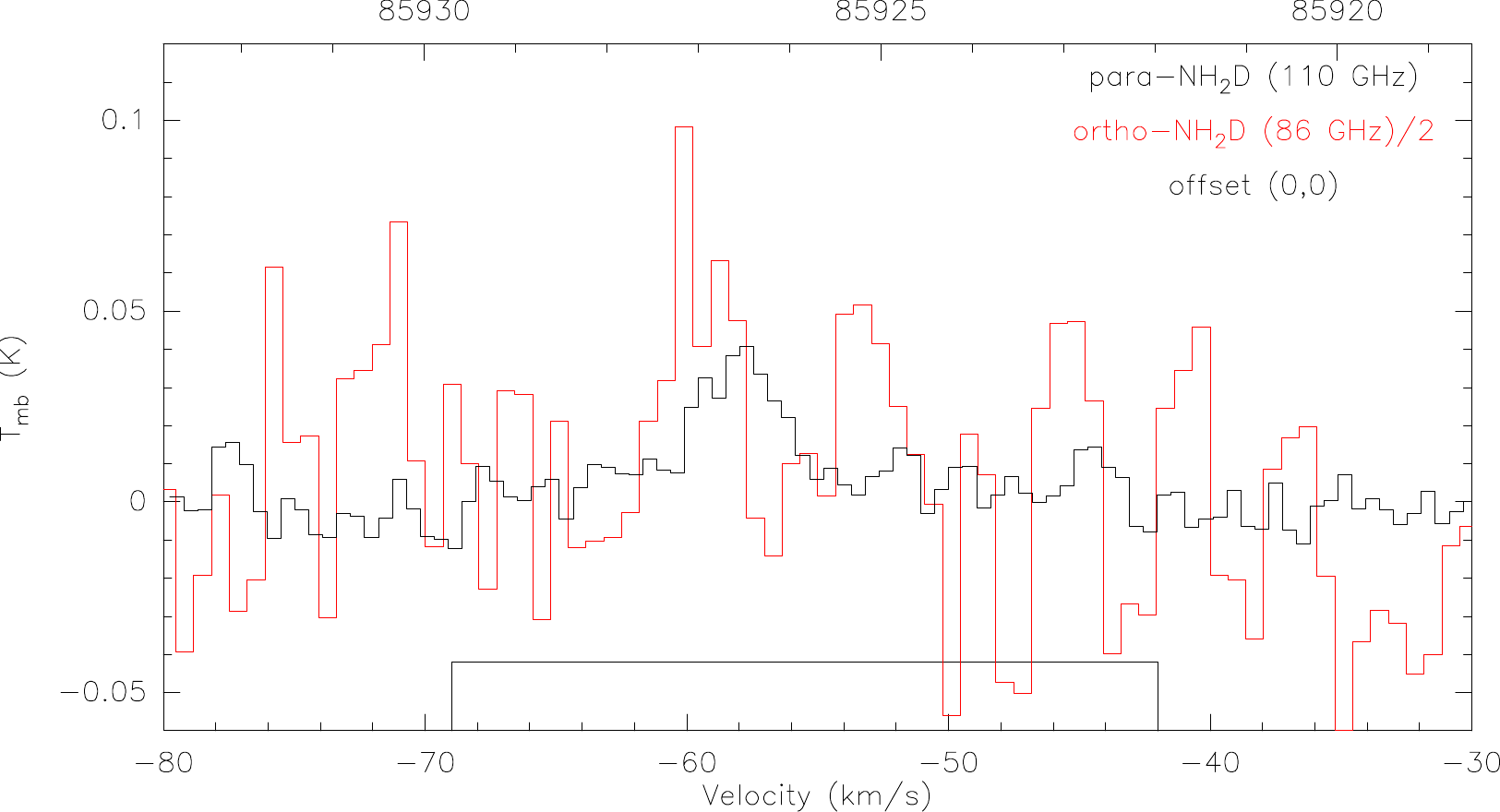}}\subfigure[]{\includegraphics[width=0.5\textwidth]{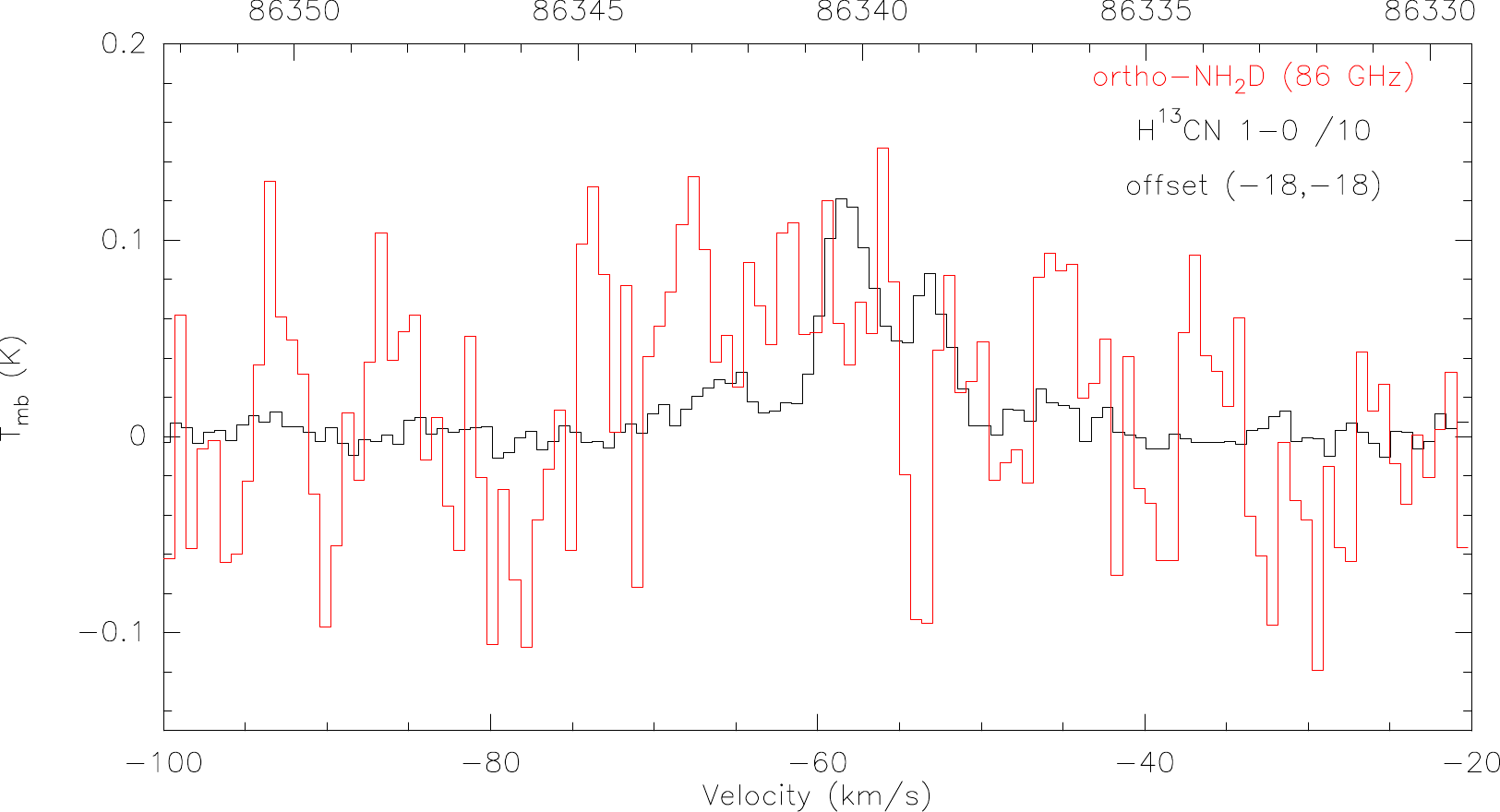}}
\subfigure[]{\includegraphics[width=0.5\textwidth]{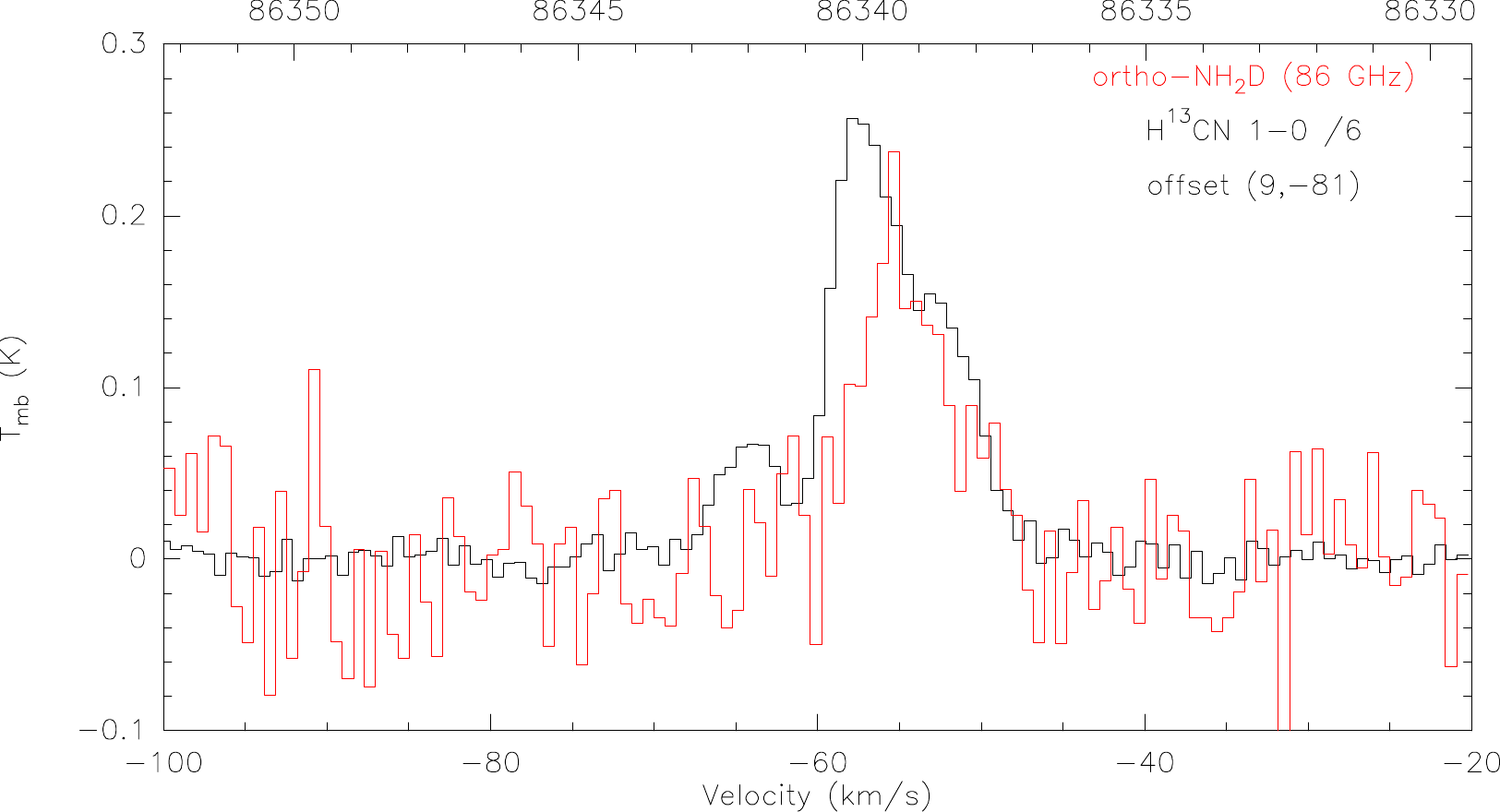}}
\caption{(a): NH$_2$D at 85.9263 GHz velocity integrated intensity contour (red contour) overlaid on H$^{13}$CN 1-0 velocity integrated intensity image (Gray scale and black contour) in G111.54+00.77. The contour levels start at 3$\sigma$ in steps of 2$\sigma$ for NH$_2$D, while the contour levels start at 15$\sigma$ in steps of 12$\sigma$ for H$^{13}$CN 1-0. The gray scale starts at 3$\sigma$. (b): NH$_2$D at 85.9263 GHz velocity integrated intensity contour (red contour) overlaid on H42$\alpha$ velocity integrated intensity image (Gray scale and black contour) in G111.54+00.77. The contour levels start at 3$\sigma$ in steps of 2$\sigma$ for NH$_2$D, while the contour levels start at 5$\sigma$ in steps of 5$\sigma$ for H42$\alpha$. The gray scale starts at 3$\sigma$. (c): Spectra of ortho-NH$_2$D at 85.9263 GHz and para-NH$_2$D at 110.1535 GHz in G111.54+00.77. The para-NH$_2$D was observed by IRAM-30m with position-switching mode. The red spectra is ortho-NH$_2$D and the black is para-NH$_2$D. The offset for two spectra is (0$^{\prime \prime}$,0$^{\prime \prime}$). The black box in the picture indicates the range of flux integration of ortho-NH$_2$D at 85.9263 GHz. (d): Spectra of ortho-NH$_2$D at 85.9263 GHz and H$^{13}$CN 1-0 in G111.54+00.77. The red spectra is ortho-NH$_2$D and the black is H$^{13}$CN 1-0. The offset for two spectra is (-18$^{\prime \prime}$,-18$^{\prime \prime}$). (e): Spectra of ortho-NH$_2$D at 85.9263 GHz and H$^{13}$CN 1-0 in G049.48-00.38.  The red spectra is ortho-NH$_2$D and the black is H$^{13}$CN 1-0. The offset for two spectra is (+9$^{\prime \prime}$,-81$^{\prime \prime}$).}
\label{app16}
\end{figure}

\begin{figure}
\centering
\subfigure[]{\includegraphics[width=0.5\textwidth]{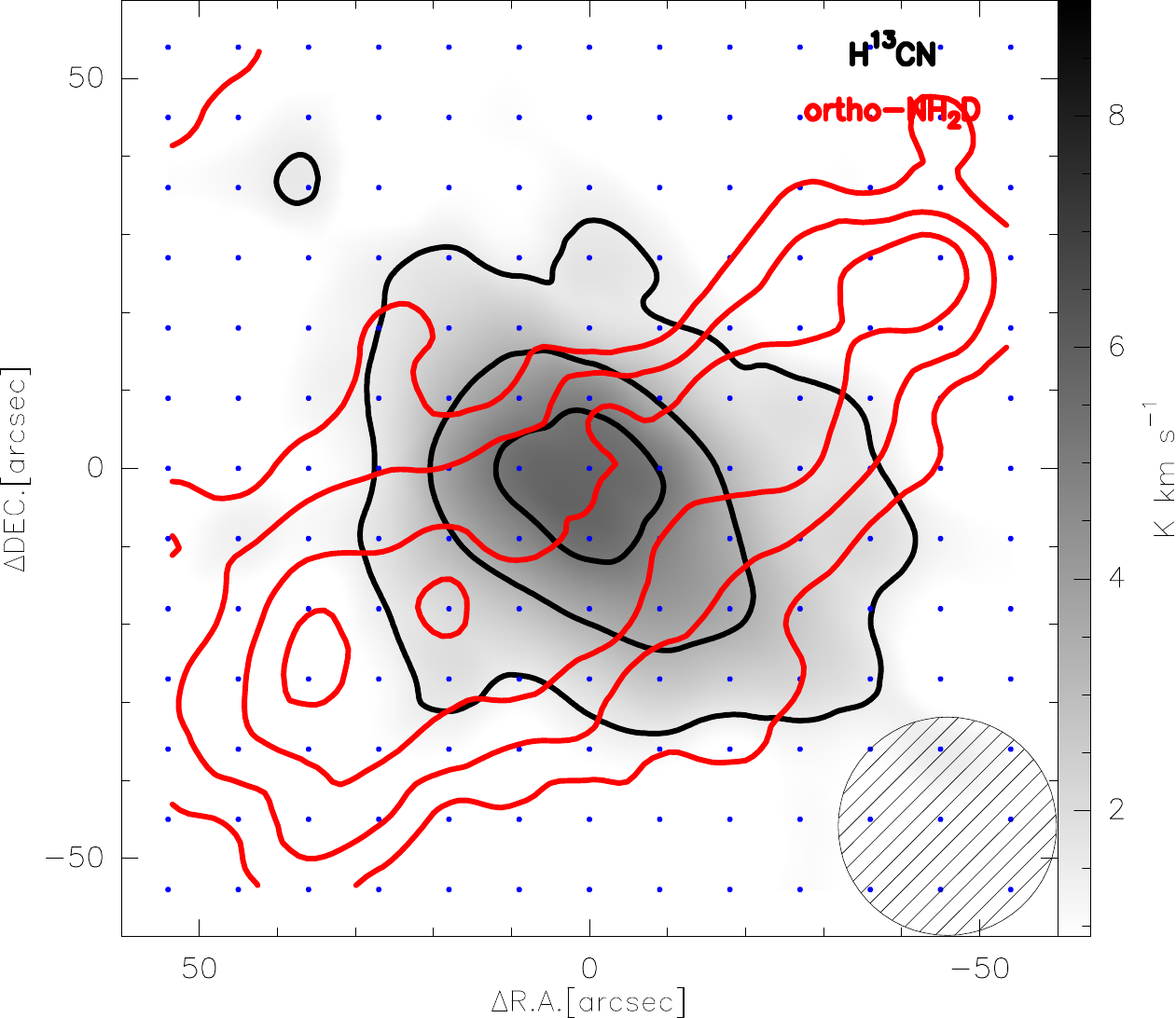}}\subfigure[]{\includegraphics[width=0.5\textwidth]{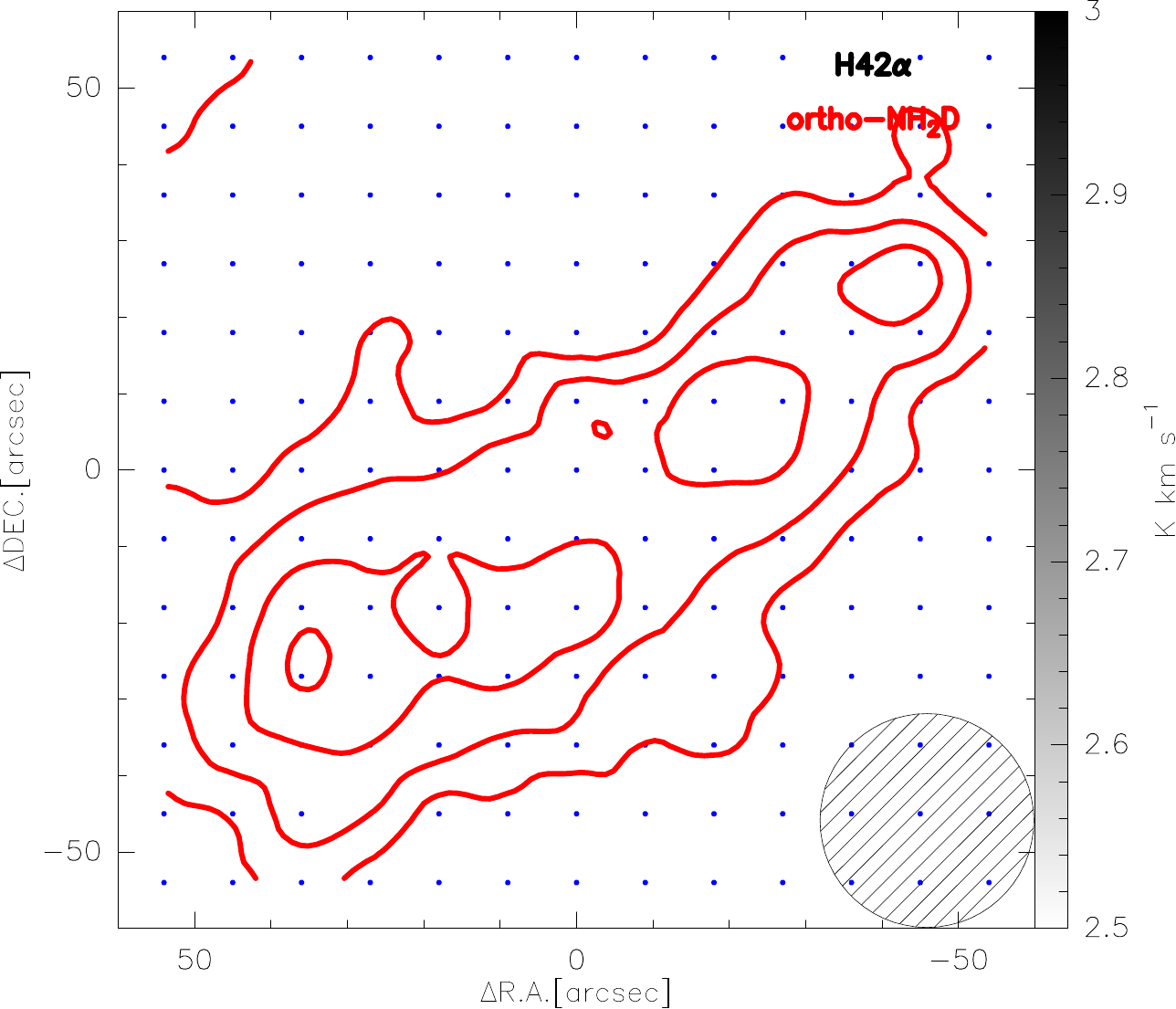}}
\subfigure[]{\includegraphics[width=0.5\textwidth]{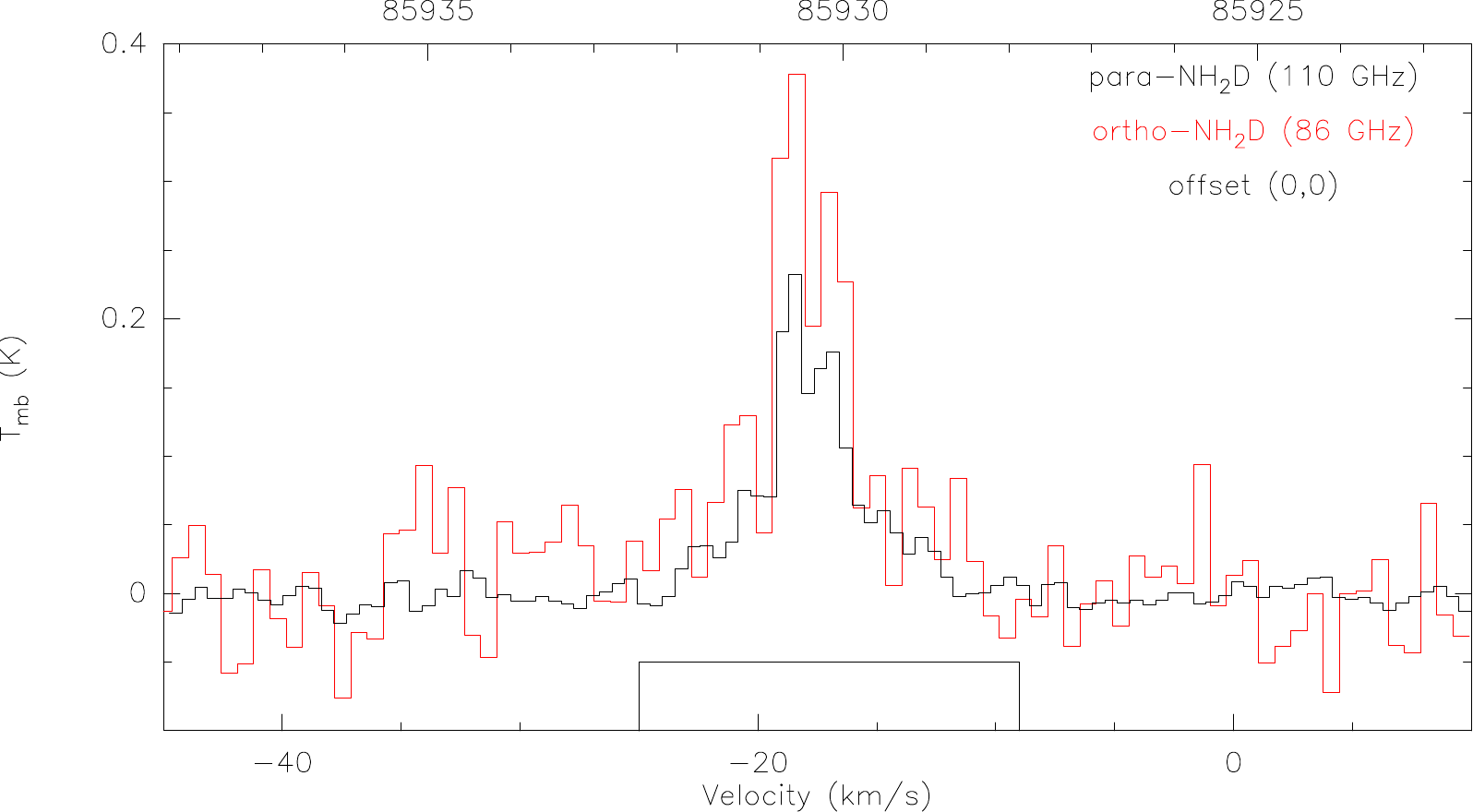}}\subfigure[]{\includegraphics[width=0.5\textwidth]{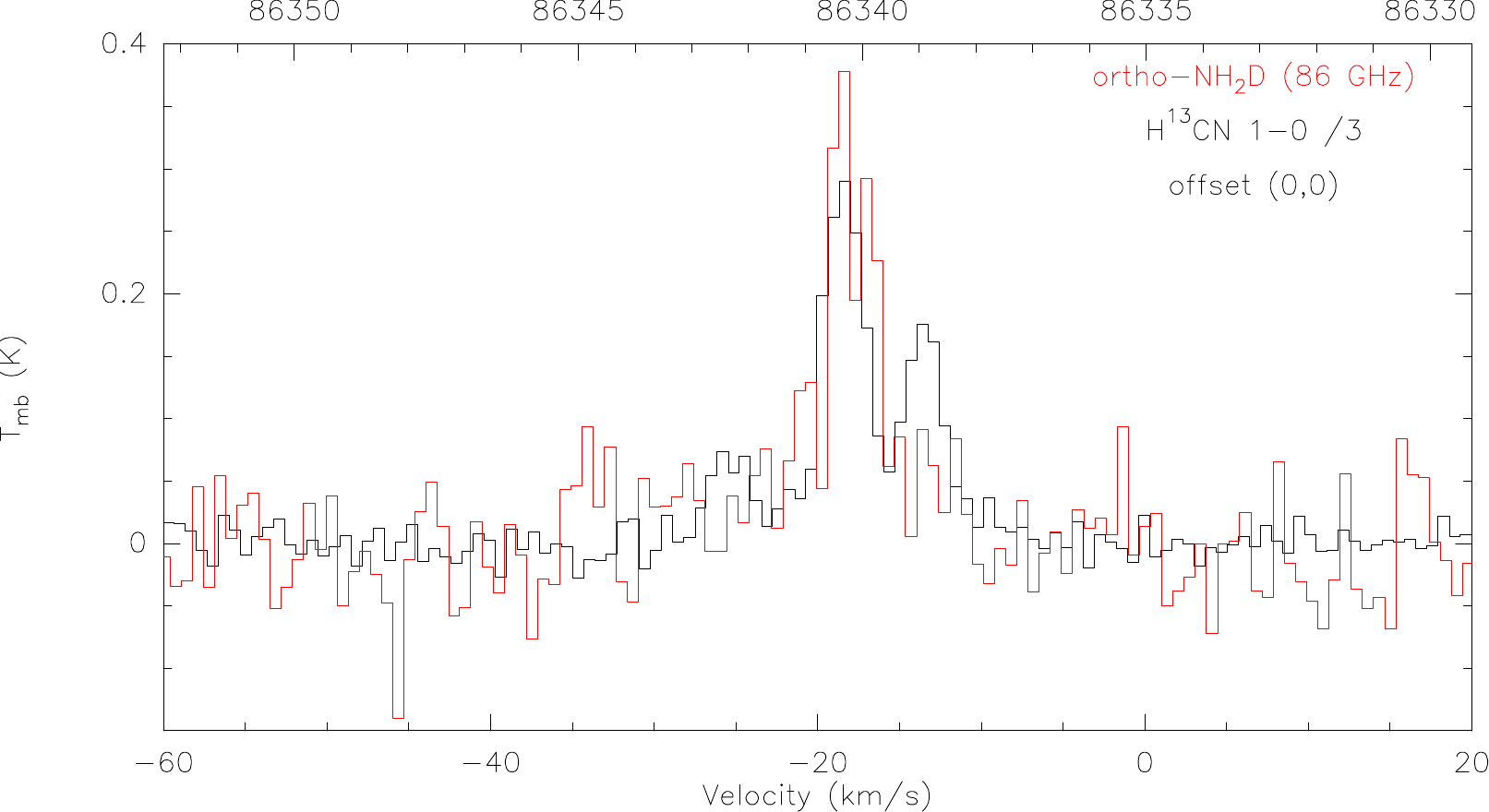}}
\subfigure[]{\includegraphics[width=0.5\textwidth]{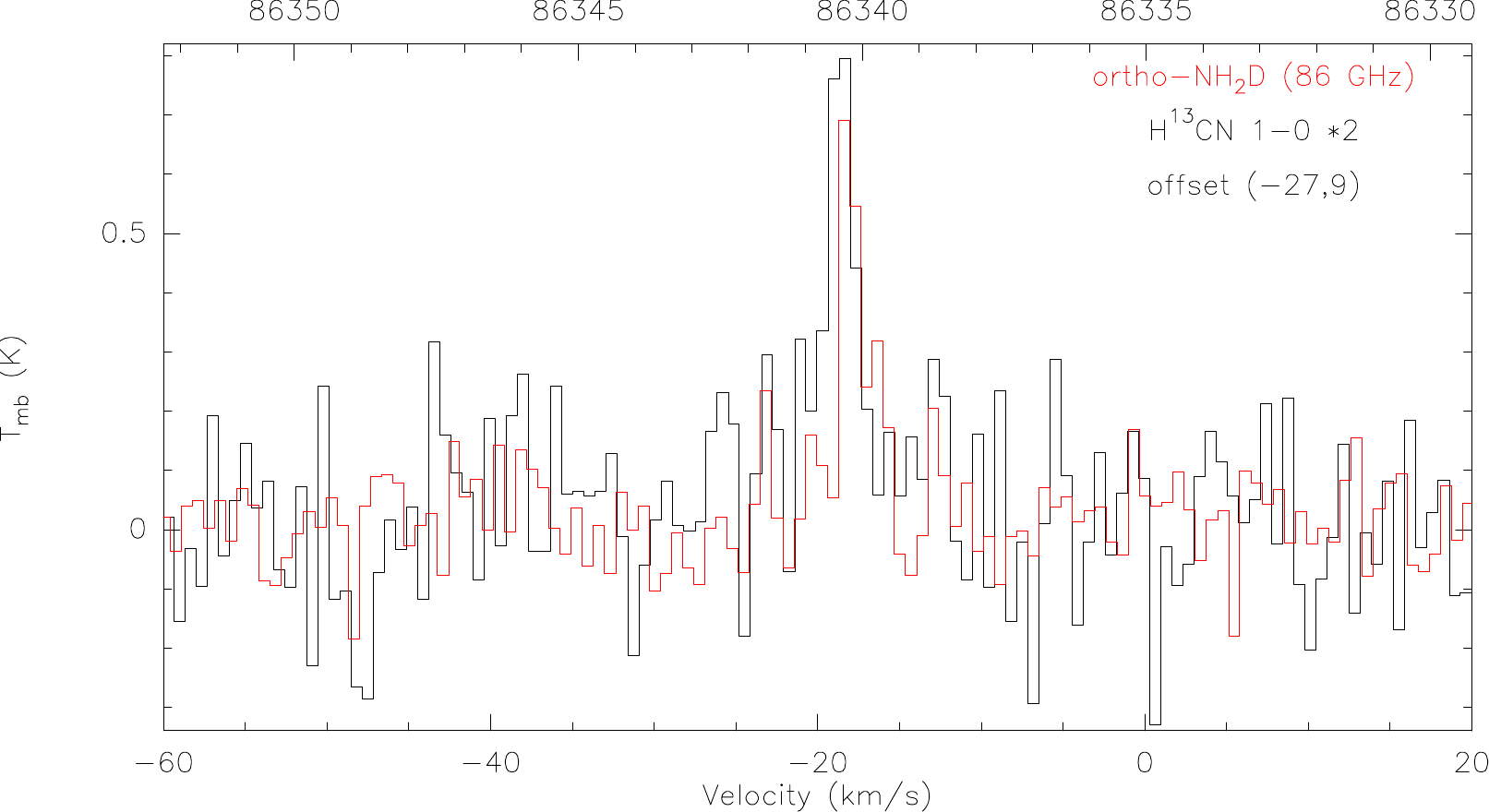}}\subfigure[]{\includegraphics[width=0.5\textwidth]{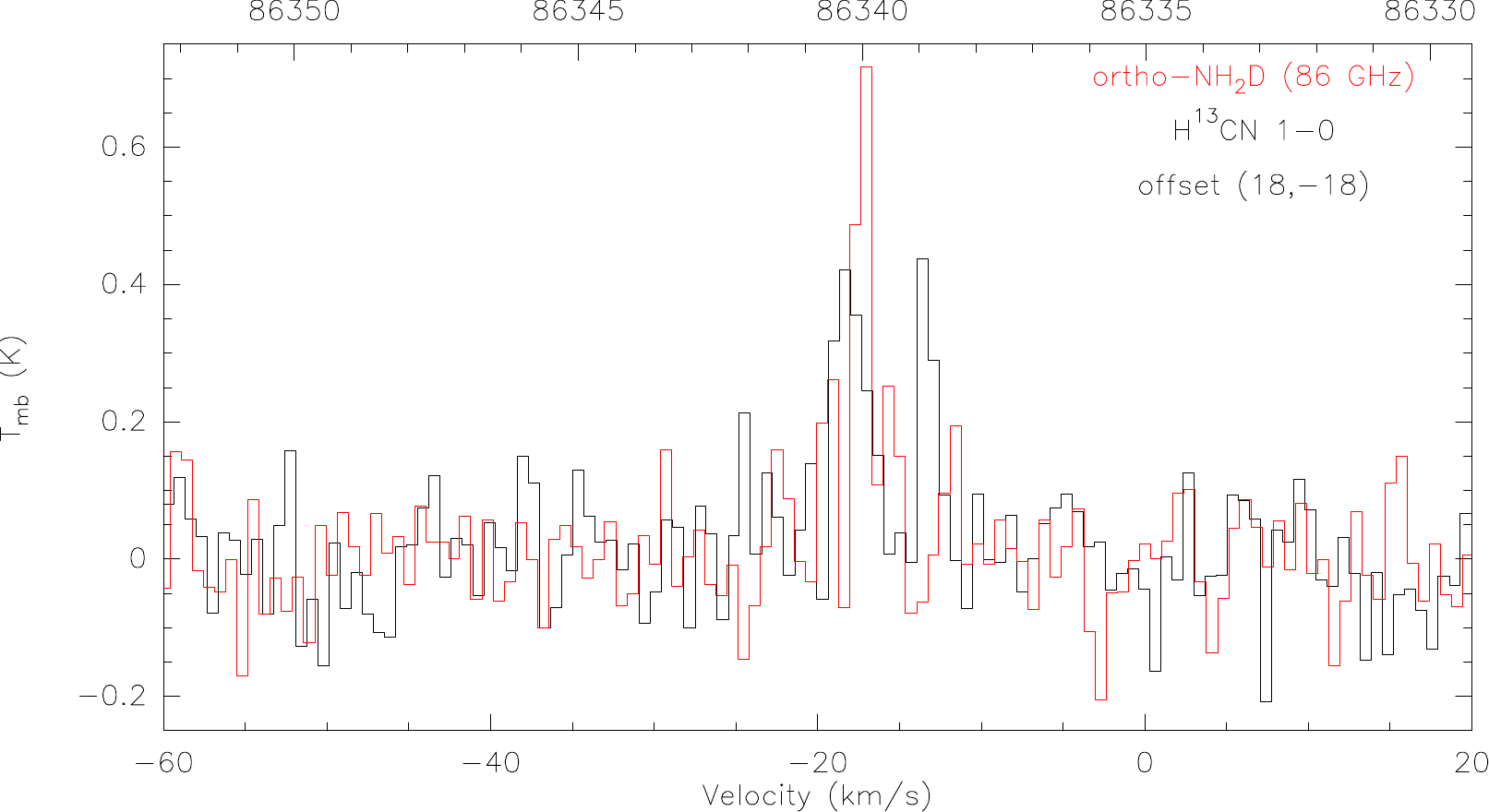}}
\caption{(a): NH$_2$D at 85.9263 GHz velocity integrated intensity contour (red contour) overlaid on H$^{13}$CN 1-0 velocity integrated intensity image (Gray scale and black contour) in G121.29+00.65. The contour levels start at 3$\sigma$ in steps of 2$\sigma$ for NH$_2$D, while the contour levels start at 5$\sigma$ in steps of 6$\sigma$ for H$^{13}$CN 1-0. The gray scale starts at 3$\sigma$. (b): NH$_2$D at 85.9263 GHz velocity integrated intensity contour (red contour) overlaid on H42$\alpha$ velocity integrated intensity image (Gray scale and black contour, while the H42$\alpha$ is not detected) in G121.29+00.65. The contour levels start at 3$\sigma$ in steps of 1$\sigma$ for NH$_2$D. (c): Spectra of ortho-NH$_2$D at 85.9263 GHz and para-NH$_2$D at 110.1535 GHz in G121.29+00.65. The para-NH$_2$D was observed by IRAM-30m with position-switching mode. The red spectra is ortho-NH$_2$D and the black is para-NH$_2$D. The offset for two spectra is (0$^{\prime \prime}$,0$^{\prime \prime}$). The black box in the picture indicates the range of flux integration of ortho-NH$_2$D at 85.9263 GHz. (d): Spectra of ortho-NH$_2$D at 85.9263 GHz and H$^{13}$CN 1-0 in G121.29+00.65. The red spectra is ortho-NH$_2$D and the black is H$^{13}$CN 1-0. The offset for two spectra is (0$^{\prime \prime}$,0$^{\prime \prime}$). (e): Spectra of ortho-NH$_2$D at 85.9263 GHz and H$^{13}$CN 1-0 in G121.29+00.65. The red spectra is ortho-NH$_2$D and the black is H$^{13}$CN 1-0. The offset for two spectra is (-27$^{\prime \prime}$,9$^{\prime \prime}$). (f): Spectra of ortho-NH$_2$D at 85.9263 GHz and H$^{13}$CN 1-0 in G121.29+00.65. The red spectra is ortho-NH$_2$D and the black is H$^{13}$CN 1-0. The offset for two spectra is (18$^{\prime \prime}$,-18$^{\prime \prime}$).}
\label{app17}
\end{figure}

\begin{figure}
\centering
\subfigure[]{\includegraphics[width=0.5\textwidth]{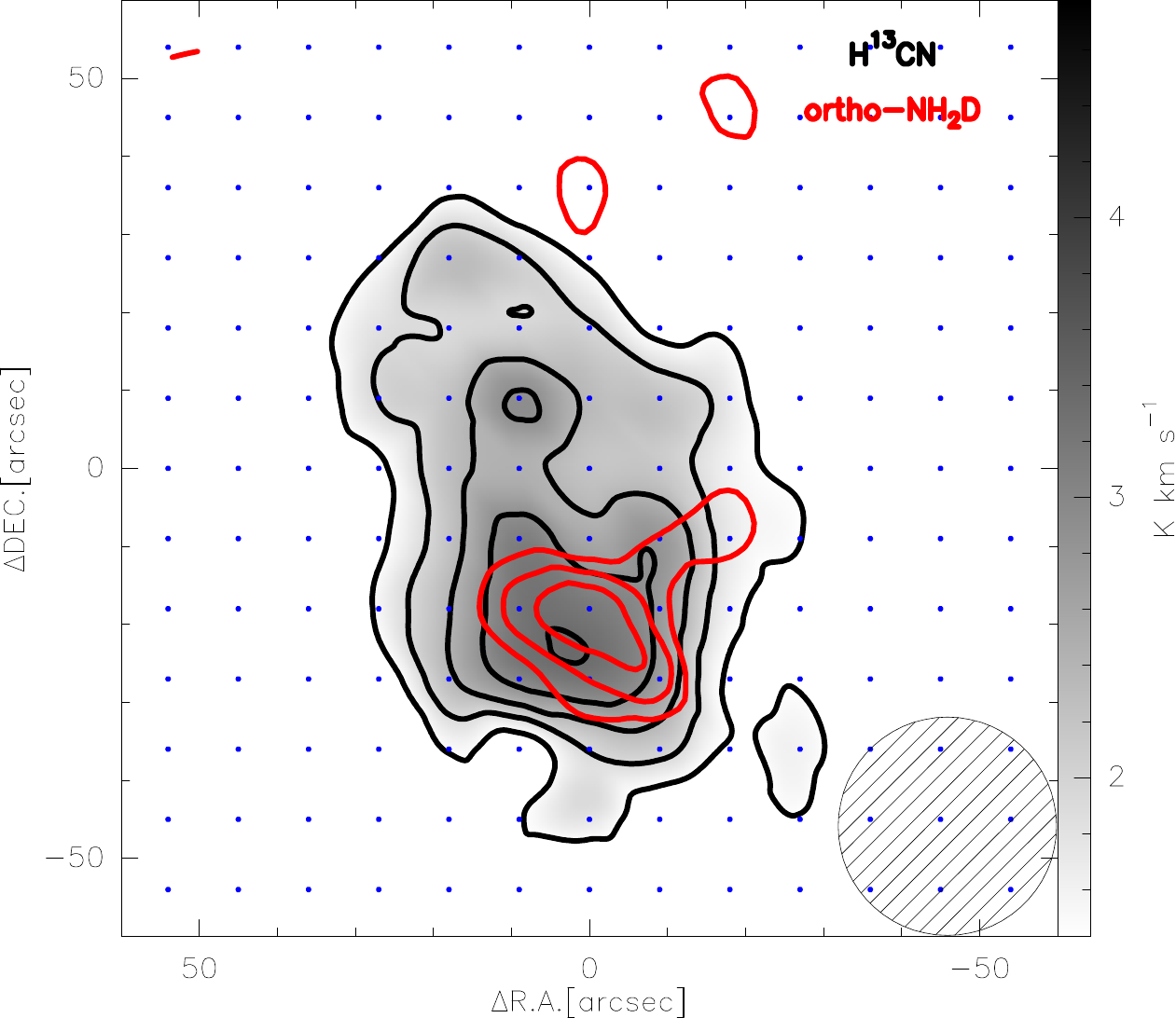}}\subfigure[]{\includegraphics[width=0.5\textwidth]{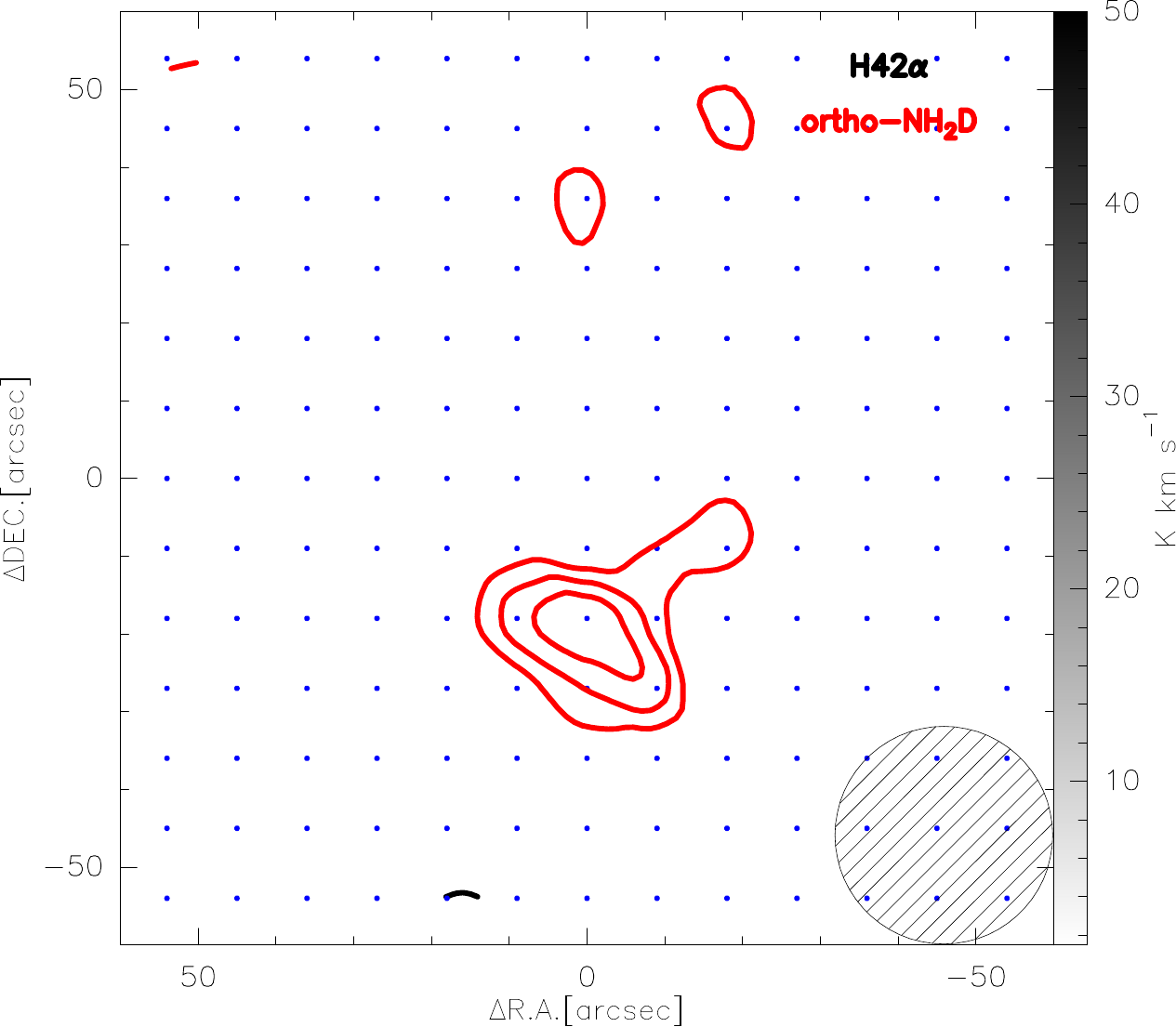}}
\subfigure[]{\includegraphics[width=0.5\textwidth]{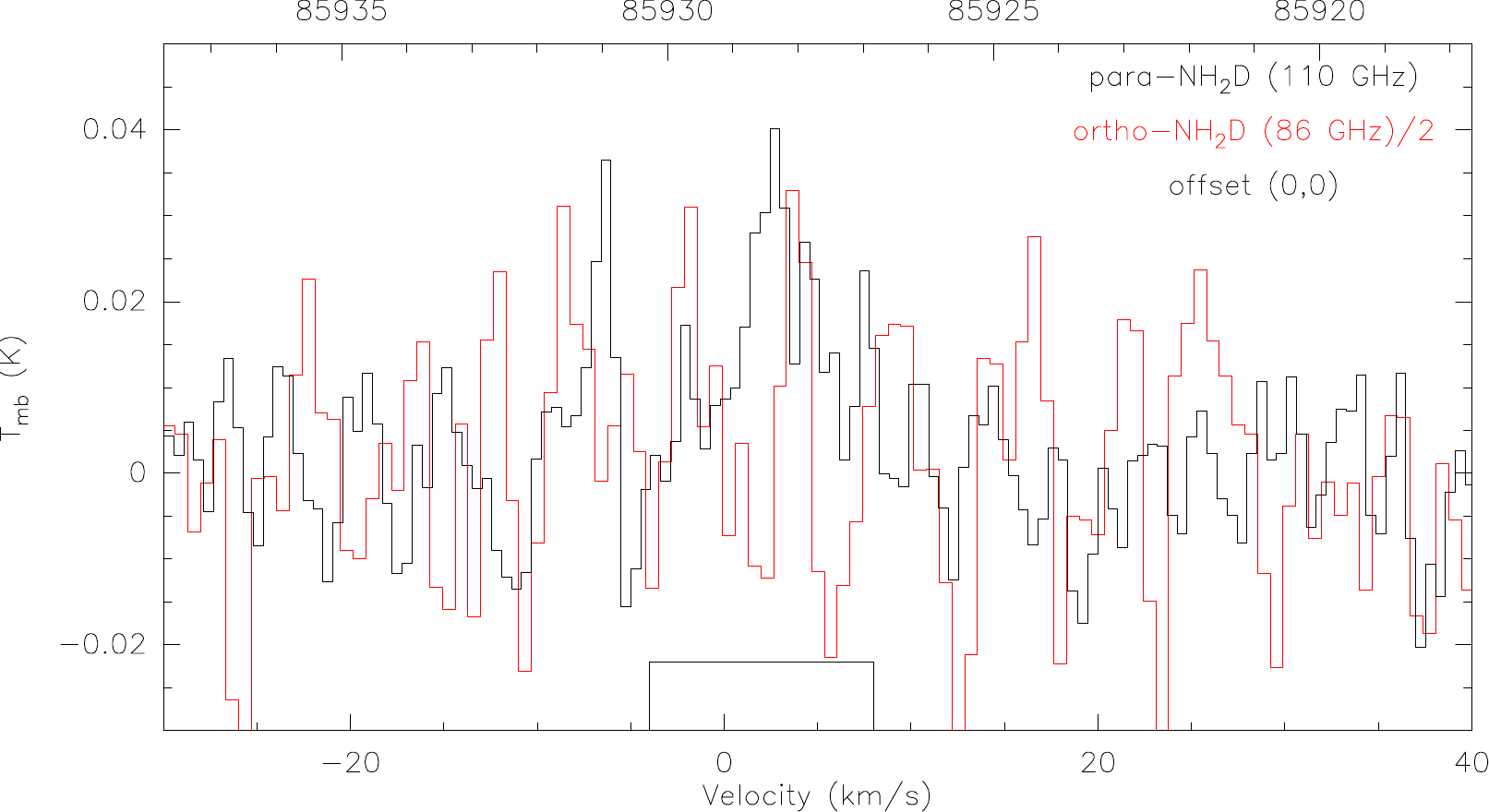}}\subfigure[]{\includegraphics[width=0.5\textwidth]{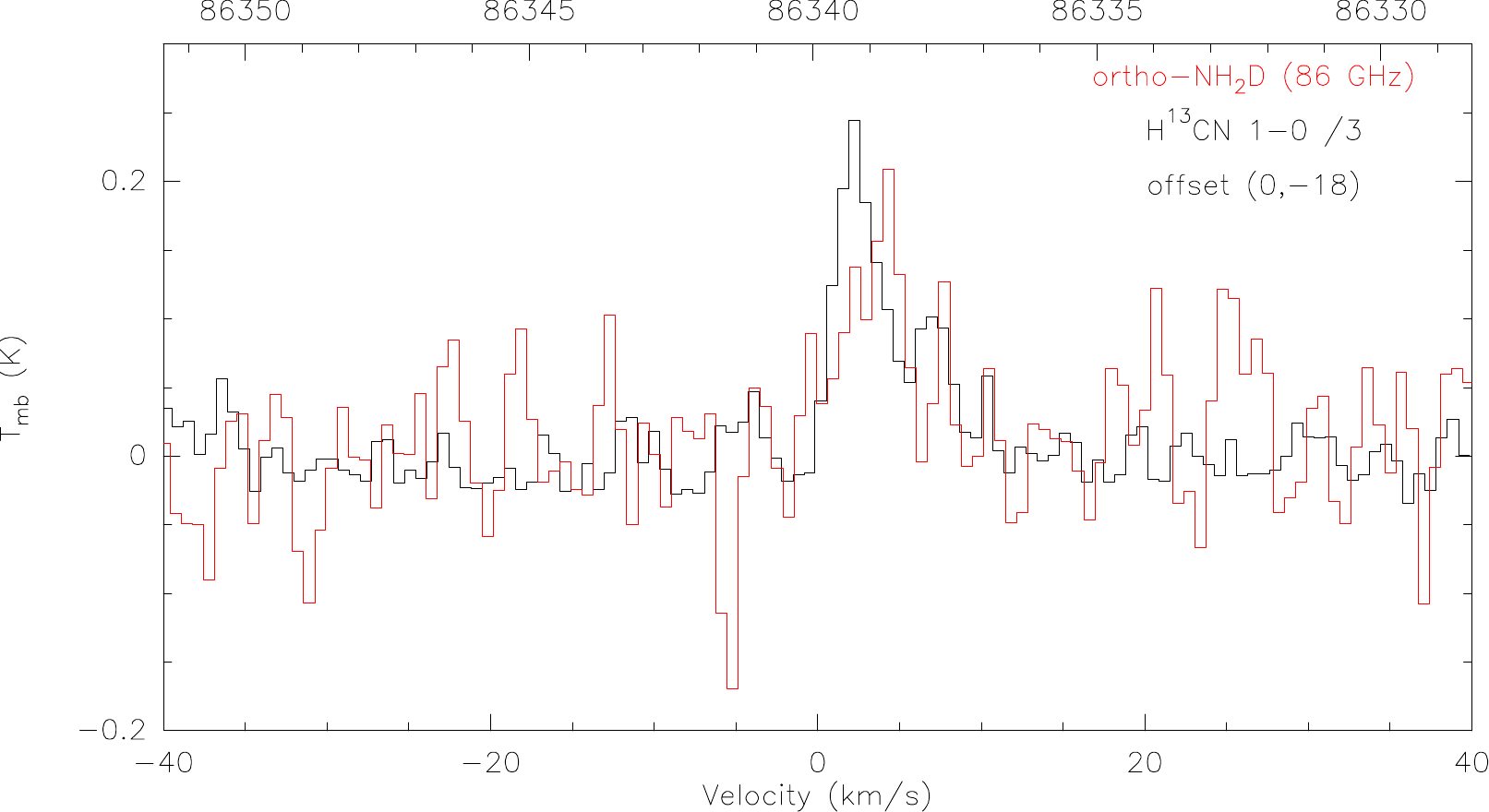}}
\caption{(a): NH$_2$D at 85.9263 GHz velocity integrated intensity contour (red contour) overlaid on H$^{13}$CN 1-0 velocity integrated intensity image (Gray scale and black contour) in G188.94+00.88. The contour levels start at 3$\sigma$ in steps of 1$\sigma$ for NH$_2$D, while the contour levels start at 5$\sigma$ in steps of 2$\sigma$ for H$^{13}$CN 1-0. The gray scale starts at 3$\sigma$. (b): NH$_2$D at 85.9263 GHz velocity integrated intensity contour (red contour) overlaid on H42$\alpha$ velocity integrated intensity image (Gray scale and black contour, while the H42$\alpha$ is not detected) in G188.94+00.88. The contour levels start at 3$\sigma$ in steps of 1$\sigma$ for NH$_2$D. (c): Spectra of ortho-NH$_2$D at 85.9263 GHz and para-NH$_2$D at 110.1535 GHz in G188.94+00.88. The para-NH$_2$D was observed by IRAM-30m with position-switching mode. The red spectra is ortho-NH$_2$D and the black is para-NH$_2$D. The offset for two spectra is (0$^{\prime \prime}$,0$^{\prime \prime}$). The black box in the picture indicates the range of flux integration of ortho-NH$_2$D at 85.9263 GHz. (d): Spectra of ortho-NH$_2$D at 85.9263 GHz and H$^{13}$CN 1-0 in G188.94+00.88. The red spectra is ortho-NH$_2$D and the black is H$^{13}$CN 1-0. The offset for two spectra is (0$^{\prime \prime}$,-18$^{\prime \prime}$).}
\label{app18}
\end{figure}

The velocity integrated intensity of ortho-NH$_2$D $1_{11}^s-1_{01}^a$ at 85.9263 GHz, observed with the IRAM-30m, are presented in Figure \ref{app1} - \ref{app18} as red contour. The spectra of ortho-NH$_2$D $1_{11}^s-1_{01}^a$ at 85.9263 GHz are also presented in Figure \ref{app1} - \ref{app18}. For overlaying with H$^{13}$CN 1-0 spectra, there are some of the sources in which several offsets are shown in Figure \ref{app1} - \ref{app18}. These several offsets are used to confirm the signal of ortho-NH$_2$D $1_{11}^s-1_{01}^a$ emission if this target has spatial structures, such as G081.75+00.59 (see Figure \ref{app13}).

\begin{figure}
\centering
\includegraphics[width=0.5\textwidth]{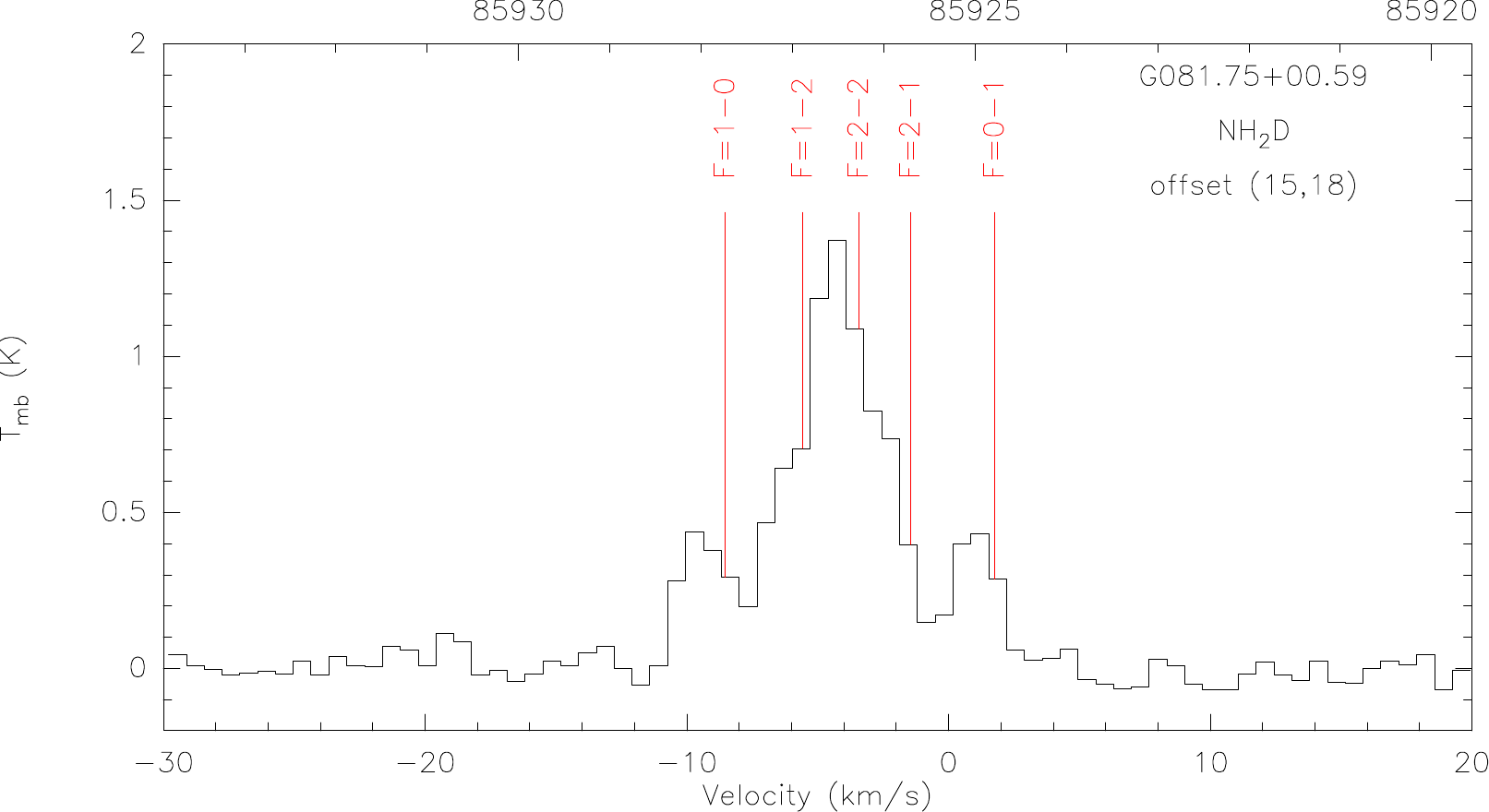}
\caption{The hyperfine structure of ortho-NH$_2$D in G081.75+00.59.}
\label{G081_HfS}
\end{figure}

Ortho-NH$_2$D $1_{11}^s-1_{01}^a$ has six hyperfine structures, with relative intensities listed in Table \ref{NH2D HfS}  \citep{2000A&A...356.1039T}. 
However, the transitions of F=2-1, F=2-2, F=1-1 and F=1-2 for ortho-NH$_2$D $1_{11}^s-1_{01}^a$ are blended due to line broadening. For example, as shown in Figure \ref{G081_HfS}, only three components can be found in G081.75+00.59 for NH$_2$D spectra. Thus, the fluxes of ortho-NH$_2$D $1_{11}^s-1_{01}^a$ were derived with the velocity integrated intensity of the emitting channels of its six hyperfine structures instead of Gaussian fitting. In the following analysis, we assume all NH$_2$D spectra are optically thin.

Four sources, G011.91-00.61, G023.44-00.18, G034.39+00.22 and G049.48-00.38, are not detected asymmetrically distributed spatial structure of ortho-NH$_2$D (see Figure \ref{app2}, Figure \ref{app4}, Figure \ref{app6} and Figure \ref{app11}, respectively). Ortho-NH$_2$D emissions were marginally detected and unresolved in two targets:  G005.88-00.39 and G188.94+00.88 (see Figure \ref{app1} and Figure \ref{app18}). In addition, although G015.03-00.67 and G037.43+01.51 (see Figure \ref{app3} and Figure \ref{app9}) can be distinguished the spatial structure, the ortho-NH$_2$D emissions are weak. Other targets can be distinguished the spatial structure and detected relatively strong ortho-NH$_2$D emission. 

The strongest velocity integrated intensity and distribution shape of ortho-NH$_2$D  for each source are listed in Table \ref{distribution}. The strongest ortho-NH$_2$D emission is 6.95$\pm$0.14 K km/s in G081.75+00.59 and the weakest ortho-NH$_2$D emission is 0.60$\pm$0.16 and 0.60$\pm$0.11 K km/s in G005.88-00.39 and G037.43+01.51, respectively. The median of ortho-NH$_2$D emission is 2.27 K km/s. For the sources whose ortho-NH$_2$D emission is lower than 1 K km/s, are considered as sources with weak NH$_2$D emission. Conversely, we define strong NH$_2$D emissions to be cases surpassing the 1 K km/s threshold. Accordingly, there are 4 sources characterized by weak NH$_2$D emissions and 14 sources are classified into strong NH$_2$D emissions.

With strong NH$_2$D emission detected, 14 sources show complex and diverse distribution, which is different from other molecule transitions, such as the dense gas tracer H$^{13}$CN 1-0 (details described in the following). 

\subsection{The ortho-to-para ratio of NH$_2$D}
\label{ortho-to-para ratio of NH$_2$D}
We presented single point observational results of  para-NH$_2$D $1_{11}^a-1_{01}^s$ at 110.1535 GHz with IRAM-30m toward 50 late-stage massive star-forming regions \citep{2022MNRAS.512.4934L},  which included all the targets of this work. The NH$_2$D spectra, both of the ortho-NH$_2$D $1_{11}^s-1_{01}^a$ at 85.9263 GHz with OTF mode and para-NH$_2$D $1_{11}^a-1_{01}^s$ at 110.1535 GHz with position-switching mode, are also shown in figure \ref{app1} - \ref{app18}. All coordinates of para-NH$_2$D $1_{11}^a-1_{01}^s$ at 110.1535 GHz are consistent with offset (0$^{\prime \prime}$,0$^{\prime \prime}$) ortho-NH$_2$D $1_{11}^s-1_{01}^a$ at 85.9263 GHz, except G011.91-00.61, G015.03-00.67 and G034.39+00.22. Para-NH$_2$D $1_{11}^a-1_{01}^s$ at 110.1535 GHz transition have offset about (-27$^{\prime \prime}$,-27$^{\prime \prime}$), (45$^{\prime \prime}$,36$^{\prime \prime}$) and (0$^{\prime \prime}$,-45$^{\prime \prime}$), corresponding to ortho-NH$_2$D $1_{11}^s-1_{01}^a$ at 85.9263 GHz in G011.91-00.61 (see Figure \ref{app2}), G015.03-00.67 (see Figure \ref{app3}) and G034.39+00.22 (see Figure \ref{app6}), respectively.

For two hydrogen atoms in NH$_2$D, different nuclear spin angular momentum can produce different states: ortho and para state. The ortho state has the largest nuclear spin statistical weight, and the para state has the largest nuclear spin statistical weight \citep{2017A&A...600A..61H}. Since normal radiative and collisional transitions can not change the spin of H nuclei, the transitions between ortho- and para-NH$_2$D are forbidden \citep{1983ARA&A..21..239H}. Therefore, obtaining ortho- and para-NH$_2$D could reflect an accurate abundance of NH$_2$D.

The corresponding nuclear statistical weights are 27 and 9 for ortho-NH$_2$D $1_{11}^s-1_{01}^a$ at 85.9263 GHz and  para-NH$_2$D $1_{11}^a-1_{01}^s$ at 110.1535 GHz, respectively (CDMS\footnote{\url{https://cdms.astro.uni-koeln.de/classic/predictions/catalog/}}, \cite{2001A&A...370L..49M,2005JMoSt.742..215M}). Therefore, the ratio of N(ortho-NH$_2$D)/N(para-NH$_2$D) is 3 in LTE and this ratio could indicate the discrepancy between total column density and ortho-NH$_2$D or para-NH$_2$D. The Einstein emission coefficient is 7.8214$\times$10$^{-6}$s$^{-1}$ and 1.6503$\times$10$^{-5}$s$^{-1}$ for ortho-NH$_2$D $1_{11}^s-1_{01}^a$ at 85.9263 GHz and  para-NH$_2$D $1_{11}^a-1_{01}^s$ at 110.1535 GHz, respectively (CDMS, \cite{2001A&A...370L..49M,2005JMoSt.742..215M}). In addition, the line ratios of velocity integrated intensities between ortho-NH$_2$D $1_{11}^s-1_{01}^a$ at 85.9263 GHz and para-NH$_2$D $1_{11}^a-1_{01}^s$ at 110.1535 GHz is also related to the excitation temperature. Considering Einstein emission coefficient, the line ratios of velocity integrated intensities between ortho-NH$_2$D $1_{11}^s-1_{01}^a$ at 85.9263 GHz and para-NH$_2$D $1_{11}^a-1_{01}^s$ at 110.1535 GHz is about 2.5 at 10 K and 2.4 at 20 K at LTE. 

The line ratios of velocity integrated intensities between ortho-NH$_2$D $1_{11}^s-1_{01}^a$ at 85.9263 GHz and  para-NH$_2$D $1_{11}^a-1_{01}^s$ at 110.1535 GHz are listed in Table \ref{NH2D}. The ortho-NH$_2$D $1_{11}^s-1_{01}^a$ at 85.9263 GHz spectra from the mapping results at the same position as that of para-NH$_2$D $1_{11}^a-1_{01}^s$ at 110.1535 GHz are used for this comparison. Upper limits are provided in six sources due to non-detection of ortho-NH$_2$D $1_{11}^s-1_{01}^a$ or para-NH$_2$D $1_{11}^a-1_{01}^s$ lines, in where coordinate is para-NH$_2$D $1_{11}^a-1_{01}^s$. The highest ratio is 4.73$\pm$1.99 in G035.20-01.73 and the lowest ratio is 0.89$\pm$0.44 in G005.88-00.39, both of them are with strong H42$\alpha$ emissions. In other sources, this ratio ranges from 1.26$\pm$0.07 to 2.95$\pm$0.12. Besides, the median of this ratio is 1.84.

\subsection{H$^{13}$CN 1-0 maps result}
\label{H13CN 1-0 result}
HCN lines can be used to trace dense gas ($\gtrsim 10^4$ cm$^{-3}$) because of high dipole moment and large Einstein \emph{A}-coefficient \citep{2006AJ....132.2398E}. With the similar dipole moment and Einstein \emph{A}-coefficient, lines of H$^{13}$CN, as an isotope molecule of HCN, can be used as a dense gas tracer.  So, H$^{13}$CN  lines can be more accurate to trace dense molecular  gas, since they are mainly optically thin. Therefore, H$^{13}$CN 1-0 is used as a dense gas tracer \citep{2004ApJ...606..271G,2010ApJS..188..313W,2011A&A...527A..88V}.

In this work, the H$^{13}$CN 1-0 emission was detected in all targets. H$^{13}$CN 1-0 has three hyperfine structures. Similar to ortho-NH$_2$D $1_{11}^s-1_{01}^a$, the velocity integrated intensity of H$^{13}$CN 1-0  was derived from its three hyperfine structures with CLASS/GILDAS package. The mapping results and spectra of H$^{13}$CN 1-0 are  shown in Figure \ref{app1} - \ref{app18} (black contour and gray scale).

Table \ref{distribution} lists whether ortho-NH$_2$D and H$^{13}$CN 1-0 show similar distribution or not in each source.  Ortho-NH$_2$D and H$^{13}$CN 1-0 are unresolved in two sources (G005.88-00.39 and G188.94+00.88,  see Figure \ref{app1} and Figure \ref{app18}), with weak ortho-NH$_2$D emissions in these two sources.  Weak NH$_2$D emissions in G015.03-00.67 and G037.43+01.51  (see Figure \ref{app3} and Figure \ref{app9}) were detected different distributions to that of H$^{13}$CN 1-0. Among other sources with relatively strong ortho-NH$_2$D emission detected, five sources show similar distributions and nine show different distributions between NH$_2$D and H$^{13}$CN 1-0 maps. G035.20-01.73, G075.76+00.33, G081.87+00.78 and G109.87+02.11 show that the distribution of strong NH$_2$D is dislocation with H$^{13}$CN (see Figure \ref{app8}, Figure \ref{app12}, Figure \ref{app14} and Figure \ref{app15}, respectively). In addition, in the targets of G031.28+00.06, G035.19-00.74, G075.76+00.33, G111.54+00.77 and G121.29+00.65, some regions present a strong NH$_2$D emission, while just weak H$^{13}$CN 1-0 emission is detected in these regions (see Figure \ref{app5}, Figure \ref{app7}, Figure \ref{app12}, Figure \ref{app16} and Figure \ref{app17}, respectively).

 In G011.91-00.61, G023.44-00.18, G034.39+00.22, G049.48-00.38 and G081.75+00.59, ortho-NH$_2$D present a similar distribution with H$^{13}$CN 1-0 (see Figure \ref{app2}, Figure \ref{app4}, Figure \ref{app6}, Figure \ref{app11} and Figure \ref{app13}, respectively). The strong ortho-NH$_2$D emission was also detected in these five sources. The strongest NH$_2$D spectra was detected in G081.75+00.59 with $\int$T$_{\rm mb}$(NH$_2$D)d$\nu$=6.95 $\pm$ 0.14 K km s$^{-1}$. However, the asymmetrically structure of ortho-NH$_2$D $1_{11}^s-1_{01}^a$ at 85.9263 GHz was not detected in G011.91-00.61, G023.44-00.18, G034.39+00.22 and G049.48-00.38. The reason is mainly due to the limited spatial resolution. Only G081.75+00.59 present asymmetrically structure of ortho-NH$_2$D $1_{11}^s-1_{01}^a$ at 85.9263 GHz.

\subsection{H42$\alpha$ maps result}
\label{H42alpha maps result}
The young massive stars, with spectral types of B or O, can produce ultra-violet photons, which  can ionize the surrounding interstellar medium and create H\uppercase\expandafter{\romannumeral2} regions. Consequently, the abundantly ionized hydrogens exist in H\uppercase\expandafter{\romannumeral2} regions and can recombine with electrons to produce the radio recombination lines (RRLs). Therefore, the RRLs can be used to trace young massive star-forming and H\uppercase\expandafter{\romannumeral2} regions.

RRL H42$\alpha$ was detected in nine sources (see also in figure \ref{app1} - \ref{app18}, the black contour and gray scale). Without turbulence, the full width at half maxima (FWHM) of RRLs is 19.1 km s$^{-1}$ at 8000 K \citep{2012MNRAS.425.2352P}. Therefore,  in the case of a faint RRLs signal observed in G031.28+00.06, the spectral resolution was smoothed to 1.4 kms$^{-1}$ to confirm whether it is H42$\alpha$ emission or not in the present work. Among these nine targets, a completely different spatial distribution between ortho-NH$_2$D and H42$\alpha$ is presented in five targets, while the spatial distribution of ortho-NH$_2$D and H42$\alpha$ have overlapping regions in the rest four targets.

 A strong H42$\alpha$ emission presents in G005.88-00.39 (see Figure \ref{app1}), while the ortho-NH$_2$D were marginally detected and almost unresolved in the same position. There are two NH$_2$D emission features and a relatively weak H42$\alpha$ structure with associated NH$_2$D emission in G011.91-00.61 (see Figure \ref{app2}). G049.48-00.36 and G049.48-00.38 are two adjacent sources (see Figure \ref{app10} and Figure \ref{app11}). Both of them, the ortho-NH$_2$D and H42$\alpha$ present different spatial distributions. However, the ortho-NH$_2$D emission peaks at the H42$\alpha$ emission peak in G049.48-00.36. Besides, in G049.48-00.36, the ortho-NH$_2$D is weaker than that in G049.48-00.38.

In G015.03-00.67 (see Figure \ref{app3}), a widely distributed and strong H42$\alpha$ emission was detected, while the ortho-NH$_2$D is scattered in other positions. In G031.28+00.06 (see Figure \ref{app5}), the ortho-NH$_2$D is isolated by H42$\alpha$ and the emission of ortho-NH$_2$D is very weak. The H42$\alpha$ are also observed in G035.20-01.73, G075.76+00.33 and G111.54+00.77 (see Figure \ref{app8}, Figure \ref{app12} and Figure \ref{app16}, respectively). The ortho-NH$_2$D and H42$\alpha$ present a completely different spatial distribution in G035.20-01.73, G075.76+00.33 and G111.54+00.77, but the NH$_2$D emission is far away from the H42$\alpha$ emission.

\subsection{Individual target detail}
\label{Individual target detail}
In our observation, most sources show different spatial distributions between ortho-NH$_2$D and H$^{13}$CN 1-0 emission, or between ortho-NH$_2$D and H42$\alpha$ emission, while a few sources show similar spatial distributions. For example, the ortho-NH$_2$D and H$^{13}$CN 1-0 emission present similar spatial distribution in G011.91-00.61. The details each source are described in following.

\subsubsection{G005.88-00.39}
G005.88-00.39 is a bright expanding shell-like UCH\uppercase\expandafter{\romannumeral2} region \citep{1998ApJ...495L.107A}. According to near-infrared NACO-VLT observations, \cite{2003ApJ...599L..91F} suggested G005.88-00.39 contains a massive young O5 V star. Furthermore, this target is one of the most powerful outflows in Galaxy \citep{1988A&A...197L..19H,2020ApJ...902L..47Z}. In the present work, strong ($\int T_{\rm mb}$d$\nu\ge$40 K km s$^{-1}$) H$^{13}$CN 1-0 and H42$\alpha$ emission were detected in G005.88-00.39, and their velocity integrated intensity maps(see figure \ref{app1}). However, ortho-NH$_2$D at 85.9 GHz was only observed at the source center with a weak emission, which is the weakest among our samples. Therefore, the ortho-NH$_2$D may be marginally detected in this source and it was plotted from 2$\sigma$. Moreover, the ortho-NH$_2$D velocity integrated intensity mapping presents less than a beam, so the spatial structure might not be distinguished in G005.88-00.39. Due to the weak ortho-NH$_2$D emission and indistinguishable spatial structure, we can not confirm whether the distribution of ortho-NH$_2$D is similar to other molecules or not. 

\subsubsection{G011.91-00.61}
With the H76$\alpha$ detected, \cite{1989ApJS...69..831W} indicated that  an UCH\uppercase\expandafter{\romannumeral2} region exists in G11.94-00.62 (RA.(J2000)=18:14:01.006, DEC.(J2000)=-18:53:24.966), and \cite{2018ApJ...869L..24I} suggested that there was a proto-O star forming in this region. The distribution of ortho-NH$_2$D, H$^{13}$CN 1-0 and H42$\alpha$ (see figure \ref{app2}) were obtained in G011.91-00.61. The ortho-NH$_2$D and H$^{13}$CN 1-0 show two structures and similar distributions. However, only the structure in the northeast (G11.94-00.62) has been detected to have H42$\alpha$ emission. Although the intensity of H$^{13}$CN 1-0 is similar in these two structures, the ortho-NH$_2$D appears weaker in the northeast structure where H42$\alpha$ is detected.

\subsubsection{G015.03-00.67}
G015.03-00.67 (M 17) is a molecular cloud complex \citep{1976ApJS...32..603L} and it is the closest (1.98 kpc, \cite{2011ApJ...733...25X}) giant H\uppercase\expandafter{\romannumeral2} region to us \citep{2020ApJ...888...98L}. Hundreds of candidate massive YSOs with spectral types have been found earlier than B9 in G015.03-00.67\citep{1991ApJ...368..432L,2009ApJ...696.1278P}. We obtained a 4$'\times4'$ size map of ortho-NH$_2$D, H$^{13}$CN 1-0 and H42$\alpha$ toward IRAM 30m observation in G015.03-00.67 (see figure \ref{app3}). The ortho-NH$_2$D was detected relatively weak emission in G015.03-00.67 ($\int$T$_{\rm mb}$(NH$_2$D)d$\nu$=1.1 K km s$^{-1}$ in the strongest pixel). The ortho-NH$_2$D emission presents a fragmented distribution. A widely distributed H42$\alpha$ distribution was detected in G015.03-00.67, indicating it is a giant H\uppercase\expandafter{\romannumeral2} region. The ortho-NH$_2$D fragments avoid obviously the widely distributed H42$\alpha$ distribution. The widely distributed H$^{13}$CN 1-0 emission was also detected in this target. One of the ortho-NH$_2$D fragments appears in the strongest H$^{13}$CN 1-0 structure, others appear with relatively weak H$^{13}$CN 1-0 emission. The strong H42$\alpha$ emission is at the northeast of G015.03-00.67 where H$^{13}$CN 1-0 is marginally or  even not detected. 

\subsubsection{G023.44-00.18}
The strong 6.7 GHz CH$_3$OH maser has been detected in the massive star-forming region G023.44-00.18 \citep{1998MNRAS.301..640W}. With Submillimeter Array (SMA) observation, \cite{2011MNRAS.415L..49R} reported that G023.44-00.18 presented two dust cores that were north and south arranged, the south one has active bipolar outflow. In addition, they suggested that these two dust cores were in the pre-UCH\uppercase\expandafter{\romannumeral2} evolutionary stage. By observing the ortho-NH$_2$D $1_{11}^s-1_{01}^a$ at 85.9263 GHz toward PdBI, \cite{2020A&A...638A.105Z} found several NH$_2$D cores in G023.44-00.18 with D$_{\rm frac}$$\sim$0.1. But these cores are offset by two dust cores which are reported in \cite{2011MNRAS.415L..49R}. However, because of the spatial resolution limitation, both ortho-NH$_2$D and H$^{13}$CN 1-0 emission can not be distinguished by two cores in our observation, and H42$\alpha$ was not detected (see Figure \ref{app4}). Nevertheless, in the northwest of G023.44-00.18, the obvious NH$_2$D signal was detected with weak H$^{13}$CN 1-0 emission.

\subsubsection{G031.28+00.06}
G031.28+00.06 is located in W43, which is a molecular cloud complex \citep{2011A&A...529A..41N}. As is shown in figure \ref{app5}, ortho-NH$_2$D, H$^{13}$CN 1-0 and H42$\alpha$ were detected in G031.28+00.06. The ortho-NH$_2$D emission presents two structures in this source, while H$^{13}$CN 1-0 and H42$\alpha$ (see figure \ref{app5}) present one structure. Compared with other targets, the ortho-NH$_2$D and H42$\alpha$ emission in G031.28+00.06 are relatively weak. Furthermore, two NH$_2$D structures seem to be isolated by the H42$\alpha$ structure. Besides, the ortho-NH$_2$D and H$^{13}$CN present a different distribution in this source.

\subsubsection{G034.39+00.22}
G034.39+00.22 is an UCH\uppercase\expandafter{\romannumeral2} region with 1000 M$_{\sun}$ mass \citep{1994ApJS...92..173M}. With IRAM 30m observation, \cite{2006ApJ...641..389R} reported 1.2 mm continuum maps in G034.39+00.22 which present north to south distribution. They indicated that nine millimeter-dense clumps existed in G034.39+00.22. In the present work, we focus on the brightest region in G034.39+00.22. The distribution of strong ortho-NH$_2$D and H$^{13}$CN 1-0 emission were detected in G034.39+00.22. These two lines present a north to south distribution (see figure \ref{app6}). The distribution of ortho-NH$_2$D is similar to H$^{13}$CN 1-0 in G034.39+00.22. Both ortho-NH$_2$D and H$^{13}$CN appear two structures and show two peaks in this target. Besides, H42$\alpha$ was not detected in G034.39+00.22.

\subsubsection{G035.19-00.74}
\cite{1984MNRAS.210..173D} reported that G035.19-00.74 was an UCH\uppercase\expandafter{\romannumeral2} region and contained a young B0 star. With James Clerk Maxwell Telescope (JCMT), Berkeley-Illinois-Maryland Association (BIMA) and VLA observation, \cite{2003MNRAS.339.1011G} found CO outflow with $\sim$45$^{\circ}$ along the northeast-southwest direction in G035.19-00.74. In our work, the distribution of strong ortho-NH$_2$D and H$^{13}$CN 1-0 emission were detected in G035.19-00.74 while H42$\alpha$ was not detected (see figure \ref{app7}). The distribution of ortho-NH$_2$D is obviously different from H$^{13}$CN 1-0. The H$^{13}$CN 1-0 distribution seems like a core. However, the distribution of ortho-NH$_2$D is oriented northwest to the southeast which is perpendicular to the CO outflow reported in \cite{2003MNRAS.339.1011G}.

\subsubsection{G035.20-01.73}
G035.20-01.73 (W 48) contains an UCH\uppercase\expandafter{\romannumeral2} region \citep{1989ApJS...69..831W}. The east to west evolutionary gradient was found in the H\uppercase\expandafter{\romannumeral2} region \citep{2014MNRAS.440..427R}. We obtained the H$^{13}$CN 1-0, distributed from east to west with two strong peaks and a weak peak (see figure \ref{app8}). \cite{2011A&A...530A.118P} present the distribution of ortho-NH$_2$D at 85.9 GHz oriented west to east in G035.20-01.73 toward PdBI observations. Combined with the 3.5 mm dust emission map, they found the dust emission peak is not the brightest NH$_2$D peak in this target. H42$\alpha$ is observed in G035.20-01.73, which is near from the center peak of H$^{13}$CN 1-0 distribution (see figure \ref{app8}). The ortho-NH$_2$D was also observed with strong emission (see figure \ref{app8}). The ortho-NH$_2$D, away from the H42$\alpha$, is located in the west of G035.20-01.73 and shows a similar distribution to the result in \cite{2011A&A...530A.118P}.  Besides, the NH$_2$D peak offsets H$^{13}$CN 1-0 peak, while the NH$_2$D distribution is similar to the 3.5 mm dust emission in \cite{2011A&A...530A.118P}.

\subsubsection{G037.43+01.51}
With MPIfR 37-element bolometer array observation, \cite{2002ApJ...566..945B} detected an extended 1.2 mm continuum core. Several deuterated molecules, HDCO and D$_2$CO \citep{2021A&A...653A..45Z} and DC$_3$N \citep{2020MNRAS.496.1990R} were detected in G037.43+01.51. In G037.43+01.51, the relatively weak ortho-NH$_2$D and H$^{13}$CN 1-0 emission were detected, while H42$\alpha$ was not detected (see figure \ref{app9}). The H$^{13}$CN 1-0 emission is similar to the 1.2 mm continuum map reported in \cite{2002ApJ...566..945B}. However, the spatial distribution of weak ortho-NH$_2$D emission appears to deviate from that of H$^{13}$CN 1-0.

\subsubsection{G049.48-00.36}
G049.48-00.36 (W 51 IRS2) is one of the massive protocluster candidates in giant molecular cloud W51 \citep{2012ApJ...758L..29G}. W 51 is a complex structure molecular cloud with several velocity components \citep{1973ApJ...180...31S}. In the present work, the ortho-NH$_2$D and H$^{13}$CN 1-0, H42$\alpha$ (see figure \ref{app10}) maps were obtained in G049.48-00.36. At the map center, the structures of the ortho-NH$_2$D, H$^{13}$CN 1-0 and H42$\alpha$ distribution are close. Although part of ortho-NH$_2$D emission is detected near one of the H42$\alpha$ structures, most ortho-NH$_2$D emission distributions deviate from the H42$\alpha$ emission. With strong emission, the ortho-NH$_2$D, H$^{13}$CN 1-0 and H42$\alpha$ were also detected in the southeast where G049.48-00.38 is located.

\subsubsection{G049.48-00.38}
G049.48-00.38 (W 51 M) is adjacent to G049.48-00.36 and it is also one of the massive protocluster candidates in W 51 \citep{2012ApJ...758L..29G}. As figure \ref{app11} shows, strong ortho-NH$_2$D and H$^{13}$CN 1-0, H42$\alpha$ were observed in this work (see figure \ref{app10}). The ortho-NH$_2$D distribution is similar to H$^{13}$CN 1-0, while the ortho-NH$_2$D and H$^{13}$CN 1-0 emission offset H42$\alpha$ peak. With the PdBI observations, \cite{2017A&A...597A..45V} reported that ortho-NH$_2$D was detected in two marginally resolved sources, while we have detected only one here due to the spatial resolution limitation.

\subsubsection{G075.76+00.33}
With BIMA interferometer observation, \cite{2010MNRAS.404.1449R} mapped HCN 1–0 in G075.76+00.33 (the south one in their observation) and presented northeast to southwest distribution. In the north of G075.76+00.33, the UCH\uppercase\expandafter{\romannumeral2} region 075.78+00.34 was mapped by \cite{1989ApJS...69..831W}. This UCH\uppercase\expandafter{\romannumeral2} region was mapped H42$\alpha$ emission toward G075.76+00.33 observation in the present work (see figure \ref{app12}). An east to west distribution of ortho-NH$_2$D was obtained in G075.76+00.33 (see figure \ref{app12}). The H$^{13}$CN 1-0 was also mapped in G075.76+00.33 (see figure \ref{app12}), and it has a similar distribution to HCN 1-0 emission, which has been observed by \cite{2010MNRAS.404.1449R}. The ortho-NH$_2$D emission is in the south of G075.76+00.33, away from the H42$\alpha$ emission in G075.76+00.33.

\subsubsection{G081.75+00.59}
G081.75+00.59 (DR 21) is a giant star-forming complex in Cygnus X molecular cloud complex \citep{1978ApJ...223..840D,2009ApJ...694.1056K}. It had been mapped to a 4 pc length ridge by Herschel observations\citep{2007A&A...476.1243M,2012A&A...543L...3H} and contains three OH 1665 MHz emissions regions DR 21M, DR 21OH and W 75S \citep{2021MNRAS.501.4825K}. With IRAM observation, \cite{2023pcsf.conf..253C} reported integrated column density map of NH$_2$D, DCN, DNC and DCO$^+$ toward Cygnus-X. In our previous work, this target has been detected the strongest velocity integrated intensity of para-NH$_2$D $1_{11}^a-1_{01}^s$ at 110.1535 GHz with IRAM-30m and the highest deuterated fraction of NH$_3$ (D$_{\rm frac}$=0.044) \citep{2022MNRAS.512.4934L}. In the present work, we mapped DR 21 (OH) and W 75S in DR 21 and we describe them separately in the following two parts.

W75S-FIR1, W75S-FIR2 and W75S-FIR3 embedded several massive YSOs in W 75S \citep{1986ApJ...300..737H,2007A&A...476.1243M}. Near W75S-FIR2 (20$^h$39$^m$02.4$s$ 42$^{\circ}$24$^{\prime}$59$^{\prime\prime}$), the strongest ortho-NH$_2$D emission was detected in the present work. In W 75S, the distribution of ortho-NH$_2$D is slightly similar to H$^{13}$CN 1-0 (see figure \ref{app13}).

DR 21OH contains several high mass protostars \citep{1991ApJ...378..576M}. It is younger than DR 21M because it was undetected in the radio continuum emission \citep{2007MNRAS.374...29D}. In the present work, we only obtained the north part in DR 21OH. The ortho-NH$_2$D and H$^{13}$CN 1-0 emissions are strong in this region (see figure \ref{app13}). The H$^{13}$CN 1-0 presents a strong peak in the southernmost toward our map and ortho-NH$_2$D offset the H$^{13}$CN peak. 

\subsubsection{G081.87+00.78}
G081.87+00.78 (W 75N) is a massive star-forming region in Cygnus X molecular cloud complex and it is in the north of G081.75+00.59 \citep{1978ApJ...223..840D,2006A&A...445..971P}. This source contains a large-scale high-velocity molecular outflow \citep{2003ApJ...584..882S} and a B-type YSO \citep{2004ApJ...601..952S}. In the present work, the ortho-NH$_2$D and H$^{13}$CN 1-0 emission were obtained in G081.87+00.78, while H42$\alpha$ was not detected (see figure \ref{app14}). In G081.87+00.78, the ortho-NH$_2$D is oriented northwest-southeast and deviates the H$^{13}$CN 1-0 peak.

\subsubsection{G109.87+02.11}
G109.87+02.11 (Cep A) is an active massive star-forming region, and it has been detected in several YSOs \citep{1984ApJ...276..204H}. The brightest radio source in G109.87+02.11 is HW2 (where locates at (0$\arcsec$,0$\arcsec$) in our maps), associated with an early-B spectral type zero-age main-sequence (ZAMS) star \citep{1984ApJ...276..204H}. With IRAM 30-m observations, \cite{2017MNRAS.466..248L} observed strong ortho-NH$_2$D emission in the northeast of G109.87+02.11. As is shown in figure \ref{app15}, the H$^{13}$CN 1-0 and strong ortho-NH$_2$D  emission were detected in G109.87+02.11. The distribution of ortho-NH$_2$D is presented in the northeast of G109.87+02.11, and it is similar to the result of \cite{2017MNRAS.466..248L}.  However, the ortho-NH$_2$D is dislocation with the strongest H$^{13}$CN 1-0 emission. The strongest H$^{13}$CN 1-0 emission is situated in proximity to HW2, and the ortho-NH$_2$D emissions appearing to be isolated by HW2. Besides, in this target, the strong ortho-NH$_2$D was detected in the northeast of G109.87+02.11 and H$^{13}$CN was detected relatively weakly.

\subsubsection{G111.54+00.77}
G111.54+00.77 (NGC 7538) is a massive star-forming region which locates in a giant molecular cloud complex \citep{2000ApJ...537..221U}. With Herschel observation, \cite{2013ApJ...773..102F} reported that G111.54+00.77 has several high-mass dense clumps candidates which may produce several intermediate- to high-mass stars in the future. A young O star with 25 M$_{\sun}$ embedded in  NGC 7538 IRS 1 has been observed with a widely distributed molecular outflow \citep{2011ApJ...728....6Q}. In G111.54+00.77, the widely distributed ortho-NH$_2$D and H$^{13}$CN 1-0 distribution were detected (see figure \ref{app16}), and H42$\alpha$ emission was detected in the north where is NGC 7538 IRS 1. A small part of ortho-NH$_2$D emission is near to the H42$\alpha$ emission while a large part appears in the south of G111.54+00.77. In the west of G111.54+00.77, the ortho-NH$_2$D emission is strong while H$^{13}$CN 1-0 emission is weak. 

\subsubsection{G121.29+00.65}
Toward SMA observations, \cite{2019A&A...621A.140J} reported that G121.29+00.65 (L 1287) have a high fragmentation level and a bipolar outflow is oriented northeast to southwest direction. G121.29+00.65 have been detected 6.7 GHz CH$_3$OH, implying that there is forming a massive star\citep{2010A&A...511A...2R}. In G121.29+00.65, the distribution of H$^{13}$CN 1-0 was obtained (see figure \ref{app17}), similar to the distribution of HCN 1-0 which is reported by \cite{1991ApJ...373..137Y}. The distribution of ortho-NH$_2$D is perpendicular to H$^{13}$CN 1-0 and the widely distributed ortho-NH$_2$D is oriented northwest to the southeast. The distribution of ortho-NH$_2$D is also perpendicular to the outflow reported in \cite{2019A&A...621A.140J}. 

\subsubsection{G188.94+00.88}
It was reported that G188.94+00.88 (S 252) includes two massive protostars and is near to the Sh 2-247 which is a complex of H\uppercase\expandafter{\romannumeral2} region and OB stars \citep{2005A&A...429..945M}. Several very weak ortho-NH$_2$D fragments were detected in G188.94+00.88 (see figure \ref{app18}). These ortho-NH$_2$D emissions are marginally detected in this target. Similar to G005.88-00.39, the spatial structure might also not be distinguished in G188.94+00.88 due to the beam size. The main ortho-NH$_2$D emission region is presented in the south of H$^{13}$CN 1-0 distribution. In G188.94+00.88, H42$\alpha$ was not detected.

\section{Discussion}
\label{discussion}

\subsection{Spatial distribution of ortho-NH$_2$D and the dense molecular gas}
\begin{figure}
\centering
\subfigure[]{\includegraphics[width=0.5\textwidth]{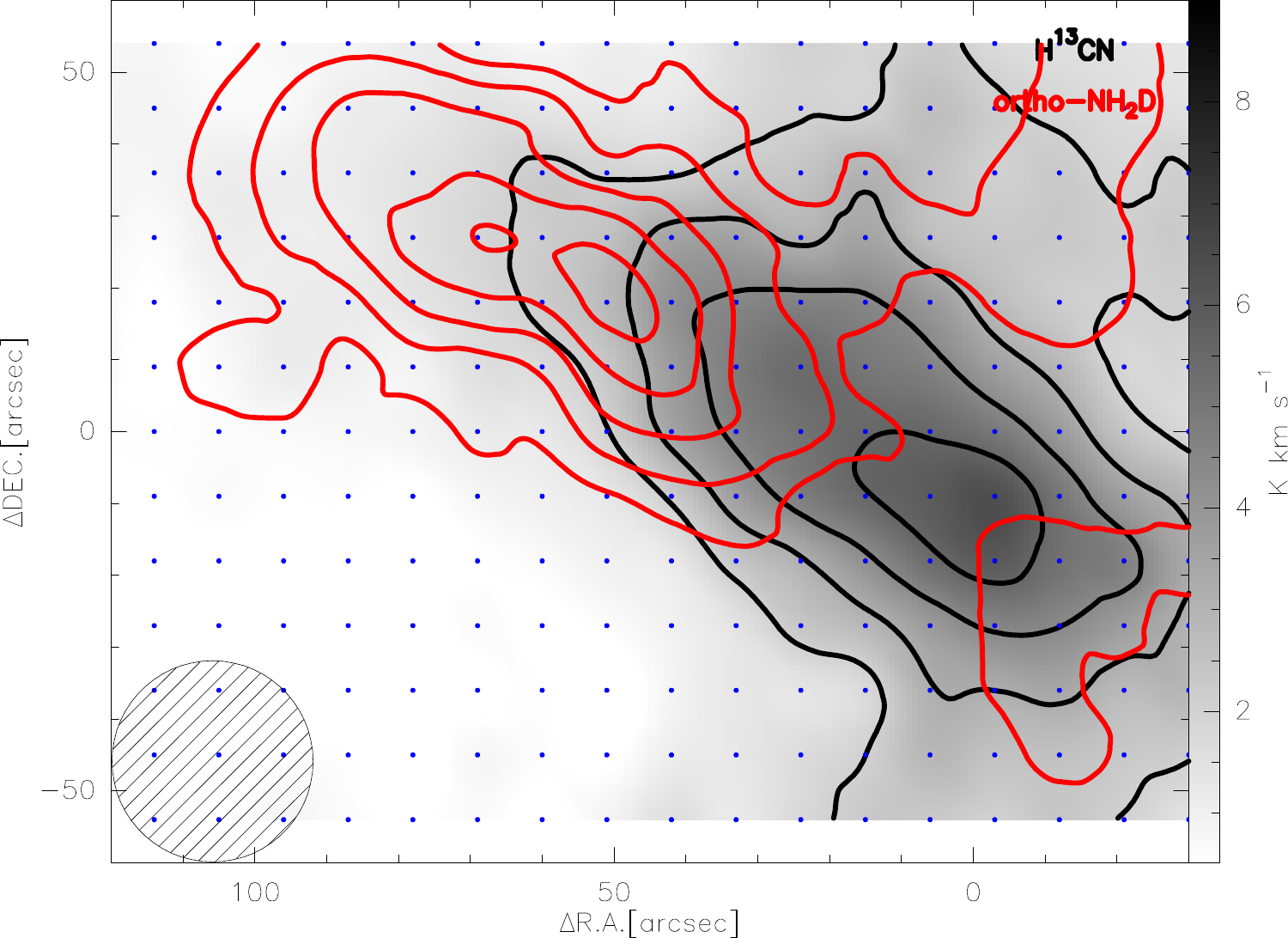}}\subfigure[]{\includegraphics[width=0.5\textwidth]{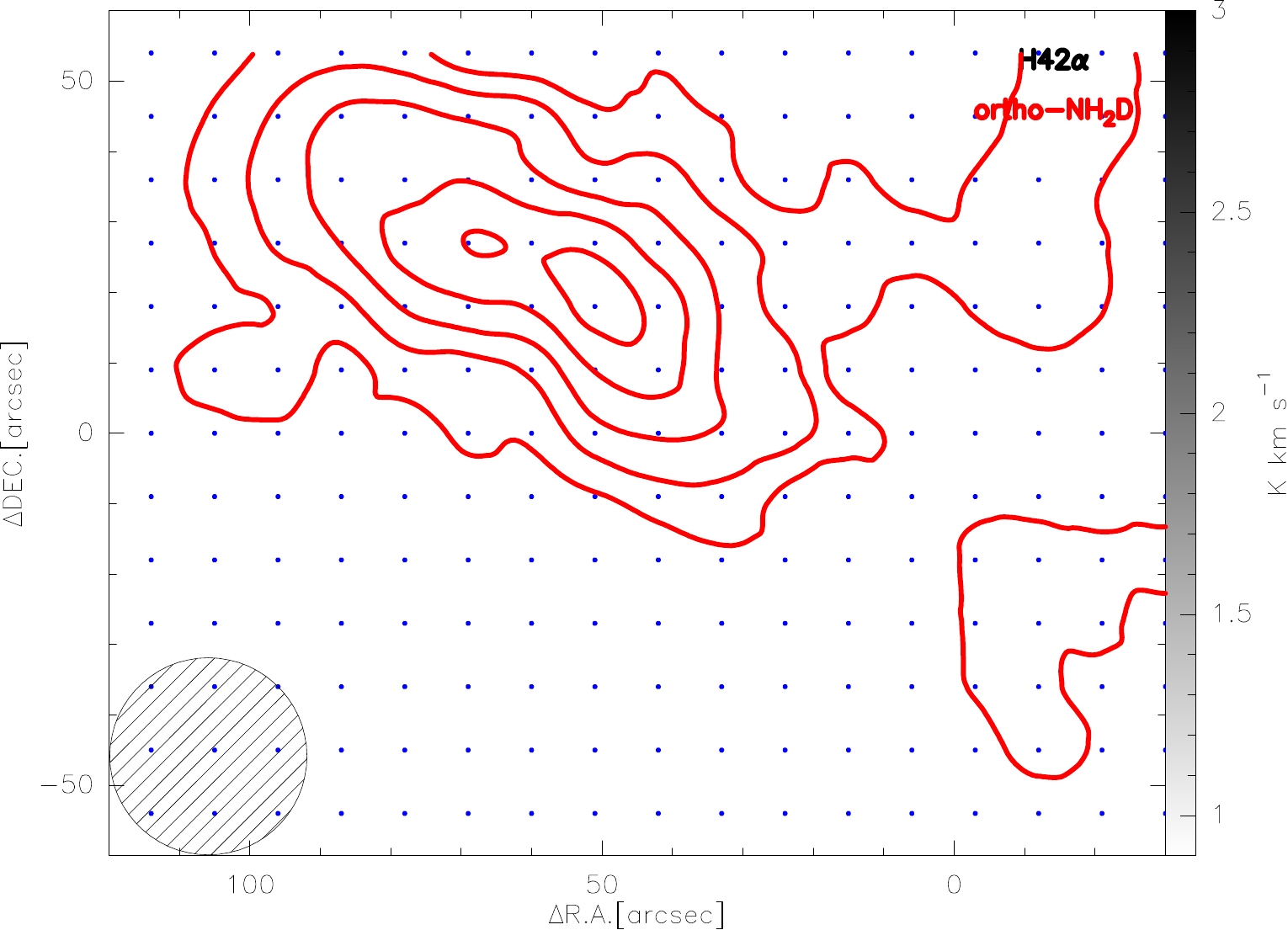}}
\subfigure[]{\includegraphics[width=0.5\textwidth]{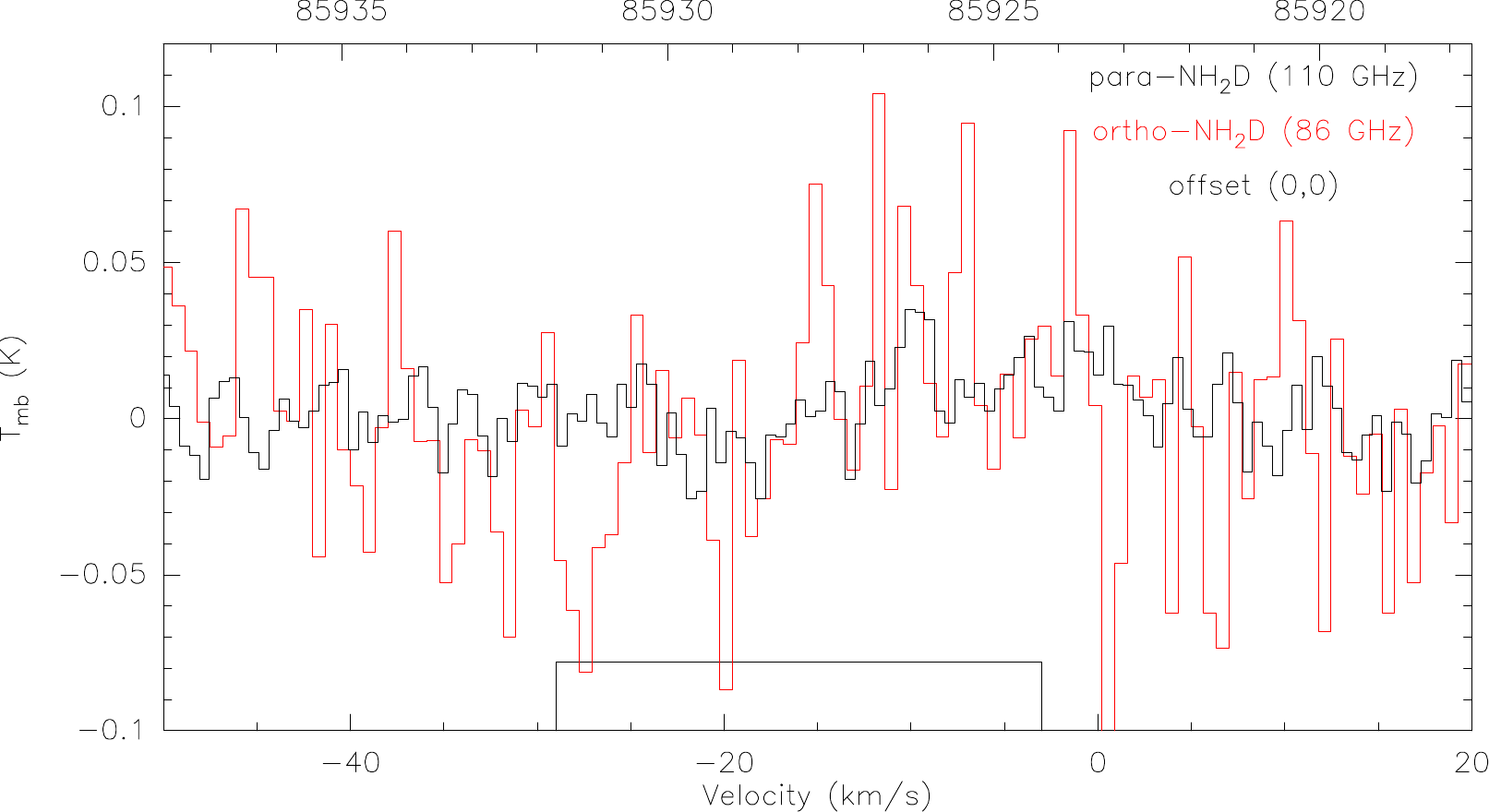}}\subfigure[]{\includegraphics[width=0.5\textwidth]{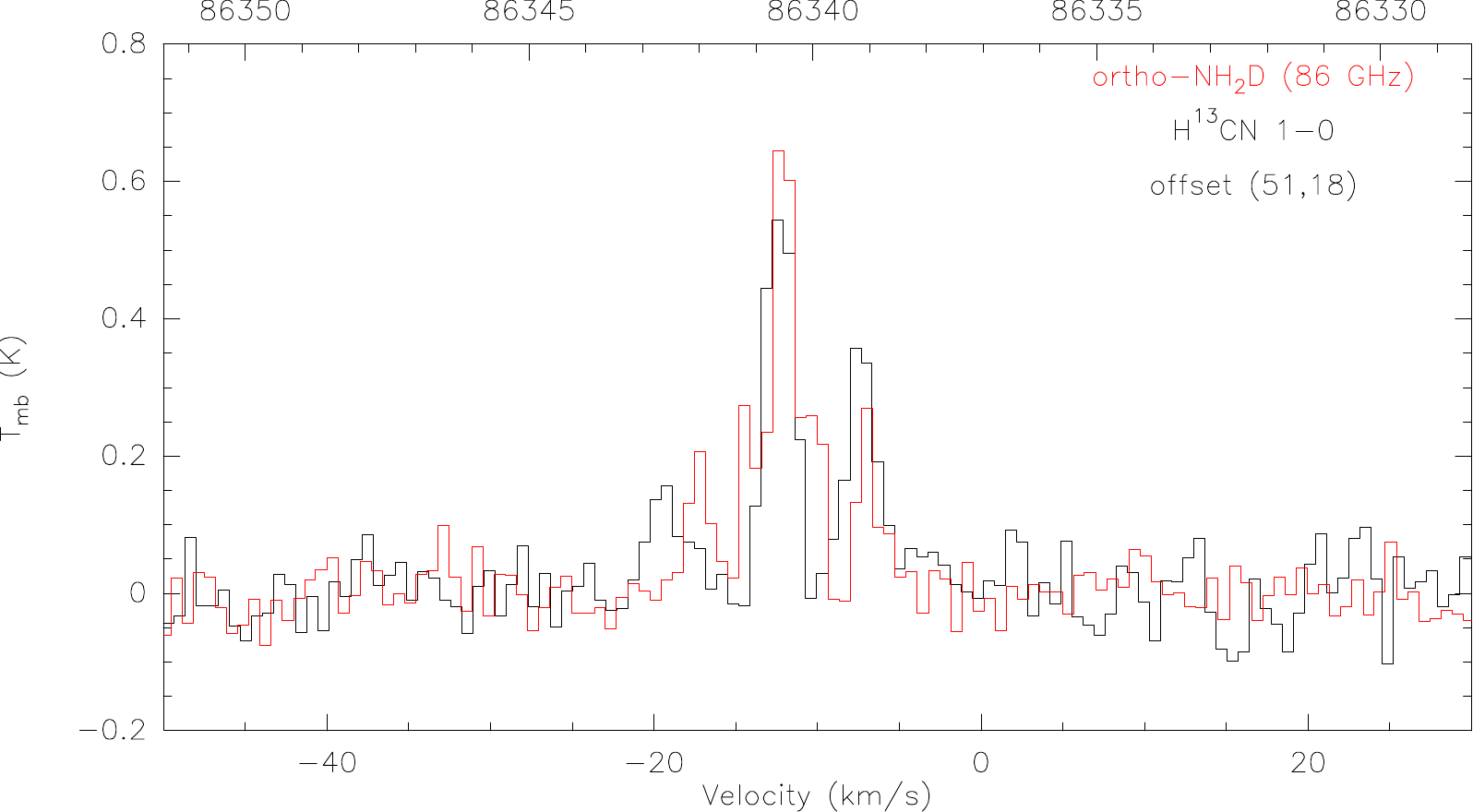}}
\caption{(a): NH$_2$D at 85.9263 GHz velocity integrated intensity contour (red contour) overlaid on H$^{13}$CN 1-0 velocity integrated intensity image (Gray scale and black contour) in G109.87+02.11. The contour levels start at 3$\sigma$ in steps of 3$\sigma$ for NH$_2$D, while the contour levels start at 12$\sigma$ in steps of 7$\sigma$ for H$^{13}$CN 1-0. The gray scale starts at 3$\sigma$. (b): NH$_2$D at 85.9263 GHz velocity integrated intensity contour (red contour) overlaid on H42$\alpha$ velocity integrated intensity image (Gray scale and black contour, while the H42$\alpha$ is not detected) in G109.87+02.11. The contour levels start at 3$\sigma$ in steps of 3$\sigma$ for NH$_2$D. (c): Spectra of ortho-NH$_2$D at 85.9263 GHz and para-NH$_2$D at 110.1535 GHz in G109.87+02.11. The para-NH$_2$D was observed by IRAM-30m with position-switching mode. The red spectra is ortho-NH$_2$D and the black is para-NH$_2$D. The offset for two spectra is (0$^{\prime \prime}$,0$^{\prime \prime}$). The black box in the picture indicates the range of flux integration of ortho-NH$_2$D at 85.9263 GHz. (d): Spectra of ortho-NH$_2$D at 85.9263 GHz and H$^{13}$CN 1-0 in G109.87+02.11. The red spectra is ortho-NH$_2$D and the black is H$^{13}$CN 1-0. The offset for two spectra is (+51$^{\prime \prime}$,+18$^{\prime \prime}$).}
\label{app15}
\end{figure}

 As shown in figure \ref{app1}-\ref{app18} and Section \ref{Individual target detail}, most of the targets manifest that the NH$_2$D emission deviates the H$^{13}$CN 1-0 in the present work, such as G109.87+00.78 (see Figure \ref{app15}). Moreover, five sources, G031.28+00.06, G035.19-00.74, G081.87+00.78, G109.87+02.11 and G121.29+00.65 (see Figure \ref{app5}, Figure \ref{app7}, Figure \ref{app14}, Figure \ref{app15} and Figure \ref{app17}, respectively), present that the strong NH$_2$D emissions appear in some regions where the H$^{13}$CN 1-0 emission is weak or even is marginally detected.

 In four sources, G011.91-00.61, G023.44-00.18, G034.39+00.22 and G049.48-00.38, the distribution of strong ortho-NH$_2$D emission is similar to H$^{13}$CN 1-0 (see Figure \ref{app2}, Figure \ref{app4}, Figure \ref{app6} and Figure \ref{app11}, respectively). However, the ortho-NH$_2$D is not detected asymmetrically distributed spatial structure in G011.91-00.61, G023.44-00.18, G034.39+00.22 and G049.48-00.38. Their substructure is unresolved. Therefore, we can not confirm whether NH$_2$D emission and H$^{13}$CN 1-0 have similar distributions in these sources or not.

Without asymmetrically distributed spatial structures detected and unresolved sources, only G081.75+00.59 presents a relatively similar distribution between NH$_2$D and H$^{13}$CN 1-0. There may exist contingencies in this distribution.

 As described in Section \ref{H13CN 1-0 result}, H$^{13}$CN 1-0 is used as a dense gas tracer \citep{2004ApJ...606..271G,2010ApJS..188..313W,2011A&A...527A..88V} in the present work to analyze the distribution of NH$_2$D and dense gas in massive star-forming late stage.

\cite{2007A&A...470..221C} performed observations of para-NH$_2$D $1_{11}^a-1_{01}^s$ at 110.1535 GHz toward low-mass starless core L 1544 with PdBI and they found that NH$_2$D intensity peaks were at 1.2 mm continuum dust peak. Therefore, they suggested that NH$_2$D is a good tracer of the high density gas in the low-mass starless core. However, toward the gas-phase model, \cite{2000A&A...364..780R} simulated that gas density seems to have no effect on D$_{\rm frac}$(NH$_3$).

 In present work, NH$_2$D seems to be enhanced in G031.28+00.06, G035.19-00.74, G081.87+00.78, G109.87+02.11 and G121.29+00.65, at relatively low gas density part (see Figure \ref{app5}, Figure \ref{app7}, Figure \ref{app14}, Figure \ref{app15} and Figure \ref{app17}, respectively).  Therefore, such results indicate that the gas density may not be an important physical parameter to influence the abundance of NH$_2$D, which is opposite to that of \cite{2007A&A...470..221C} and agrees with the gas-phase model of \cite{2000A&A...364..780R}.
 Compared with other physical parameters (such as temperature), the gas density might not be the important physical parameter to affect the abundance of NH$_2$D in the late-stage massive star-forming regions. But without any information of NH$_3$, we can not confirm that the D$_{\rm frac}$(NH$_3$) is affected by dense gas. More detail should be obtained with the information of NH$_3$.

\subsection{Spatial distribution of ortho-NH$_2$D and massive star formation}
\label{Spatial distribution of ortho-NH$_2$D and massive star forming}
To analyze whether NH$_2$D is affected by the formed massive young stellar objects or not, we compare the distribution between ortho-NH$_2$D and H42$\alpha$. In the present work,  ortho-NH$_2$D and H\uppercase\expandafter{\romannumeral2} regions present a different distribution in most targets. Both ortho-NH$_2$D and H42$\alpha$ emissions were detected in nine targets. In five sources, G015.03-00.67, G031.28+00.06, G035.20-01.73, G075.76+00.33 and G111.54+00.77, the NH$_2$D emission are away from H\uppercase\expandafter{\romannumeral2} regions (see Figure \ref{app3}, Figure \ref{app5}, Figure \ref{app8}, Figure \ref{app12} and Figure \ref{app16}, respectively). In these five H\uppercase\expandafter{\romannumeral2} regions, the NH$_2$D may be depleted during the massive stars forming. 

 For G049.48-00.36 and G049.48-00.38, the relatively strong ortho-NH$_2$D emission was detected in H\uppercase\expandafter{\romannumeral2} regions (see Figure \ref{app10} and Figure \ref{app11}). The structure in W 51 is complex \citep{1973ApJ...180...31S}, we can not confirm if that the ortho-NH$_2$D and H42$\alpha$ emission come from the same structure. But it is obvious that ortho-NH$_2$D emission is away from H42$\alpha$ emission, especially the strong H42$\alpha$ emission and the center of the H\uppercase\expandafter{\romannumeral2} region.

The ortho-NH$_2$D presents two structures in G011.91-00.61 (see Figure \ref{app2}). One is the northeast structure which is similar to the distribution of H42$\alpha$ emission; another is the southwest structure without H42$\alpha$ emission. However, the ortho-NH$_2$D emissions in the southwest structure are stronger and more extensive than emissions in the northeast structure. The reason may be that the massive star-forming stage in the northeast part is later than the southwest structure, and as a result, NH$_2$D in the northeast part is much more depleted than NH$_2$D in the southwest part.

In cold molecular gas, there is an active process of deuterated molecules forming \citep{2012A&ARv..20...56C}. And during this process, several neutral species are frozen-out onto dust grains and molecules are deuterated on the surface of dust grains \citep{2012A&ARv..20...56C}. Moreover, according to the simulation, \cite{2008A&A...492..703C} suggested that the deuterium fractionation of H$^+_3$ (which means the abundance ratio of H$^+_3$ to H$_2$D$^+$) dropped rapidly when the gas kinetic temperature is above $\sim$ 15 K. The formation process of NH$_2$D is related to H$_2$D$^+$ \citep{2009A&A...493...89E}. It may indicate that the $D_{\rm frac}$(NH$_3$) decreases with temperature increasing when the temperature is relatively high. Besides, with PdBI observation toward the massive star-forming region IRAS 20293+3952, \cite{2010A&A...517L...6B} reported that they found strong NH$_2$D emission in starless cores and did not detect NH$_2$D near YSOs. Therefore, according to previous research of observation and theory \citep[e.g.][]{2008A&A...492..703C,2009A&A...493...89E,2010A&A...517L...6B,2012A&ARv..20...56C}, the abundance of NH$_2$D should decrease with the evolution of massive star formation.

Without H42$\alpha$ detected, these nine sources present both strong and weak NH$_2$D emission. In two sources, G037.43+01.51 and G188.94+00.88, the velocity integrated intensity of ortho-NH$_2$D is lower than 1 K km/s (see Figure \ref{app9} and Figure \ref{app18}). In G037.43+01.51, the NH$_3$ column density has been derived with VLA observation \citep{2014ApJ...790...84L}. Compared with other sources, this target presents low column density, such as G034.39+00.22 whose column density is one order of magnitude higher than G037.43+01.51 \citep{2014ApJ...790...84L}. Similarly, the NH$_3$ column density of G188.94+00.88 is also lower than other sources, such as G081.75+00.59 whose column density is one order of magnitude higher than G188.94+00.88 with Effeslsberg observation \citep{1999ApJS..125..161J}. With low NH$_3$ column density, the NH$_2$D might generate relatively less in these two sources. 

The NH$_2$D seems to be depleted near the H\uppercase\expandafter{\romannumeral2} regions which indicates a later evolution stage. Furthermore, in the region without H42$\alpha$ emission, the strong ortho-NH$_2$D emission is detected. Therefore, combining with previous researches \citep[e.g.][]{2008A&A...492..703C,2009A&A...493...89E,2010A&A...517L...6B,2012A&ARv..20...56C}, we suggest that the abundance of NH$_2$D may be affected by the evolution of massive star formation.

\subsection{NH$_2$D enhancement}
\begin{figure}
\centering
\includegraphics[width=0.5\textwidth]{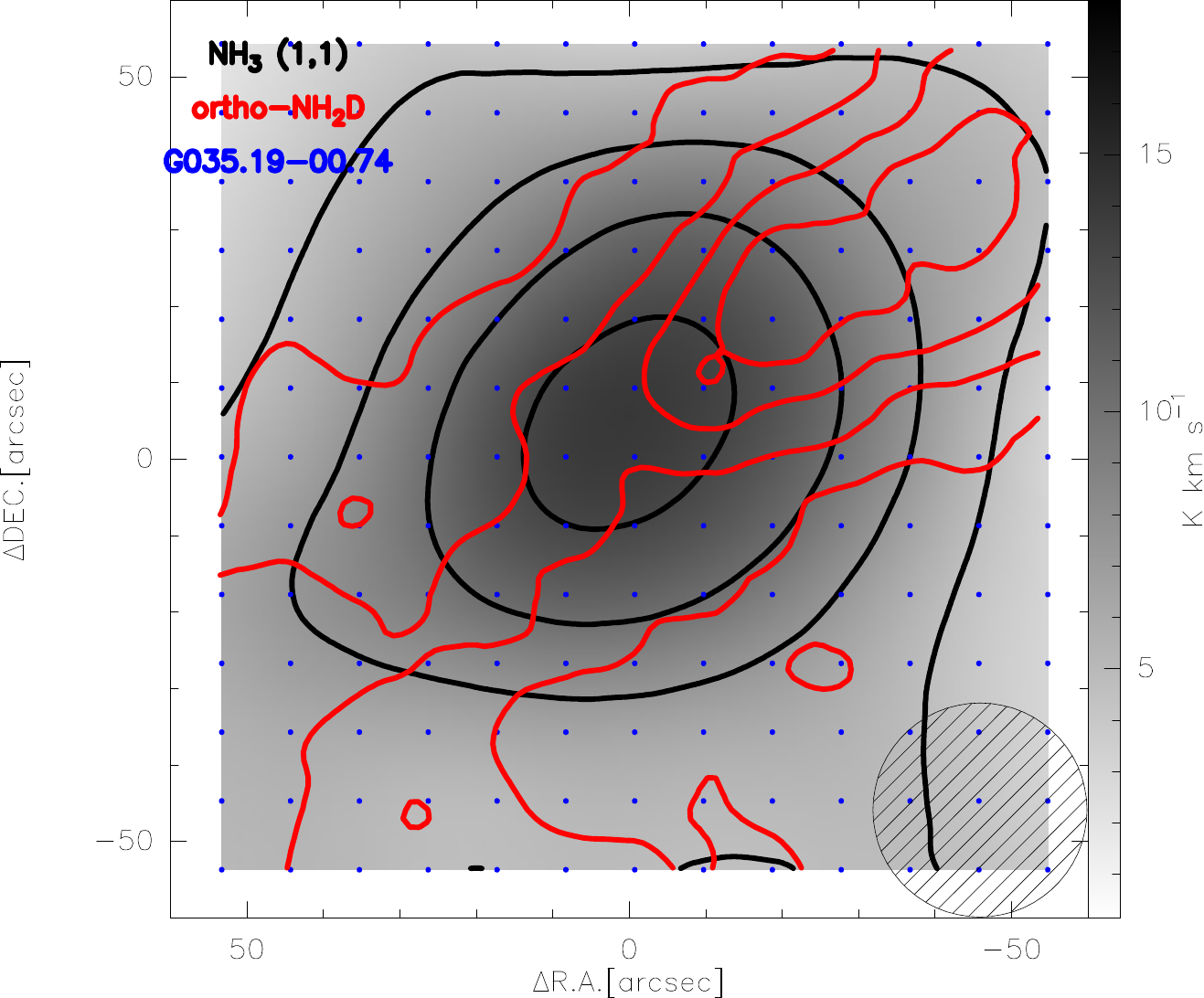}\includegraphics[width=0.5\textwidth]{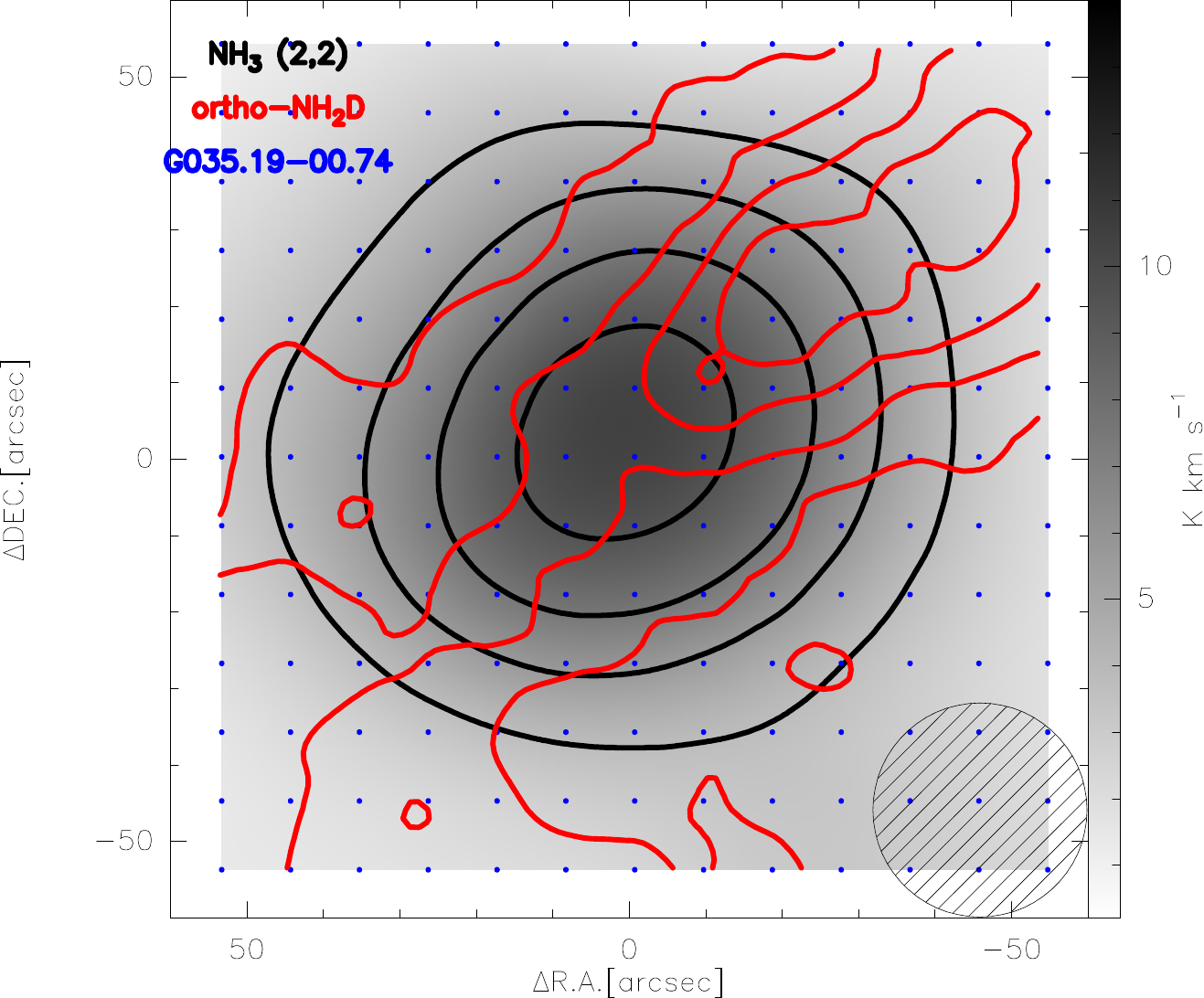}
\caption{(a): NH$_2$D at 85.926 GHz velocity integrated intensity contour (red contour) overlaid on NH$_3$(1,1) main group velocity integrated intensity image (Gray scale and black contour) in G035.19-00.74. The contour levels start at 5$\sigma$ in steps of 5$\sigma$ for NH$_2$D, while the contour levels start at 90$\sigma$ in steps of 54$\sigma$ for NH$_3$(1,1) main group. The gray scale starts at 3$\sigma$. (b): NH$_2$D at 85.926 GHz velocity integrated intensity contour (red contour) overlaid on NH$_3$(2,2) main group velocity integrated intensity image (Gray scale and black contour) in G035.19-00.74. The contour levels start at 5$\sigma$ in steps of 5$\sigma$ for NH$_2$D, while the contour levels start at 54$\sigma$ in steps of 27$\sigma$ for NH$_3$(2,2) main group. The gray scale starts at 3$\sigma$.}
\label{G035_NH3}
\end{figure}

With the Green Bank Telescope and the K-Band Focal Plane Array observation, \cite{2015MNRAS.452.4029U} have mapped the NH$_3$ (1,1) and NH$_3$ (2,2) inversion emission toward 66 massive star-forming regions, identified by the Red Midcourse Space Experiment Source survey \citep{2013ApJS..208...11L}. These 66 massive star-forming regions include several of our targets, such as G035.19-00.74 (see figure \ref{G035_NH3}). In G035.19-00.74, strong NH$_2$D emissions present a northwest to southeast distribution, while the northwest and the southeast regions present weak H$^{13}$CN 1-0 emission (see figure \ref{app7}), these regions might be devoid of dense molecular gases. However, in the weak H$^{13}$CN 1-0 region, NH$_3$(1,1) and NH$_3$(2,2) were detected. Therefore, the molecular gases that exist in these regions and their density is relatively low.

The obvious NH$_2$D enrichment is presented in the southeast and northwest of G035.19-00.74. In particular, in the northwest of G035.19-00.74, the very high deuterium fractionation of NH$_3$ is detected. Compared with the H$^{13}$CN emission in G035.19-00.74, the deuterium fractionation of NH$_3$ is either not affected by gas density. Moreover, combining with the ammonia data, we could obtain the distribution of the deuterium fractionation in the following work.

\section{Summary and conclusion remarks}
\label{summary}
Using of the IRAM-30m, we carried the ortho-NH$_2$D $1_{11}^s-1_{01}^a$ at 85.9263 GHz mapping toward a sample of 24 late stage massive star-forming. The ortho-NH$_2$D $1_{11}^s-1_{01}^a$ at 85.9263 GHz was detected in 18 targets. H$^{13}$CN 1-0 at 86.3387 GHz was also detected in these 18 targets, while H42$\alpha$ at 85.6884 GHz was detected in nine targets.

 11 of 18 sources, present a different distribution between ortho-NH$_2$D $1_{11}^s-1_{01}^a$ and H$^{13}$CN 1-0 with asymmetrically and resolvable distributed spatial structure. For the other 7 sources, no significant difference between these two lines can be found. This is mainly due to the limited spatial resolution and sensitivity.

Our main results include:

\begin{enumerate}[1.]
	\item The ortho-NH$_2$D $1_{11}^s-1_{01}^a$ emissions present a complex distribution, which is different from other molecule transitions, such as H$^{13}$CN 1-0. 
	\item Most of the targets show the different distribution between ortho-NH$_2$D $1_{11}^s-1_{01}^a$ and H$^{13}$CN 1-0.  Compared with other physical parameters (such as temperature), therefore, the dense gas (or gas density) may not be an important physical parameter to affect the NH$_2$D enhancement in these targets. 
	\item The ortho-NH$_2$D $1_{11}^s-1_{01}^a$ emissions are away from the H42$\alpha$. Combining with previous researches \citep[e.g.][]{2008A&A...492..703C,2009A&A...493...89E,2010A&A...517L...6B,2012A&ARv..20...56C}, the abundance of NH$_2$D may be affected during the massive star-forming.
\end{enumerate}

\section*{Acknowledgements}
This work is supported by the National Key R$\&$D Program of China (No. 2022YFA1603101)  and the National Natural Science Foundation of China (NSFC, Grant No. 12173067). This study is based on observations carried out under project numbers 042-19, 147-19, and 127-20 with the IRAM 30-m telescope. IRAM is supported by INSU/CNRS (France), MPG (Germany) and IGN (Spain).  

\section*{Data availability}
The original data observed with IRAM 30 meter can be accessed by IRAM archive system at \url{https://www.iram-institute.org/EN/content-page-386-7-386-0-0-0.html}. If anyone is interested in the reduced data presented in this paper, please contact  Junzhi Wang at junzhiwang@gxu.edu.cn.

\clearpage
\onecolumn 
\centering

\begin{longtable}{lcccccccc}
\caption{Source List}
\label{source}\\
\hline
\hline
source name      & RA(J2000)     & DEC(J2000)       & D$_{\rm GC}$  & $v_{\rm LSR}$   &Size\\
                 & hh:mm:ss      & dd:mm:ss         & kpc       & km s$^{-1}$ &    \\
\hline
\endfirsthead
\endhead
\hline
\endfoot
\endlastfoot
G005.88-00.39  & 18:00:30.31  & -24:04:04.50    &  5.3 & 9     & 2$'\times2'$ \\
G009.62+00.19 &  18:06:14.66 & -20:31:31.70 &  3.3 & 2 & 2$'\times2'$ \\
G011.49-014.8 & 18:16:22.13 &  -19:41:27.20 & 7.1 & 11 & 2$'\times2'$ \\
G011.91-00.61  & 18:13:59.72  & -18:53:50.30    &  5.1 & 37    & 3$'\times3'$ \\
G012.80-00.20 & 18:14:14.23 & -17:55:40.50 & 5.5 & 34 & 4$'\times4'$ \\
G015.03-00.67  & 18:20:22.01  & -16:12:11.30    &  6.4 & 22    & 4$'\times4'$ \\
G023.44-00.18  & 18:34:39.29  & -08:31:25.40    &  3.7 & 97    & 2$'\times2'$ \\
G031.28+00.06  & 18:48:12.39  & -01:26:30.70   &  5.2 & 109   & 2$'\times2'$ \\
G034.39+00.22  & 18:53:19.00  & +01:24:50.80    &  7.1 & 57    & 2$'\times2'$ \\
G035.19-00.74  & 18:58:13.05  & +01:40:35.70   &  6.6 & 30    & 2$'\times2'$ \\
G035.20-01.73  & 19:01:45.54  & +01:13:32.50   &  5.9 & 42    & 3$'\times2'$ \\
G043.16+00.01 & 19:10:13.41 & +09:06:12.80 & 7.6 & 10 & 2$'\times2'$ \\
G037.43+01.51  & 18:54:14.35  & +04:41:41.70   &  6.9 & 41    & 2$'\times2'$ \\
G049.48-00.36  & 19:23:39.82  & +14:31:05.00   &  6.3 & 56    & 2$'\times2'$ \\
G049.48-00.38  & 19:23:43.87  & +14:30:29.50   &  6.3 & 58    & 2$'\times2'$ \\
G075.76+00.33  & 20:21:41.09  & +37:25:29.30   &  8.2 & -9    & 2$'\times2'$ \\
G081.75+00.59  & 20:39:01.99  & +42:24:59.30   &  8.2 & -3    & 2$'\times3'$ \\
G081.87+00.78  & 20:38:36.43  & +42:37:34.80   &  8.2 & 7     & 2$'\times2'$ \\
G109.87+02.11  & 22:56:18.10  & +62:01:49.50   &  8.6 & -7    & 2.5$'\times2'$ \\
G111.54+00.77  & 23:13:45.36  & +61:28:10.60    &  9.6 & -57   & 2$'\times4'$ \\
G121.29+00.65  & 00:36:47.35  & +63:29:02.20   &  8.8 & -23   & 2$'\times2'$ \\
G133.94+01.06 & 02:27:03.82 & +61:52:25.20 & 9.8 & -47  & 2$'\times2'$ \\
G188.94+00.88  & 06:08:53.35  & +21:38:28.70    &  10.4 & 8     & 2$'\times2'$ \\
G192.60-00.04  & 06:12:54.02  & +17:59:47.3    &  9.9 & 6     & 2$'\times2'$ \\
\hline
\end{longtable}

\begin{longtable}{lccc}
\caption{Parameters of ortho-NH$_2$D $1_{11}^s-1_{01}^a$ hyperfine structure}
\label{NH2D HfS}\\
\hline
\hline
F$^{\prime\prime}$-F$^{\prime\prime}$ & frequency (MHz) & relative intensity \\
\hline
0-1 & 85924.7829 & 0.111 \\
2-1 & 85925.7031 & 0.139 \\
2-2 & 85926.2703 & 0.417 \\
1-1 & 85926.3165 & 0.083 \\
1-2 & 85926.8837 & 0.139 \\
1-0 & 85927.7345 & 0.111
\\ \hline
\end{longtable}

\newpage

\begin{longtable}{lccc}
\caption{NH$_2$D velocity integrated intensity}
\label{NH2D}\\
\hline
\hline
\multirow{2}{*}{Source} & $\int T_{\rm mb}$(85.9 GHz)d$\nu$ & $\int T_{\rm mb}$(110 GHz)d$\nu$ &  \multirow{2}{*}{$\frac{\int T_{\rm mb}(\rm 85.9 GHz)d\nu}{\int T_{\rm mb}(\rm 110 GHz)d\nu}$}\\
                        & K km s$^{-1}$                     & K km s$^{-1}$                                                                  \\ 
\hline
\endfirsthead
\endhead
\hline
\endfoot
\endlastfoot
G005.88-00.39           & 0.33$\pm$0.15                     & 0.37$\pm$0.07        & 0.89$\pm$0.44 \\
G011.91-00.61           & 2.24$\pm$0.14                     & 1.08$\pm$0.05        & 2.07$\pm$0.16 \\
G015.03-00.67           & $\le$0.509                            & $\le$0.157               & ...\\
G023.44-00.18           & 1.52$\pm$ 0.14                    & 0.75$\pm$0.05        & 2.03$\pm$0.23 \\
G031.28+00.06          & $\le$0.280                            & 0.18$\pm$0.04        & ...\\
G034.39+00.22          & 5.11$\pm$ 0.18                    & 1.73$\pm$0.03        & 2.95$\pm$0.12 \\
G035.19-00.74           & 1.29$\pm$0.16                     & 0.57$\pm$0.04        & 2.26$\pm$0.32 \\
G035.20-01.73           & 0.52$\pm$0.11                      & 0.11$\pm$0.04        & 4.73$\pm$1.99 \\
G037.43+01.51          & 0.31$\pm$0.11                      & 0.18$\pm$0.02        & 1.72$\pm$0.64 \\
G049.48-00.36           & 1.06$\pm$0.17                      & 0.55$\pm$0.05        & 1.93$\pm$0.36 \\
G049.48-00.38           & 2.92$\pm$0.14                      & 2.31$\pm$0.06        & 1.26$\pm$0.07 \\
G075.76+00.33          & $\le$0.305                             & $\le$0.096                & ...\\
G081.75+00.59          & 5.98$\pm$0.07                      & 4.10$\pm$0.03        & 1.45$\pm$0.02 \\
G081.87+00.78          & 0.41$\pm$0.14                      & 0.31$\pm$0.03        & 1.32$\pm$0.47 \\
G109.87+02.11           & $\le$0.472                            & 0.08$\pm$0.02        & ...\\
G111.54+00.77           & $\le$0.881                             & 0.20$\pm$0.02        &  ...\\
G121.29+00.65          & 1.59$\pm$0.14                       & 0.91$\pm$0.02        & 1.75$\pm$0.16 \\
G188.94+00.88          & $\le$0.257                              & 0.14$\pm$0.02        & ...\\ \hline
\end{longtable}

\begin{longtable}{llllll}
\caption{NH$_2$D, H$^{13}$CN and H42$\alpha$ distribution information}
\label{distribution}\\
\hline
\hline
source        &  \begin{tabular}[c]{@{}l@{}}$\int$T$_{\rm mb}$(NH$_2$D)d$\nu$  \\ K km s$^{-1}$\end{tabular}    & NH$_2$D direction    & \begin{tabular}[c]{@{}l@{}}obvious \\ structure\end{tabular} & \begin{tabular}[c]{@{}l@{}}similar with \\ H$^{13}$CN\end{tabular} & \begin{tabular}[c]{@{}l@{}}detected \\ H42$\alpha$\end{tabular} \\ \hline
G005.88-00.39 & 0.60$\pm$0.16              &            &                                                              &                                                               & \checkmark                                                          \\
G011.91-00.61 & 2.33$\pm$0.15              & northeast-southwest & \checkmark                                                          & \checkmark                                                           & \checkmark                                                          \\
G015.03-00.67 & 1.11$\pm$0.18              & fragmentization     &                                                              &                                                               & \checkmark                                                          \\
G023.44-00.18 & 2.31$\pm$0.13              & disciform      & \checkmark                                                          &  \checkmark                                                     &                                                                     \\
G031.28+00.06 & 0.86$\pm$0.11              & northeast-southwest & \checkmark                                                          &                                                               & \checkmark                                                          \\
G034.39+00.22 & 5.11$\pm$0.18             & north-south         & \checkmark                                                          & \checkmark                                                           &                                                              \\
G035.19-00.74 & 3.05$\pm$0.11              & northwest-southeast & \checkmark                                                          &                                                               &                                                              \\
G035.20-01.73 & 4.06$\pm$0.11              & west-east           & \checkmark                                                          &                                                               & \checkmark                                                          \\
G037.43+01.51 & 0.60$\pm$0.11              & irregular           &                                                                     &                                                               &                                                              \\
G049.48-00.36 & 2.45$\pm$0.17              & complex structure   & \checkmark                                                          &                                                               & \checkmark                                                          \\
G049.48-00.38 & 2.92$\pm$0.14              & complex structure   & \checkmark                                                          & \checkmark                                                           & \checkmark                                                          \\
G075.76+00.33 & 1.27$\pm$0.17              & west-east           &     \checkmark                                                                  &                                                               & \checkmark                                                          \\
G081.75+00.59 & 6.95$\pm$0.14              & north-south         & \checkmark                                                          & \checkmark                                                           &                                                              \\
G081.87+00.78 & 1.70$\pm$0.13              & northwest-southeast & \checkmark                                                          &                                                               &                                                              \\
G109.87+02.11 & 2.51$\pm$0.17              & northeast-southwest & \checkmark                                                          &                                                               &                                                              \\
G111.54+00.77 & 1.86$\pm$0.16              & complex structure   & \checkmark                                                          &                                                               & \checkmark                                                          \\
G121.29+00.65 & 2.22$\pm$0.26              & northwest-southeast & \checkmark                                                          &                                                               &                                                              \\
G188.94+00.88 & 0.86$\pm$0.16              &                     &                                                              &                                                               &                                                              \\ \hline
\end{longtable}

\bsp	

\label{lastpage}
\end{document}